\documentclass[traditabstract]{aa} 

\usepackage{graphicx}
\usepackage{txfonts}
\usepackage{longtable}
\usepackage{lscape}
\usepackage{hyperref}
\hypersetup{colorlinks=true, citecolor=blue}
\begin{document}

   \title{A Virgo Environmental Survey Tracing Ionised Gas Emission (VESTIGE).XIV. The main sequence relation in a rich environment down to $M_{star}$ $\simeq$ 10$^6$ M$_{\odot}$\thanks{Based on observations obtained with
   MegaPrime/MegaCam, a joint project of CFHT and CEA/DAPNIA, at the Canada-French-Hawaii Telescope
   (CFHT) which is operated by the National Research Council (NRC) of Canada, the Institut National
   des Sciences de l'Univers of the Centre National de la Recherche Scientifique (CNRS) of France and
   the University of Hawaii.}
      }
   \subtitle{}
  \author{A. Boselli\inst{1,**},  
 	  M. Fossati\inst{2},
	  J. Roediger\inst{3},
	  M. Boquien\inst{4},
	  M. Fumagalli\inst{2},
	  M. Balogh\inst{5},
	  S. Boissier\inst{1}, 
	  J. Braine\inst{6},
	  L. Ciesla\inst{1},
          P. C{\^o}t{\'e}\inst{3},
          J.C. Cuillandre\inst{7},
          L. Ferrarese\inst{3},
	  G. Gavazzi\inst{2},
          S. Gwyn\inst{3}, 
	  Junais\inst{8},
	  G. Hensler\inst{9},
          A. Longobardi\inst{2},
	  M. Sun\inst{10}
       }

\institute{     
                Aix Marseille Univ, CNRS, CNES, LAM, Marseille, France\thanks{Scientific associate INAF at the Osservatorio Astronomico di Brera, Italy}
                \email{alessandro.boselli@lam.fr}
        \and
		Universit\'a di Milano-Bicocca, piazza della scienza 3, 20100 Milano, Italy
	\and
                National Research Council of Canada, Herzberg Astronomy and Astrophysics, 5071 West Saanich Road, Victoria, BC, V9E 2E7, Canada
	\and	
		Centro de Astronomi\'a (CITEVA), Universidad de Antofagasta, Avenida Angamos 601, Antofagasta, Chile
	\and
		Waterloo Centre for Astrophysics, University of Waterloo, Waterloo, Ontario, N2L3G1, Canada
	\and
		Laboratoire d'Astrophysique de Bordeaux, Univ. Bordeaux, CNRS, B18N, all\'ee Geoffroy Saint-Hilaire, 33615 Pessac, France	
        \and        
		AIM, CEA, CNRS, Universit\'e Paris-Saclay, Universit\'e Paris Diderot, Sorbonne Paris Cit\'e, Observatoire de Paris, PSL University, F-91191 Gif-sur-Yvette Cedex, France
	\and
		National Centre for Nuclear Research, Pasteura 7, PL-02-093 Warsaw, Poland
	\and
		Department of Astrophysics, University of Vienna, T\"urkenschanzstrasse 17, 1180 Vienna, Austria
	\and
		Department of Physics \& Astronomy, University of Alabama in Huntsville, 3001 Sparkman Drive, Huntsville, AL 35899, USA
                 }

\authorrunning{Boselli et al.}
\titlerunning{VESTIGE}

   \date{}

 
  \abstract  
{Using a compilation of H$\alpha$ fluxes for 384 star forming galaxies detected during the Virgo Environmental Survey Tracing Ionised Gas Emission (VESTIGE), 
we study several important scaling relations linking
the star formation rate, specific star formation rate, stellar mass, stellar mass surface density, and atomic gas depletion timescale for a complete sample of galaxies in a rich environment.  
The extraordinary sensitivity of the narrow-band imaging data allows us to sample the whole dynamic range of the H$\alpha$ luminosity function, from massive galaxies 
($M_{star}$ $\simeq$ 10$^{11}$ M$_{\odot}$) to dwarf systems ($M_{star}$ $\simeq$ 10$^{6}$ M$_{\odot}$) where the ionised gas emission is due to the emission of single 
O-early B stars. This extends previous works to a dynamic range in stellar mass and star formation rate (10$^{-4}$ $\lesssim$ $SFR$ $\lesssim$ 10 M$_{\odot}$ yr$^{-1}$)
never explored so far. The main sequence relation derived for all star forming
galaxies within one virial radius of the Virgo cluster has a slope comparable to that observed in other nearby samples of isolated objects, but has a dispersion $\sim$ 3 times 
larger ($\sim$ 1 dex). The dispersion is tightly connected to the available amount of HI gas, with gas-poor systems located far below objects of similar
stellar mass but with a normal HI content. When measured on unperturbed galaxies with a normal HI gas content ($HI-def$ $\leq$ 0.4), the relation has a slope $a$=0.92$\pm$0.06, 
an intercept $b$ = -1.57$\pm$0.06 (at a pivot point of log $M_{star}$ = 8.451 M$_{\odot}$), and a scatter $\sigma$ $\simeq$ 0.40, and has a constant slope in the stellar mass range 10$^{6}$ $\lesssim$ $M_{star}$ $\lesssim$ 3 $\times$ 10$^{11}$ M$_{\odot}$. 
The specific star formation rate of HI-poor galaxies is significantly reduced with respect to that of HI-rich systems of similar stellar mass,
while their atomic gas consumption timescale $\tau_{HI}$ is fairly similar in particular for objects of stellar mass 10$^7$ $\lesssim$ $M_{star}$ $\lesssim$ 10$^9$ M$_{\odot}$.
We compare these observational results to the prediction of models expressly tuned to reproduce the effects induced by the interaction of galaxies with their surrounding environment.
The observed scatter in the main sequence relation can be reproduced only after a violent and active stripping process such as ram-pressure that removes gas from the disc 
(outer parts first) and quenches star formation on short ($<1$ Gyr) timescales.  This rules out milder processes such as starvation.
This interpretation is also consistent with the position of galaxies of different star formation activity and gas content within the phase-space diagram.
We also show that the star forming regions formed in the stripped material outside perturbed galaxies are located well above the main sequence relation drawn by unperturbed systems.
These extraplanar HII regions, which might be at the origin of ultra compact dwarf galaxies (UCDs) and other compact sources typical in rich environments, are living a starburst 
phase lasting only $\lesssim$ 50 Myr, later becoming quiescent systems.
 }
   {}
   {}
   {}
   {}
   {}


   \maketitle
%

\section{Introduction}

The physical, structural, and kinematical properties of galaxies are tightly connected via several scaling relations between important parameters such as 
stellar masses, diameters, luminosities, rotational velocities, etc. which can be
accurately measured using multifrequency observations. These important relations are often studied in the literature since they retain the imprint of different processes
which shaped galaxy formation and evolution. For this reason, they are often used to constrain free parameters in galaxy models, either semi-analytic or in full cosmological simulations. 
Among these scaling relations, the so-called ``main sequence" relation linking the star formation activity of galaxies
to their stellar mass (e.g. Brinchmann et al. 2004; Daddi et al. 2007; Elbaz et al. 2007; Noeske et al. 2007; Speagle et al. 2014) has a critical importance. Although its origin
is driven by the range in scales of the galaxy formation process (``bigger galaxies have more of everything", Kennicutt 1990), the main sequence relation has been
the subject of hundreds of dedicated works due to its tight connection to the history of star formation of galaxies. Indeed, the evidence that most star forming galaxies
populate a tight sequence in SFR vs stellar mass across more than 12 billion years of cosmic time (e.g. Whitaker et al. 2014, Schreiber et al. 2015) places strong constraints 
on the regulation of the star formation process in galaxies. In the so called ``bathtub'' models (e.g. Bouchè et al. 2010, Lilly et al. 2013), the balance of gas inflows, star formation 
and outflows due to stellar feedback, naturally explain the presence of a star forming main sequence with a small intrinsic scatter of $\sim 0.3$ dex (Noeske et al. 2007, Whitaker et al. 2012).
As a result, a particular attention has been devoted to the study of the main sequence parameters in different stellar mass regimes, including the origin of the scatter, and its possible 
evolution with redshift (e.g. Rodighiero et al. 2011; Whitaker et al. 2012, 2014; Ilbert et al. 2013; Speagle et al. 2014; Gavazzi et al. 2015; Tasca et al. 2015; Schreiber et al. 2015;
Erfanianfar et al. 2016; Pearson et al. 2018; Popesso et al. 2019, 2022). For instance, it has been shown that above a given stellar mass the slope of the relation
decreases, indicating a partial quenching of the activity of star formation attributed either to AGN feedback (Bouche et al. 2010, Lilly et al. 2013)
or to a secular evolution due to the presence of prominent bulges or bars (e.g. Gavazzi et al. 2015, Erfanianfar et al. 2016).

The main sequence relation has been also used to study the effects of the environment on galaxy evolution. It is now well established that galaxies inhabiting
rich environments undergo a different evolution than their counterparts in the field. Indeed, high density regions are characterised by a dominant quiescent population 
(morphology segregation effect, Dressler 1980, Whitmore et al. 1993), and their spiral population is generally composed of gas-poor objects 
(Haynes \& Giovanelli 1984; Cayatte et al. 1990; Solanes et al. 2001; Gavazzi et al. 2005, 2006a, 2013; Boselli et al. 2014a,b) with a reduced star formation 
activity (Gavazzi et al. 1998, 2002, 2006b, 2010, 2013; Boselli et al. 2015). Different physical mechanisms have been proposed in the literature to explain these differences,
including galaxy harassment (e.g. Moore et al. 1998), starvation (e.g. Larson et al. 1980; Balogh et al. 2000), and ram pressure stripping (Gunn \& Gott 1972; Boselli et al. 2022a), as reviewed in Boselli \& Gavazzi (2006, 2014). The identification of the dominant mechanism as a function of galaxy and host halo mass, however, is still hotly debated. Multifrequency
observations of large statistical samples in the nearby Universe, or of selected representative objects in nearby clusters, indicate ram pressure
stripping as the most important phenomenon responsible for the gas removal and for the following quenching of the star formation activity (e.g. Boselli et al. 2022a).
On the contrary, observations of large statistical samples compared to the prediction of cosmological simulations suggest that other milder mechanisms
such as starvation are at the origin of the observed suppression of the star formation activity in galaxies inhabiting high density regions (McGee et al. 2009; 
Wolf et al. 2009; von der Linden et al. 2010; DeLucia et al. 2012; Wheeler et al. 2014; Taranu et al. 2014; Haines et al. 2015). 

The study of the main sequence relation can be of great help in understanding the effects of the environment on the activity of star formation in cluster galaxies, for instance
the scatter in the relation can be modulated by an increased or a reduced activity occurring in perturbed systems.
The main sequence relation in rich environments has been studied in details in local (e.g. Tyler et al. 2013, 2014; Boselli et al. 2015, 2022a; Paccagnella et al. 2016; Vulcani et al. 2018) 
and more distant samples (e.g. Vulcani et al. 2010; Tran et al. 2010, 2017; Koyama et al. 2013; 
Zeimann et al. 2013; Lin et al. 2014; Erfanianfar et al. 2016; Old et al. 2020; Nantais et al. 2020), and compared to what predicted by hydrodynamic simulations 
(e.g. Sparre et al. 2015; Wang et al. 2018; Donnari et al. 2019; Matthee \& Schaye 2019). 
The results obtained so far often give inconsistent results. Some of them suggest that at intermediate redshift ($z$$\sim$0.4) the activity of star formation in cluster 
galaxies is enhanced (e.g. Koyama et al. 2013) with respect to field objects. Others, on the contrary, indicate the presence of galaxies with
a reduced star formation activity (e.g. Vulcani et al. 2010; Lin et al. 2014; Erfanianfar et al. 2016). Systematic differences in the results are also present in 
local studies, some of which indicate an enhanced 
activity in ram pressure stripped galaxies (e.g. Vulcani et al. 2018), other rather suggesting a comparable (e.g. Tyler et al. 2013, 2014)
or reduced activity (e.g. Boselli et al. 2015, Paccagnella et al. 2016).
Furthermore, whenever the observational evidence was consistent, the results have been explained with different perturbing mechanisms, such as a slow quenching process 
in galaxies with a reduced activity of star formation by Paccagnella et al. (2016) or a rather rapid quenching due to a ram pressure stripping episode 
(e.g. Boselli et al. 2014ab, 2015, 2016a).
 
The difference in these observational results and/or in their interpretation might be strongly related to observational biases, to selection criteria, 
or to the limited dynamic range in the analysed samples. 

The Virgo Environmental Survey Tracing Ionised Gas Emission (VESTIGE, Boselli et al. 2018a) is a deep H$\alpha$ narrow-band imaging survey of the Virgo cluster.
The H$\alpha$ emission line is a hydrogen recombination line produced in the gas ionised by young ($\lesssim$ 10 Myr) and massive ($M_{star}$ $\geq$ 10 M$_{\odot}$) O and 
early B stars, and is thus considered as the most direct tracer of recent star formation activity in galaxies (e.g. Kennicutt 1998, Boselli et al. 2009).
Given its untargeted nature, VESTIGE is providing us with a unique sample of galaxies perfectly defined for a complete census of the star formation activity in 
a rich nearby cluster of galaxies. Designed to cover the whole Virgo cluster up to its virial radius, the survey is sufficiently deep to detect all H$\alpha$ emitting sources 
brighter than $L(H\alpha)$ 10$^{36.5}$ erg s$^{-1}$, corresponding to star formation rates of $SFR$ $\geq$ 2 $\times$ 10$^{-5}$ M$_{\odot}$ yr$^{-1}$ and stellar masses
$M_{star}$ $\geq$ 10$^6$ M$_{\odot}$, values never reached in other nearby or high redshift clusters. Furthermore, the proximity of the cluster (16.5 Mpc, Mei et al. 2007) 
and the availability of multifrequency data covering the whole electromagnetic spectrum obtained with other untargeted surveys of comparable sensitivity (e.g. UV - GUViCS,
Boselli et al. 2011; visible - NGVS, Ferrarese et al. 2012; far-IR - HeViCS, Davies et al. 2010; ALFALFA - HI, Giovanelli et al. 2005) are crucial for a coherent study 
aimed at identifing the dominant perturbing mechanism at the origin of galaxy transformation in a rich environment. The galaxies analysed in this work detected during the
VESTIGE survey are used to reconstruct the main sequence relation and a few other scaling relations of galaxies within the Virgo cluster. 
The narrow-band imaging observations are described in Sec. 2, along with the multifrequency
observations used in the analysis. In Sec. 3 we describe how the data are corrected for [NII] contamination and dust attenuation, and how they are transformed into star formation rates.
The most important scaling relations requiring an accurate determination of the star formation activity of galaxies including the main sequence relation are 
described, analysed, and compared to the predictions of tuned models in Sec. 4 and 5. The discussion and conclusion are presented in Sec. 6 and 7.

\section{Observations and data reduction}

\subsection{VESTIGE narrow-band imaging}

The data analysed in this work have been gathered during the VESTIGE survey, an untargeted H$\alpha$ narrow-band (NB) imaging survey covering the Virgo cluster
up to its virial radius (104${\degr}$$^2$). The observations have been carried out using MegaCam at the CFHT in the NB filter MP9603 ($\lambda_c$ = 6591 \AA; $\Delta\lambda$ = 106 \AA).
At the redshift of the galaxies (-300 $\leq$ $v_{hel}$ $\leq$ 3000 km s$^{-1}$), the filter includes the emission of the Balmer H$\alpha$ line ($\lambda$ = 6563 \AA)
and of the two [NII] lines ($\lambda$ = 6548, 6583 \AA)\footnote{Hereafter we refer to the H$\alpha$+[NII] band simply as H$\alpha$, unless otherwise stated.}.
The survey is now $\sim$ 76\% complete, with an integration time of 2~h in the NB filter and 12 min in the broad-band $r$ filter necessary to subtract the stellar 
continuum emission. Full depth of the survey has been gathered over most of the cluster, with eventually shorter exposures at the periphery.
The observing strategy, which has been fully described in Boselli et al. (2018a), consists in mapping the full cluster with MegaCam, a wide 
field detector composed of 40 CCDs with a pixel scale of 0.187 arcsec pixel$^{-1}$. The full cluster has been mapped following a specific observing sequence optimised
for the determination of the sky background necessary for the detection of low surface brightness, extended features. For this purpose, the images have been gathered 
using a large dithering (15 arcmin in R.A. and 20 arcmin in Dec.) necessary to minimise any possible contribution of unwanted reflections of bright stars in the 
construction of the sky flat fields. The observations have been obtained under excellent seeing conditions ($\theta$=0.76\arcsec\ $\pm$ 0.07\arcsec)
for both the narrow- and broad-band images. The sensitivity of the survey
is $f(H\alpha)$ $\simeq$ 4 $\times$ 10$^{-17}$ erg s$^{-1}$ cm$^{-2}$ (5$\sigma$) for point sources and $\Sigma(H\alpha)$ $\simeq$ 2 $\times$ 10$^{-18}$ erg s$^{-1}$ cm$^{-2}$ arcsec$^{-2}$
(1$\sigma$ after smoothing the data to $\sim$ 3\arcsec resolution) for extended sources.

Consistently with previous works, the images have been reduced using Elixir-LSB (Ferrarese et al. 2012), a data reduction pipeline expressly designed to detect extended low surface
brightness structures as those expected in interacting systems. The data have been photometrically calibrated and corrected for astrometry using the standard MegaCam procedures 
described in Gwyn (2008), with a typical photometric accuracy in the two bands of $\lesssim$ 0.02-0.03 mag.

The subtraction of the stellar continuum emission has been secured as described in Boselli et al. (2019). This process is particularly critical in the centre of early-type galaxies, 
where the emission in the NB filter is highly dominated by that of the stars. The contribution of the stellar continuum within the NB filter is estimated using the $r$-band frame 
combined with a $g-r$ colour to take into account any possible variation of the spectral energy distribution within the broad-band image. For this purpose we used the $g$-band images
gathered at the CFHT under similar conditions during the NGVS survey (Ferrarese et al. 2012).

\subsection{Galaxy identification}

\begin{figure}
\centering
\includegraphics[width=0.5\textwidth]{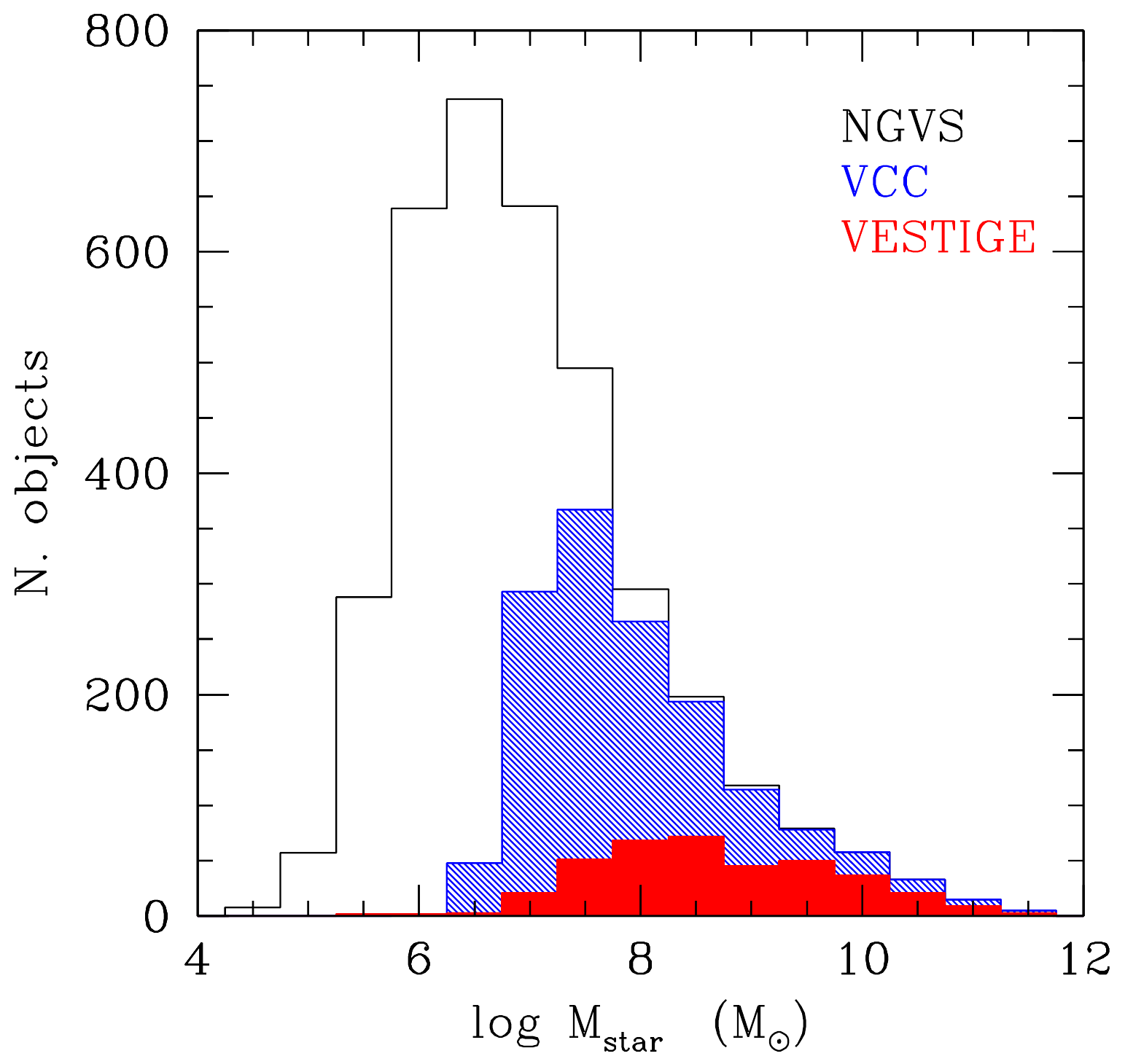}
\caption{Stellar mass distribution of the H$\alpha$ detected galaxies (red filled histogram), of the VCC (blu hatched histogram)
and of the NGVS galaxies (black empty histogram) classified as Virgo cluster members. Stellar masses are derived as described in 
Sec. (3.1).}
\label{censoring}%
\end{figure}

The H$\alpha$ emitting sources have been identified after visual inspection of all the continuum-subtracted images as follows. We first looked for the presence of any line emitting source
associated to all of the 2096 galaxies included in the VCC catalogue (Binggeli et al. 1985), and we identified 307 galaxies with emission at the redshift of the cluster.
Given the tight relationship between the presence of cold gas and the activity of star formation, we then looked for possible counterparts to the HI detected sources in the 
ALFALFA survey (Giovanelli et al. 2005), a HI blind survey carried out with the Arecibo radio telescope and covering the whole Virgo cluster region mapped by VESTIGE. This
HI survey, which has a typical sensitivity of $rms$ = 2.3 mJy at 10 km s$^{-1}$ spectral resolution and 3.2\arcmin\ angular resolution, is able to detect galaxies 
at the distance of the Virgo cluster with HI masses of $M_{HI}$ $\simeq$ 10$^{7.5}$ M$_{\odot}$. For this purpose we used the ALFALFA catalogue of Haynes et al. (2018), and included 
only those H$\alpha$ sources with a HI redshift $\leq$ 3000 km s$^{-1}$ (37 objects). We then looked for any possible counterpart of the 3689 galaxies identified in the NGVS 
survey as Virgo cluster members by the use of different scaling relations, as extensively described in Ferrarese et al. (2012, 2020; Lim et al. 2020; 31 objects). 
Four extra bright galaxies with clear H$\alpha$ 
emission located outside the NGVS and VCC footprint have been included. Finally, we visually inspected all the H$\alpha$
continuum-subtracted images gathered during the VESTIGE survey and identified a few other extended emitting sources not included in the previous catalogues (5 objects).
To avoid any possible contamination of background line emitters, we excluded all those sources without any clear structured and extended emission typical of compact nearby dwarf systems,
such as multiple HII regions, extended filaments etc. We recall that the trasmissivity curve of the NB filter perfectly matches that of line emitters in the Virgo cluster, thus
the detection of any line emitting source with these properties grants the membership to the cluster. The final sample of H$\alpha$ emitting galaxies analysed in this work 
thus includes 384 Virgo cluster objects. Figure \ref{censoring} shows the stellar mass distribution of the H$\alpha$ detected objects 
compared to that of the VCC and NGVS galaxies identified as Virgo cluster members. Most of the morphologically classified late-type galaxies
included in the VCC (spirals, Magellanic Irregulars, BCDs) are detected in H$\alpha$, with a few early-type systems mainly characterised by a
nuclear or circumnuclear star formation activity (e.g. Boselli et al. 2008a, 2022b). The large majority of the Virgo cluster members undetected by VESTIGE
are early-type systems (E, S0, dE, dS0). Since the sensitivity of the survey is able to detect a single early-B star at the distance of the cluster (see Sec. 3.4),
Fig. \ref{censoring} indicates that the galaxies forming massive, ionising stars are only $\sim$ 10\%\ of the Virgo cluster population.
For this reason, censoring should not significantly affect the results. 

\subsection{Flux extraction}

The H$\alpha$ total flux of each source has been computed using the same procedure used in previous VESTIGE works and described in Fossati et al. (2018) and Boselli et al. (2018b,c).
Fluxes and uncertainties were derived by measuring both the galaxy emission and the sky background within the same elliptical aperture randomly located on the sky after masking
other contaminating sources. The elliptical apertures were optimised to encompass the total H$\alpha$ emission on the galaxy disc and at the same time minimise the sky contribution 
within the aperture. This was not always optimal in a few low surface brightness extended galaxies with the H$\alpha$ emission dominated by a few HII regions located at different edges
of the disc. Furthermore, to minimise possible effects due to large scale residual gradients in the continuum-subtracted frames, the sky background has been measured 1000 times within $\sim$ 5 times the 
diameter of the target. The uncertainties on the fluxes were obtainbed as the quadratic sum of the uncertainties on the flux counts and the 
uncertainties on the background (rms of the bootsrap iterations). The uncertainties on the flux counts were derived assuming a Poissonian distribution for the source photo-electrons.
In a few objects the resulting signal-to-noise (S/N) is very low, close to 1. Despite their weak emission, however, these are bona fide detections since all consisting in several 
HII regions with a correlated diffuse signal. As mentioned above, their low S/N is due to the fact that their few HII regions are located at large distances within the stellar disc,
thus the large uncertainty on the measurement is principally due to the uncertain measure of the sky in the large aperture adopted for the flux extraction.

\subsection{Multifrequency data}

The analysis presented in this work was made possible by the large amount of multifrequency data available for the Virgo cluster region (see 
Table \ref{photom}).
Spectroscopic data in the optical domain are necessary to correct the fluxes obtained in the NB continuum-subtracted images for [NII] contamination and
dust attenuation. Integrated spectroscopy gathered by drifting the slit of the spectrograph over the stellar disc is available for the brightest galaxies
included in the \textit{Herschel} Reference Survey (HRS, Boselli et al. 2013), and for several other cluster members in Gavazzi et al. (2004). Optical spectroscopy
of the nuclear regions is also available from the SDSS. A few objects have also dedicated FORS and MUSE observations of excellent quality (e.g. Fossati et al. 2018,
Boselli et al. 2018c, 2021, 2022b).

Dust attenuation corrections of the H$\alpha$ fluxes can also be done using mid-IR data. For this purpose we used WISE (Wright et al. 2010) 22 $\mu$m fluxes
obtained for the brightest galaxies included in the HRS by Ciesla et al. (2014), or for the remaining Virgo cluster members by Boselli et al. (2014a).
For a few galaxies not included in these catalogues, we extracted WISE 22 $\mu$m fluxes from the images as described in Boselli et al. (2014a).

We also used optical images obtained during the NGVS survey (Ferrarese et al. 2012) for securing an accurate subtraction of the stellar continuum emission, 
for the optical identification of the H$\alpha$ emitting sources, and for the determination of their total stellar mass. 
Total magnitudes in the $u,g,i,z$ photometric bands were measured following one of two 
approaches: fitting elliptical isophotes with a bespoke code based on 
IRAF/ELLIPSE or fitting 2D Sersic models with GALFIT (see Ferrarese et al. 2020).
For galaxies with a photographic magnitude in the VCC calatogue $B_{VCC}$ $<$16 mag we draw growth 
curves from IRAF/ELLIPSE, and measured fluxes within the first 
isophote where the $g$-band growth curve flattens. For all other galaxies we used 
the total fluxes (integrated to infinity) of the best-fit model found by 
GALFIT. Errors in the integrated fluxes were estimated by summing the per-pixel 
contributions from Poisson noise of the source and sky, and read noise, while 
enforcing lower limits equal to the precision of the NGVS photometric 
calibration.

Finally, we used different sets of HI data for the identification of the H$\alpha$ emitting source (ALFALFA, Giovanelli et al. 2005, Haynes et al. 2018)
and for the determination of the HI-deficiency parameter, a parameter generally used in the literature to identify galaxies with a low atomic gas
content probably due to a recent interaction with their surrounding cluster environment (e.g. Cortese et al. 2021, Boselli et al. 2022a). 
This parameter is defined as the difference between the expected and the observed HI gas mass of
each single galaxy on logarithmic scale (Haynes \& Giovanelli 1984), where the expected atomic gas mass is the mean HI mass of
a galaxy of a given optical size and morphological type determined in a complete reference sample of isolated objects. For this purpose we
used here the recent calibration of Cattorini et al. (2022) based on a sample of $\sim$ 8000 galaxies in the nearby universe. We remark that this calibration
is optimised for massive and dwarf systems, these last generally undersampled in previous determinations (e.g. Haynes \& Giovanelli 1984; Solanes et al. 1996; Boselli \& Gavazzi 2009). 
It is highly uncertain for early-type galaxies, where the atomic gas content does not necessary follow well defined scaling relations (e.g. Serra et al. 2012). 
Throughout this work we consider as unperturbed objects those with a HI-deficiency parameter $HI-def$ $\leq$ 0.4, which is the typical
dispersion in the scaling relation used to calibrate this parameter (e.g. Haynes \& Giovanelli 1984; Cattorini et al. 2022).
For the determination of the HI-deficiency parameter we used in first priority the HI data collected in the GoldMine database (Gavazzi et al. 2003)
generally gathered through deep pointed observations mainly with the Arecibo radio telescope (typical rms $\lesssim$ 1 mJy), otherwise the ALFALFA data (Haynes et al. 2018).
The ALFALFA data, which cover at an homogeneous sensitivity the whole cluster region, have been also used to derive stringent upper limits to the total HI mass 
and lower limits to the HI-deficiency parameter of H$\alpha$ detected sources without any available HI pointed observation. 
Upper limits in the HI mass (in solar units) are derived using the relation (e.g. Boselli 2011):

\begin{equation}
{M_{HI_{ul}} = 2.356 \times 10^{5} d^2 ~ {2 \times rms} \sqrt{200 \times \delta V_{HI} \times \rm{sin}(i)}}
\end{equation}

\noindent
where $d$ is the distance of the galaxy (in Mpc, see Sec. 3.4), $rms$ is the rms of the HI data (in Jy, here assumed to be $rms$ = 2.3 mJy) measured for a spectral resolution $\delta V_{HI}$ 
(10 km s$^{-1}$) and $i$ is the inclination of the galaxy 
on the plane of the sky. Since all the HI undetected galaxies are dwarf systems, we assume that their rotational velocity is of 200 km s$^{-1}$.
The determination of the HI-deficiency parameter requires an estimate of the $B$-band 
25 mag arcsec$^{-2}$ isophotal diameter $D_{25}(B)$. Whenever this value was not available in the VCC, we derived it using the effective radius $R_{e}(g)$ available in the NGVS 
data catalogue measured as described in Ferrarese et al. (2020). Effective radii have been transformed into $B$-band isophotal diameters through the relation:

\begin{equation}
{D_{25}(B) ~~\rm{[arcmin]}} = \frac{R_{e}(g) ~~\rm{[arcsec]}}{15.043}
\end{equation}

\noindent
that we calibrated on a large sample of galaxies with both sets of data available. The inclination of each single galaxy $i$ has been derived as described in Haynes \& Giovanelli (1984)
with the axial ratios given in the VCC whenever available, or from the NGVS catalogue for the remaining cases. 
Although important in the star formation process, we do not consider here the molecular gas phase because of an evident lack of data, available only 
for the most massive $\sim$15\%\ galaxies of the sample (Boselli et al. 2014a; Brown et al. 2021).

As indicated in Table \ref{photom}, the photometric and spectroscopic coverage of the sample is optimal in the optical, UV, and HI bands,
while limited to $\sim$ 45-50\%\ in the infrared bands ($\sim$ 70-80\%\ when including stringent upper limits). Given the all sky coverage of the WISE survey,
at 22$\mu$m the undetected galaxies are all very faint sources, while those lacking are too faint to give stringent upper limits for the SED 
fitting analysis. Those lacking in the 5 \textit{Herschel} bands are located outside the footprint of the HeViCS survey (Davies et al. 2010).

\begin{table}
\caption{Multifrequency data}
\label{photom}
{
\[
\begin{tabular}{ccc}
\hline
\noalign{\smallskip}
\hline
Data			& Detections	& \%		\\
\hline	
H$\alpha$		& 384		& 100		\\
$FUV$			& 319		& 83		\\
$NUV$			& 368		& 96		\\
$ugriz$			& 384		& 100		\\
WISE 22$\mu$m		& 197(310)	& 51(81)	\\
$\textit{Herschel}$ 100$\mu$m	& 165(250)	& 43(65)	\\
$\textit{Herschel}$ 160$\mu$m	& 177(250)	& 46(65)	\\
$\textit{Herschel}$ 250$\mu$m	& 197(257)	& 51(67)	\\
$\textit{Herschel}$ 350$\mu$m	& 182(254)	& 47(66)	\\
$\textit{Herschel}$ 500$\mu$m	& 170(255)	& 44(66)	\\
Spectra			& 276		& 72		\\
HI			& 286(384)	& 74(100)	\\
\noalign{\smallskip}
\hline
\end{tabular}
\]
Notes: Number of galaxies with available multifrequency data. The number in parenthesis
includes upper limits.\\
}
\end{table}

\section{Derived parameters}

\subsection{Stellar masses}

To estimate the stellar mass, we have modelled the SED of our sample with the latest version of the CIGALE code (Boquien et al. 2019). For this we have considered a flexible 
star formation history (SFH) consisting of a delayed SFH ($\propto t\times\exp(-t/\tau)$), with a recent burst or quench (module \texttt{sfhdelayedbq}). We have set the age 
of all the galaxies to common value of 13~Gyr. The $e$-folding time free, sampling 10 values from 1~Myr to 8~Gyr. The most recent variation of the SFH is set to have happen 
between 10~Myr and 1~Gyr ago, sampling 9 values. This strength, which is defined as the ratio of the SFR after and before the beginning of the burst, ranges 
from 0 (complete quench) to 10 (strong burst), with a sampling of 10 values. This particular 
shape of the SFH has been chosen to take into account the possible abrupt variations of the star formation activity observed in perturbed cluster galaxies
(e.g. Boselli et al. 2016b, 2021; Fossati et al. 2018).
The stellar spectrum is computed from the Bruzual \& Charlot (2003) single stellar populations with a Chabrier (2003) IMF with a fixed metallicity 
$Z=0.02$. We include the nebular emission, with a gas metallicity following that of the stars, an ionisation parameter $\log U=-3$, and an electron density 
$n_e=100$~cm$^{-3}$. The emission is attenuated with a modified starburst law (Calzetti et al. 1994, 2000), with a line reddening covering 16 values from 0.005~mag to 0.6~mag. 
The differential reddening is allowed to vary with the continuum having a lower reddening by a factor 0.25, 0.5, or 0.75. An optional bump is included with a strength up to that of the 
Milky Way. For more flexibility, the attenuation curve slope is multiplied by a power law of index $\delta$ ranging from 0 (bona fide starburst) to $-1$ (steeper than the LMC). Finally, the 
dust emission is modelled with the Dale et al. (2014) dust template, with $\alpha$ taking 8 values from 0.5 to 4. Overall we compute a grid of 15.206.400 models.

All the models are fitted to the observations, using the GALEX FUV and NUV bands, Megacam $u, g, i, z$\footnote{For the galaxies VCC 322 and VCC 331, which do not have available NGVS data, we used
the SDSS magnitudes given in the EVCC of Kim et al. (2014).}, WISE 22~$\mu$m, and \textit{Herschel} at 100, 160, 250, 350, and 500~$\mu$m. 
Upper limits are handled in the computation of the $\chi^2$ through Eq. 15 from Boquien et al. (2019).
The stellar mass and the related uncertainties are estimated from the likelihood-weighted mean and standard deviation from the model.
The mean uncertainty on the stellar mass given by the SED fitting analysis is 0.07 dex, and increases to 0.08 dex in the 
low mass systems ($M_{star}$ $<$ 10$^8$ M$_{\odot}$) where the photometric coverage of the sample is less complete. The uncertainty on the stellar mass, however,
is always $<$ 0.15 dex, and is larger than 0.1 dex in only 12\%\ of the sample. These low uncertainties are due to the fact
that i) an infrared flux or stringent upper limit necessary to constrain the dust attenuation is available for $>$ 80\%\ of the sample, and ii)
those with lacking IR data are the lowest mass systems of the sample, where dust attenuation, if present, is minimal (see Sec. 3.3 and Fig. \ref{AHaNII}). 
To further check the accuracy of these values we compared the stellar masses derived using the CIGALE SED fitting code with those derived for the whole NGVS sample
using PROSPECTOR, a SED fitting code based on the Flexible Stellar Population Synthesis model suite of Conroy et al. (2009), using similar parameters 
for the IMF and the star formation history of the sample galaxies, but using only the four NGVS photometric bands ($ugiz$) and excluding any dust attenuation correction (J. Roediger, priv. comm.).
The agreement between the two estimates is remarkable, with a mean ratio of 0.05 dex and a dispersion of $\sim$ 0.13 dex. This difference is similar to the one generally obtained
when comparing stellar mass estiamtes derived using different codes (e.g. Conroy 2013) which depend on the assumed star formation histories, population
synthesis codes, dust attenuation laws etc. We estimate that the uncertainty on $M_{star}$ is somewhere in between the uncertainty given by the CIGALE SED fitting code
and the one derived by adding in quadrature an extra source of uncertainty due to the adopted model (assumed IMF, stellar population synthesis model, SFH, metallicity, SED fitting code...)
that we define as modelling uncertainty and assume as 0.1 dex (Conroy et al. 2009; Hunt et al. 2019). The resulting mean
uncertainty on $M_{star}$ is thus 0.07 $\lesssim$ $\sigma(logM_{star})$ $\lesssim$ 0.14 dex.  

We also derived stellar mass surface densities defined as in Boselli et al. (2014c), i.e.:

\begin{equation}
{\mu_{star} ~~\rm{[M_{\odot} kpc^{-2}]} = \it{\frac{M_{star}}{2\pi R_e(i)^2}}  }
\end{equation}

\noindent
where $R_e(i)$ is the effective radius in the $i$-band in kpc\footnote{The factor 2 at the denominator of eq. 3 is adopted here since $R_e(i)$ is an effective radius,
the radius including half of the total $i$-band light, thus presumably also $\sim$ 50\%\ of the total stellar mass.}. 
This quantity is available for 216 objects of the sample. For those galaxies without a growth curve in 
the NGVS catalogue (168 objects), $R_{e}(i)$ has been derived from the $B$-band isophotal diameter using the relation:

\begin{equation}
{R_{e}(i) ~~\rm{[arcsec]} }= {16.189 \times D_{25}(B) ~~\rm{[arcmin]}}
\end{equation}
 
\noindent
measured on a large sample of galaxies with both sets of data available.

\subsection{[NII] contamination}

\begin{figure}
\centering
\includegraphics[width=0.5\textwidth]{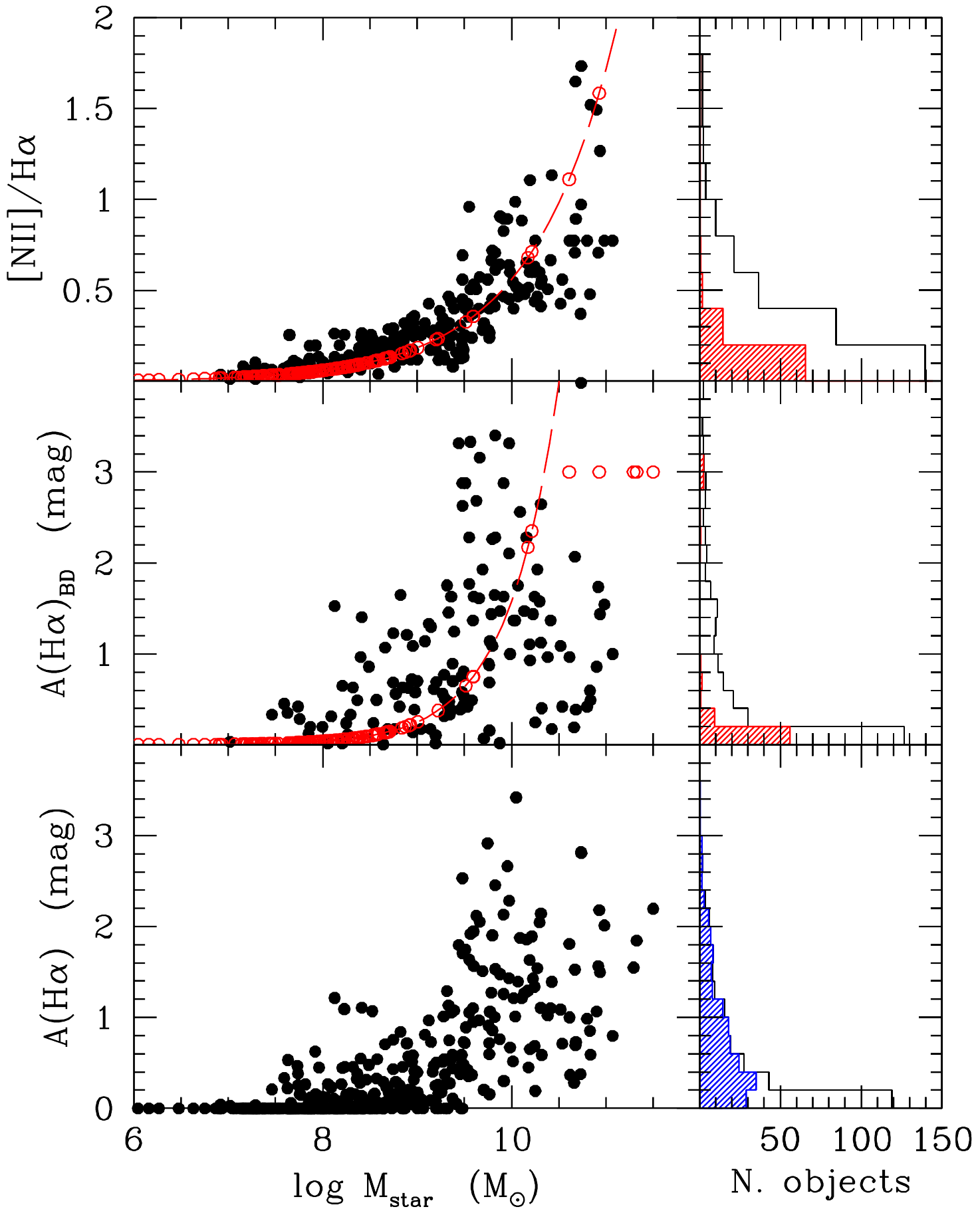}
\caption{Left panels: relationship between the [NII]$\lambda$6548 + 6583\AA/H$\alpha$ line ratio (upper panel), the dust attenuation measured using the Balmer decrement
$A(H\alpha)_{BD}$ (middle panel), and the dust attenuation $A(H\alpha)$ derived by averaging the Balmer decrement and the far infrared determinations (see Sec. 3.3) (lower panel)
and the stellar mass of the selected galaxies. The red long-dashed lines
indicate the mean relations derived in this work. Right panels: [NII]$\lambda$6548 + 6583\AA/H$\alpha$, $A(H\alpha)_{BD}$, and $A(H\alpha)$ distributions. 
The red and blue hatched histograms give the distribution of [NII]$\lambda$6548 + 6583\AA/H$\alpha$, $A(H\alpha)_{BD}$, and $A(H\alpha)$ derived from
the mean scaling relations given in the text and adopted in the H$\alpha$ flux correction for those galaxies without any spectroscopic data. 
}
\label{AHaNII}
\end{figure}

The H$\alpha$ fluxes extracted from the images can be converted into SFRs once corrected for the contamination of the [NII] emission lines in the NB filter
and for dust attenuation. The contribution of the [NII] lines to the total line emission of each source within the NB filter is measured, in order of priority, 
using targeted MUSE or long slit high quality observations published in Boselli et al. (2018c, 2021, 2022b) and Fossati et al. (2018) (12 galaxies), long slit integrated spectra of 
Boselli et al. (2013) and published in Boselli et al. (2015) (72 objects), similar integrated spectra of Virgo galaxies published 
in Gavazzi et al. (2004) (44 objects), public SDSS spectra obtained within a circular aperture of 3\arcsec\ diameter centred on the nucleus of the galaxies (114 galaxies),
or assuming a standard scaling relation linking the emission line ratio [NII]/H$\alpha$ to the total stellar mass of galaxies as described in 
Boselli et al. (2009) (142 objects). This standard relation has been recalibrated using this VESTIGE sample, as indicated in Fig. \ref{AHaNII}:

\begin{equation}
{\frac{[NII]}{H\alpha} = 10^{0.485 \times log M_{star} - 5.1}}
\end{equation}

\noindent
with $M_{star}$ in solar units, where [NII]/H$\alpha$ here indicates the flux ratio of the [NII] doublet (6548+6583\AA) over H$\alpha$.

Since integrated long slit spectra are available for most of the massive systems of the sample, SDSS spectra are used mainly for dwarfs
with flat aboundance gradients, where the emission within a 3\arcsec\ aperture centred on the nucleus is taken as representative of the whole galaxy. This assumption
has been verified and confirmed by comparing the [NII]$\lambda$6583\AA/H$\alpha$ ratios in galaxies with both sets of data available. Indeed, these are all 
low-mass systems without any nuclear activity, where [NII]$\lambda$6583\AA/H$\alpha$ is generally close to zero (the [NII]/H$\alpha$ ratio is $<$ 0.2 in 60\%\ of the sample). 

\subsection{Dust attenuation}

\subsubsection{Balmer decrement}

The H$\alpha$ emission must also be corrected for dust attenuation if luminosities want to be converted into star formation rates. For this purpose we corrected H$\alpha$ fluxes
using the same prescriptions presented in Boselli et al. (2015). The full set of data has been first corrected for Galactic extinction using the Schlegel et al. (1998) 
dust attenuation map combined with the Cardelli et al. (1989) extinction curve, then for internal attenuation using the mean value derived from the Balmer decrement and the 
22$\mu$m WISE emission. As for the [NII] contamination, the Balmer decrement is derived using spectroscopic data in order of preference from targeted MUSE observations whenever available,
extracted from the HRS catalogue of Boselli et al. (2013, 2015), the integrated spectra of Gavazzi et al. (2004), SDSS nuclear spectra, or the typical scaling relation linking 
the H$\alpha$ attenuation derived from the Balmer decrement $A(H\alpha)_{BD}$ and the stellar mass of galaxies, here derived from the sample galaxies: 

\begin{equation}
{A(H\alpha)_{BD} [\rm{mag}]= 10^{0.8 \times log M_{star} -7.8}}
\end{equation}

\noindent
with the Balmer attenuation assumed $A(H\alpha)_{BD}$=3 mag for log $M_{star}$ $\gtrsim$ 10.35 M$_{\odot}$.
In a few massive
lenticular galaxies without any available spectroscopic data, the attenuation has been derived as described in Boselli et al. (2022b).
The integrated spectra used for this work have been gathered mainly with CARELEC at the Observatoire de Haute Provance, and have the sufficient spectral 
resolution ($R$ $\sim$ 1000) to resolve H$\alpha$ from the two satellite lines [NII]$\lambda\lambda$6548,6583\AA\ and to accurately measure the contribution 
of the underlying Balmer absorption at H$\beta$. As extensively discussed in Boselli et al. (2015), the contribution of the underlying Balmer absorption
at H$\beta$ is critical for determining an accurate value for the Balmer decrement in massive, fairly quiescent spiral galaxies, where the attenuated H$\beta$ 
emission line is weak compared to the absorption line. The H$\beta$ emission, however, is accurately measured using the GANDALF code 
in most of the massive galaxies which are included in the HRS (Boselli et al. 2015). In the remaining galaxies, which are mainly dwarf systems with a limited
attenuation, the contribution of the underlying Balmer absorption is directly measured on the spectra as described in Gavazzi et al. (2004), or from the public SDSS
spectra using the relation:

\begin{equation}
{H\alpha,\beta_{cor} = H\alpha,\beta_{obs} + H\alpha,\beta_{cont} \times (H\alpha,\beta_{reqw} - H\alpha,\beta_{eqw})}
\end{equation}

\noindent
where H$\alpha,\beta_{cor}$ are the H$\alpha$ and H$\beta$ corrected fluxes, H$\alpha,\beta_{cont}$ are the stellar continuum emission at H$\alpha$ and H$\beta$, and H$\alpha,\beta_{reqw}$
and H$\alpha,\beta_{eqw}$ the measured and stellar continuum corrected equivalent widths of the two lines from the {\it galspecline} Table of the SDSS DR16 database.

\subsubsection{Dust attenuation from far-IR data}

Dust attenuation of the H$\alpha$ emission line is also derived using the prescription of Calzetti et al. (2010) based on the WISE 22 $\mu$m emission,
as extensively described in Boselli et al. (2015). Since a large fraction of the sample is composed of dust-poor dwarf systems, WISE 22 $\mu$m detections are available 
only for 197 objects. For the remaining galaxies we estimate that $A(H\alpha)_{22\mu m}$ $\simeq$ 0.67 $\times$ $A(H\alpha)_{BD}$ as derived for the whole HRS sample.
Finally, to reduce an possible systematic effect in the two dust attenuation estimates, we adopt:

\begin{equation}
{A(H\alpha) = \frac{A(H\alpha)_{BD} + A(H\alpha)_{22\mu m}}{2}}
\end{equation}

The distribution of the dust attenuation derived for the target galaxies is shown in Fig. \ref{AHaNII}. 
As expected, $A(H\alpha)$ has a weak dependence on the stellar mass of the target galaxies, as observed in other samples (e.g. Boselli et al. 2009, 2015).
Given their dwarf nature, in a large fraction of the galaxies the attenuation is $A(H\alpha)$ = 0 mag (39\%) or limited (0 $<$ $A(H\alpha)$ $\leq$ 0.2 mag, 18\%).

\subsection{H$\alpha$ luminosities and SFRs}

H$\alpha$ fluxes corrected for dust attenuation and [NII] contamination are used to derive H$\alpha$ luminosities. These are derived assuming that galaxies are located at the average distance of 
the cluster substructure to which they belong, consistently with (Boselli et al. 2014a)\footnote{In Boselli et al. (2014a) the distance of cluster A, C, and the LVC are all 
assumed at 17 Mpc. For consistency with other NGVS works, we now adopt 16.5 Mpc for these substructures.}. The distance of each substructure are assumed at: 16.5 Mpc for cluster A, 
cluster C, and the low velocity cloud (LVC); 23 Mpc for cluster B and for the W\arcmin\ cloud; 32 Mpc for the W and M clouds (see Gavazzi et al. 1999; Mei et al. 2007). 

H$\alpha$ luminosities are converted into star formation rates ($SFR$, in units of M$_{\odot}$ yr$^{-1}$) using the calibration of Calzetti et al. (2010) converted to 
a Chabrier IMF, thus consistent with the one used to derive stellar masses. It is worth mentioning that the impressive depth of the VESTIGE survey,
which is complete to $L(H\alpha)$ $\geq$ 10$^{36}$ erg s$^{-1}$, allowed us to 
detect sources with H$\alpha$ luminosities as low as $L(H\alpha)$ $\simeq$ 2 $\times$ 10$^{37}$ erg s$^{-1}$, corresponding to star formation rates of a few 
10$^{-5}$ M$_{\odot}$ yr$^{-1}$ when derived using this calibration. These extremely low H$\alpha$ luminosities are produced by a number of ionising 
photons smaller than those emitted by a single early O star and are comparable to those emitted by a single early B star (see Fig. \ref{mainHI}). 
Among the ionising stellar population, early B stars are those with the lowest mass and temperature 
producing the lowest number of ionising photons (Sternberg et al. 2003).
This means that the lowest H$\alpha$ luminosities measured within this sample can be attributed to the ionising photons produced by a single star. This important 
result has several major consequences which should be considered in the following analysis: a) we are sampling the whole dynamic range of the H$\alpha$ luminosity 
function of galaxies (see Sec. 6.1), b) the stationary regime required to transform H$\alpha$ luminosities into
star formation rates using the standard relation: 

\begin{equation}
{SFR  ~~ [\rm{M_{\odot} yr^{-1}}] = 5.01 \times 10^{-42}  \it{L(H\alpha)} ~~[\rm{erg~s^{-1}}]} 
\end{equation}

\noindent
as the one mentioned above is not necessary satisfied, and c) as a consequence, the observed H$\alpha$ 
luminosity of some dwarf systems can strongly vary for age effects (e.g. Boselli et al. 2009). Finally, d) in the lowest luminosity regime
the sampling of the IMF might be stochastic, leading to an inaccurate estimate of the $SFR$ (e.g. Fumagalli et al. 2011, see Sec. 6.2). Despite these important caveats, we 
decided to transform H$\alpha$ luminosities into star formation rates to compare our results with those obtained in other works. To stress this point, we
decided to plot in the Y-axis of the main sequence relation diagrams presented in this work both the star formation rates
(in M$_{\odot}$ yr$^{-1}$) and the H$\alpha$ luminosities (in erg s$^{-1}$). We further assume that a) the escape fraction of ionising photons and b)
the contribution to the gas ionisation by other sources such as evolved stars and AGN are zero.
The first assumption is reasonable since there is converging evidence that in the local Universe the escape fraction of the ionising radiation from star forming galaxies
as those observed in this work, including metal-poor, dwarf systems, is low, of a few \%\ (Izotov et al. 2016; Leitherer et al. 2016; Chisholm et al. 2018; 
see however Choi et al. 2020)\footnote{The disagreement with the recent PHANGS-MUSE results (escape fraction $f_{esc}$ $\sim$ 40\%; Belfiore et al. 2022) is 
only apparent since this high value of $f_{esc}$ indicates the fraction of ionisng photons escaping from individual HII regions, not the one from the disc
of star forming galaxies. Indeed, as stated in Belfiore et al. (2022), part of the ionising radiation escaping from the HII regions can participate to the ionisation of the 
diffuse gas (DIG), whose emission is included in the integrated H$\alpha$ fluxes measured in the VESTIGE data (e.g. Oey et al. 2007).}. 

The second assumption is also reasonable: the number of AGN derived using the nuclear spectra available for the large majority
of the sample (Cattorini et al. 2022) is of only 4 when active galaxies are identified using the BPT diagram (Baldwin et al. 1981) and 8 using the WHAN 
classification scheme (Cid Fernandes et al. 2011). These are all well known bright galaxies where the continuum-subtracted H$\alpha$ image clearly indicates that
the total H$\alpha$ emission is largely dominated by numerous bright HII regions within the disc.

The lack of IFU spectroscopic data prevents us from quantifying the contribution of 
evolved stars such as post-AGBs to the ionisation of the gas. We can, however, estimate an upper limit to this contribution using the output of the CIGALE SED fitting code,
which provides us with the fraction of ionising photons produced by stars of age older than 10 Myr. When estimated within the same aperture used for the H$\alpha$ flux extraction
(see Sec. 2.3), the median contribution of stars of age $>$ 10 Myr
to the total H$\alpha$ flux emission is of $\sim$ 2\%, with a significant fraction in only an handfull of objects\footnote{We stress the fact that this 
number is just an upper limit since it includes also the contribution of young stars of age 10 to 100 Myr (late-B - F stars)
responsible for the emission in the UV domain.}. This very low contribution is in agreement with
the lack of any diffuse smooth emission in the images of all the detected objects, where the H$\alpha$ flux comes principally form clumpy and very structured featurs 
typical of star forming regions. This is also the case in the few early-type systems where the H$\alpha$ emission has been detected (e.g. Boselli et al. 2022b)
and where the contribution of the evolved stellar population should be the strongest. Multifrequency observations consistently indicate
that in these extreme objects the gas is ionised by young stars.

The uncertainty on the star formation rate parameter is hardly quantifiable since it depends on serveral (not always well constrained) parameters: 1)
the uncertainty on the observed flux measurement, 2) on the correction for the [NII] contamination and 3) dust attenuation, and finally 4) on the conversion of H$\alpha$
luminosities into star formation rates using standard recipes (eq. 9). They also depend on the uncertainty on the distance of galaxies, but since this affects 
in the same way star formation rates and stellar masses, we do not consider it here. The way uncertainties on the observed flux (H$\alpha$ and WISE 22$\mu$m)
are measured are described in Sec. 2.3.
Those on the [NII] contamination are directly derived from the spectroscopic data when uncertainties on the flux of the H$\alpha$ and [NII] lines are available 
(SDSS). Otherwise, if an estimate of the [NII]/H$\alpha$ ratio is not available in the literature we assume $\sigma([NII]/H\alpha)/[NII]/H\alpha$ = 0.11, the mean value derived for the SDSS subsample.
In those objects where [NII]/H$\alpha$ is estimated by the standard relation given in eq. 5 we assume $\sigma([NII]/H\alpha)/[NII]/H\alpha$ = 0.15, the typical scatter of this relation.
The uncrtainty on the Balmer decrement is available for the HRS sample of galaxies (Boselli et al. 2015) and can be derived from the SDSS spectra. For the remaining galaxies, 
we assume $\sigma(A(H\alpha))/H\alpha$ = 0.4, a value close to the dispersion of eq. 6 and to the mean value derived for the SDSS and integrated spectra ($\sim$ 0.34).
We consider that the use of the SDSS nuclear spectra to quantify the typical line ratios of the observed galaxies do not introduce any further major uncertainty (aperture
correction). The mean uncertainty of the sample on the H$\alpha$ luminosity derived in this way is $\sigma(L(H\alpha))$ = 0.16 dex, and should be considered as a lower limit
in the uncertainty on the star formation rate. Indeed, as mentioned above, a further source of uncertainty is the conversion of H$\alpha$ luminosities into star formation rates
which depends on the adopted model (assumed IMF, stellar population synthesis model, metallicity, star formation history etc.).  
This modelling uncertainty is $\simeq$ 0.10 dex, and as been estimated by comparing the output of different SED fitting codes on various samples of galaxies 
(Conroy et al. 2009; Hunt et al. 2019). Since this 0.10 dex modeling uncertainty also includes the uncertainty on the photometric data, we can estimate an upper limit
on the uncertainty of the star formation rate by adding in quadrature the modelling uncertainty (0.10 dex) to the uncertainty on the H$\alpha$ luminosities derived from the data.
The resulting mean uncertainty on $log SFR$ is thus 0.16 $\lesssim$ $\sigma(log SFR)$ $\lesssim$ 0.20 dex.  

Combined with stellar masses, these values are used to estimate specific star formation rates defined as:

\begin{equation}
{SSFR ~~[\rm{yr^{-1}}]= \it{\frac{SFR ~~[\rm{M_{\odot} yr^{-1}}]}{M_{star} ~~[\rm{M_{\odot}}]} = \frac{b}{t_0(1-R)}}  }
\end{equation}

\noindent
where $b$ is the birthrate parameter (e.g. Sandage 1986), $R$ the returned gas fraction, generally taken equal to $R$=0.3 (Boselli et al. 2001)
and $t_0$ the age of galaxies ($t_0$ = 13.5 Gyr).
Star formation rates and atomic gas masses are also used to estimate HI gas depletion timescales, defined as:

\begin{equation}
{\tau_{HI} ~~[\rm{yr}]= \it{\frac{M_{HI} ~~[\rm{M_{\odot}}]}{SFR ~~[\rm{M_{\odot} yr^{-1}}]} = \frac{1}{SFE_{HI} ~~[\rm{yr^{-1}}]}}}
\end{equation} 

\noindent
where $SFE_{HI}$ is the efficiency with which the HI gas is transformed into stars. As remarked in Boselli et al. (2014c), the gas depletion timescale 
corresponds to the Roberts time whenever the recycled gas fraction is also taken into account (Roberts 1963; Boselli et al. 2001). This is an approximative estimate
of the time necessary to stop the star formation activity. 

\section{Scaling relations}

We use this unique set of H$\alpha$ narrow-band imaging data to derive the main scaling relations involving the star formation rate of unperturbed and cluster galaxies down to 
stellar masses of a few 10$^6$ M$_{\odot}$. As done in previous works, we identify unperturbed and perturbed systems using the HI-deficiency parameter as explained in Sec. 2.4. 
We describe here the main scaling relations linking the specific star formation rate and the HI gas consumption timescale with the galaxy stellar mass and stellar mass surface density.
Those relating these parameters with the morphological type are given in Appendix A.

\subsection{Specific star formation rate}

\begin{figure*}
\centering
\includegraphics[width=0.8\textwidth]{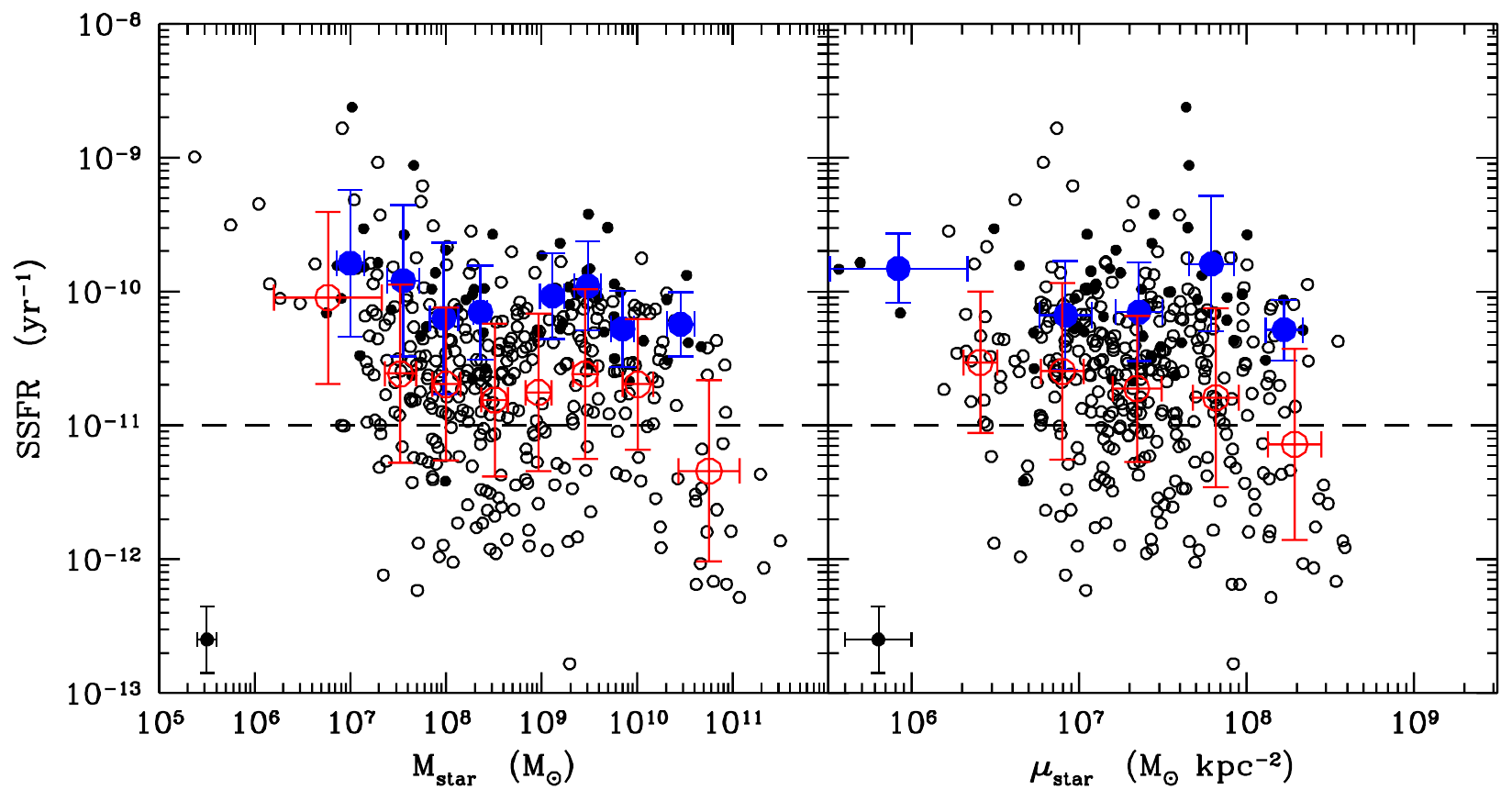}
\caption{Relationship between the specific star formation rate and the stellar mass (left) and stellar mass surface density (right) for HI-normal ($HI-def$ $\leq$0.4; filled dots) and 
HI-deficient ($HI-def$ $>$0.4; empty circles) galaxies. The large filled blue dots indicate the mean values for normal 
gas-rich systems and the empty red ones for cluster HI-deficient galaxies. For the large symbols, the error bar shows the standard deviation of the distribution.
The dashed line shows the limit between star forming and quiescent galaxies. The typical uncertainty in the data is shown at the lower left corner of each panel.
}
\label{scaling}%
\end{figure*}

Figure \ref{scaling} shows the variation of the specific star formation rate as a function of the stellar mass and the
stellar mass surface density. The same figure can be compared to those previously derived for massive objects in the HRS, where the gas-poor systems were mainly 
Virgo cluster galaxies (Boselli et al. 2015).
Figure \ref{scaling} shows that the specific star formation rate of gas-rich galaxies is fairly constant with stellar mass up to $\sim$ 10$^{10.5}$ M$_{\odot}$, 
as already noticed in previous works (e.g. Boselli et al. 2001; Gavazzi et al. 2015). 
The $SSFR$ of HI-rich systems is also fairly constant with the stellar mass surface density.
There is, however, a clear segregation with the amount of atomic gas since all gas-rich galaxies have specific star formation rates higher than those of gas-poor systems of 
similar stellar mass or stellar mass surface density. This systematic difference was already noticed in Boselli et al. (2015) for galaxies with $M_{star}$ $\geq$ 10$^9$ M$_{\odot}$, 
but it is now extended down to $M_{star}$ $\lesssim$ 10$^7$ M$_{\odot}$.

\begin{table*}
\caption{Coefficients of the main sequence relation}
\label{fit}
{
\[
\begin{tabular}{ccccccccc}
\hline
\noalign{\smallskip}
\hline
y		&x			& 	sample         	& slope     	& intercept     & pivot~point 	& $\rho$& intrinsic~scatter 	&	N.obj\\
\hline	
$SFR$		& $M_{star}$		& $HI-def$ $\leq$0.4, a	& 0.92$\pm$0.06 & -1.57$\pm$0.06& 8.451		& 0.92	& 0.42$\pm$0.05		& 56	\\
		&			& $HI-def$ $\leq$0.4, b	& 0.93$\pm$0.06 & -1.58$\pm$0.06& 8.451		& 0.92	& 0.38$\pm$0.05		& 56	\\
\hline	
		&			& $HI-def$ $\leq$0.3, a	& 0.89$\pm$0.07 & -1.01$\pm$0.09& 9.001		& 0.91	& 0.45$\pm$0.07		& 33	\\
		&			& $HI-def$ $\leq$0.3, b	& 0.89$\pm$0.07 & -1.01$\pm$0.08& 9.001		& 0.91	& 0.41$\pm$0.08		& 33	\\
		&			& $HI-def$ $\leq$0.5, a	& 0.90$\pm$0.05 & -1.59$\pm$0.06& 8.488		& 0.90	& 0.43$\pm$0.04		& 73	\\
		&			& $HI-def$ $\leq$0.5, b	& 0.91$\pm$0.05 & -1.59$\pm$0.05& 8.489		& 0.91	& 0.40$\pm$0.05		& 73	\\
\hline
		&			& $HI-def$ $>$0.4,a	& 0.81$\pm$0.04	& -2.13$\pm$0.04& 8.557		& 0.82	& 0.60$\pm$0.03		& 328 \\
		&			& $HI-def$ $>$0.4,b	& 0.82$\pm$0.03	& -2.14$\pm$0.03& 8.557		& 0.83	& 0.57$\pm$0.03		& 328 \\
\noalign{\smallskip}
\hline
\end{tabular}
\]
Notes: in sample a the uncertainties on the two variables are those derived through error propagation ($SFR$) or given by the SED fitting analysis ($M_{star}$), 
in sample b they also include modeling uncertainties. 
The intercept is measured at the pivot point; $\rho$ is the Spearman correlation coefficient. The best fit is shown as a solid line in Fig. \ref{mainHIavg}.\\
}
\end{table*}

\begin{table}
\caption{Average scaling relations }
\label{Tabscaling}
{\scriptsize
\[
\begin{tabular}{cccccc}
\hline
\hline
\noalign{\smallskip}
\multicolumn{3}{c}{}&\multicolumn{1}{c}{}&\multicolumn{1}{c}{}&\multicolumn{1}{c}{}\\
y		&	x	&  sample	&$<x>$ 		&	 $<y>$    &  N  \\
\hline
log$SFR$      	& log$M_{star}$	&$HI-def \leq$0.4&10.57$\pm$0.08	&  0.35$\pm$0.26 & 3    \\ %
		&		&		 &10.27$\pm$0.14	& -0.02$\pm$0.27 & 4	\\ %
		&		&		 & 9.72$\pm$0.12	& -0.39$\pm$0.30 & 9	\\ %
		&		&		 & 9.30$\pm$0.16	& -0.68$\pm$0.36 & 10	\\ %
		&		&		 & 8.95$\pm$-   	& -1.33$\pm$-    & 1	\\ %
		&		&		 & 8.32$\pm$0.11	& -1.82$\pm$0.35 & 10	\\ 
		&		&		 & 7.79$\pm$0.15	& -2.30$\pm$0.64 & 9	\\
		&		&		 & 7.17$\pm$0.14	& -2.57$\pm$0.53 & 7	\\
\hline
log$SFR$      	& log$M_{star}$	&$HI-def >$0.4   &10.87$\pm$0.25	& -0.61$\pm$0.62 & 21   \\ %
		&		&		 &10.19$\pm$0.11	& -0.55$\pm$0.59 & 22	\\ %
		&		&		 & 9.73$\pm$0.16	& -0.82$\pm$0.44 & 30	\\ %
		&		&		 & 9.29$\pm$0.14	& -1.44$\pm$0.71 & 39	\\ %
		&		&		 & 8.74$\pm$0.14	& -2.02$\pm$0.60 & 59	\\ %
		&		&		 & 8.26$\pm$0.15	& -2.51$\pm$0.53 & 62	\\
		&		&		 & 7.78$\pm$0.13	& -2.94$\pm$0.61 & 55	\\
		&		&		 & 7.30$\pm$0.11	& -3.12$\pm$0.63 & 30	\\		
\hline
log$SSFR$      	& log$M_{star}$	&$HI-def \leq$0.4&10.46$\pm$0.14	&-10.24$\pm$0.24 & 6   \\ 
		&		&		 & 9.85$\pm$0.12	&-10.28$\pm$0.29 & 6	\\
		&		&		 & 9.49$\pm$0.14	& -9.96$\pm$0.33 & 10	\\
		&		&		 & 9.11$\pm$0.13	&-10.03$\pm$0.32 & 5	\\
		&		&		 & 8.36$\pm$0.08	&-10.16$\pm$0.35 & 8	\\
		&		&		 & 7.98$\pm$0.15	&-10.20$\pm$0.57 & 7	\\
		&		&		 & 7.55$\pm$0.17	& -9.92$\pm$0.57 & 6	\\
		&		&		 & 7.00$\pm$0.14	& -9.79$\pm$0.55 & 8	\\
		& $\mu_{star}$	&		 & 5.92$\pm$0.41	& -9.83$\pm$0.26 &  4	\\
		&		&		 & 6.92$\pm$0.16	&-10.18$\pm$0.41 & 16	\\
		&		&		 & 7.36$\pm$0.14 	&-10.15$\pm$0.37 & 20	\\
		&		&		 & 7.79$\pm$0.13 	& -9.79$\pm$0.51 & 13	\\
		&		&		 & 8.22$\pm$0.11	&-10.29$\pm$0.23 &  3	\\
\hline
log$SSFR$      	& log$M_{star}$	&$HI-def >$0.4   &10.75$\pm$0.32	&-11.34$\pm$0.68 & 27   \\ 
		&		&		 &10.01$\pm$0.16	&-10.69$\pm$0.49 & 31	\\
		&		&		 & 9.46$\pm$0.13	&-10.62$\pm$0.64 & 40	\\
		&		&		 & 8.97$\pm$0.14	&-10.76$\pm$0.59 & 40	\\
		&		&		 & 8.51$\pm$0.14	&-10.81$\pm$0.57 & 64	\\
		&		&		 & 8.00$\pm$0.15	&-10.69$\pm$0.57 & 61	\\
		&		&		 & 7.52$\pm$0.16	&-10.61$\pm$0.66 & 45	\\
		&		&		 & 6.77$\pm$0.57	&-10.05$\pm$0.64 & 20	\\
		& $\mu_{star}$	&		 & 6.41$\pm$0.10	&-10.53$\pm$0.53 & 21	\\
		&		&		 & 6.90$\pm$0.13 	&-10.60$\pm$0.66 & 78	\\
		&		&		 & 7.35$\pm$0.15 	&-10.73$\pm$0.55 &120	\\
		&		&		 & 7.82$\pm$0.14	&-10.79$\pm$0.67 & 71	\\
		&		&		 & 8.29$\pm$0.16	&-11.14$\pm$0.71 & 30	\\
\hline
log$\tau_{HI}$	& log$M_{star}$	&$HI-def \leq$0.4&10.46$\pm$0.14	&  9.44$\pm$0.24 & 6    \\ 
		&		&		 & 9.85$\pm$0.12	&  9.81$\pm$0.34 & 6	\\
		&		&		 & 9.49$\pm$0.14	&  9.78$\pm$0.36 & 10	\\
		&		&		 & 9.11$\pm$0.13	& 10.08$\pm$0.33 & 5	\\
		&		&		 & 8.36$\pm$0.08	& 10.35$\pm$0.50 & 8	\\
		&		&		 & 7.98$\pm$0.15	& 10.50$\pm$0.53 & 7	\\
		&		&		 & 7.55$\pm$0.17	& 10.44$\pm$0.55 & 6	\\
		&		&		 & 7.00$\pm$0.14	& 10.81$\pm$0.67 & 8	\\
		& $\mu_{star}$	&		 & 5.92$\pm$0.41	& 11.10$\pm$0.39 &  4	\\
		&		&		 & 6.92$\pm$0.16	& 10.53$\pm$0.51 & 16	\\
		&		&		 & 7.36$\pm$0.14 	& 10.08$\pm$0.50 & 20	\\
		&		&		 & 7.79$\pm$0.13 	&  9.61$\pm$0.36 & 13	\\
		&		&		 & 8.22$\pm$0.11	&  9.94$\pm$0.31 &  3	\\
		& log$SSFR$     &		 &-9.44$\pm$0.34	&  9.77$\pm$0.43 & 10   \\
		&		&		 &-10.00$\pm$0.13	& 10.16$\pm$0.64 & 28	\\
		&		&		 &-10.44$\pm$0.14	& 10.32$\pm$0.54 & 17	\\	
\hline
log$\tau_{HI}$  & log$M_{star}$	&$HI-def >$0.4   &10.75$\pm$0.32	&  8.67$\pm$0.85 & 27   \\ 
		&		&		 &10.01$\pm$0.16	&  9.01$\pm$0.50 & 31	\\
		&		&		 & 9.46$\pm$0.13	&  9.36$\pm$0.66 & 40	\\
		&		&		 & 8.97$\pm$0.14	&  9.80$\pm$0.45 & 40	\\
		&		&		 & 8.51$\pm$0.14	& 10.02$\pm$0.45 & 64	\\
		&		&		 & 8.00$\pm$0.15	& 10.34$\pm$0.52 & 61	\\
		&		&		 & 7.52$\pm$0.16	& 10.46$\pm$0.65 & 45	\\
		&		&		 & 6.77$\pm$0.57	& 10.54$\pm$0.52 & 20	\\
		& $\mu_{star}$	&		 & 6.41$\pm$0.10	& 10.70$\pm$0.43 & 21	\\
		&		&		 & 6.90$\pm$0.13 	& 10.25$\pm$0.55 & 78	\\
		&		&		 & 7.35$\pm$0.15 	&  9.91$\pm$0.66 &120	\\
		&		&		 & 7.82$\pm$0.14	&  9.26$\pm$0.73 & 71	\\
		&		&		 & 8.29$\pm$0.16	&  9.22$\pm$0.94 & 30	\\
		&log$SSFR$	&		 &-9.36$\pm$0.28	&  9.70$\pm$0.68 & 14	\\
		&		&		 &-10.06$\pm$0.14	&  9.63$\pm$0.66 & 63	\\
		&		&		 &-10.50$\pm$0.13	&  9.92$\pm$0.70 &108	\\
		&		&		 &-10.97$\pm$0.13	&  9.97$\pm$0.75 & 74	\\
		&   		&		 &-11.67$\pm$0.31	&  9.87$\pm$1.12 & 69	\\
\noalign{\smallskip}
\hline
\end{tabular}
\]
Note: mean values and standard deviations for the scaling relations (big symbols in Fig. \ref{scaling}, \ref{scalinggas}, and \ref{mainHIavg}). }
\end{table}

\subsection{Gas depletion timescale}

\begin{figure*}
\centering
\includegraphics[width=1.0\textwidth]{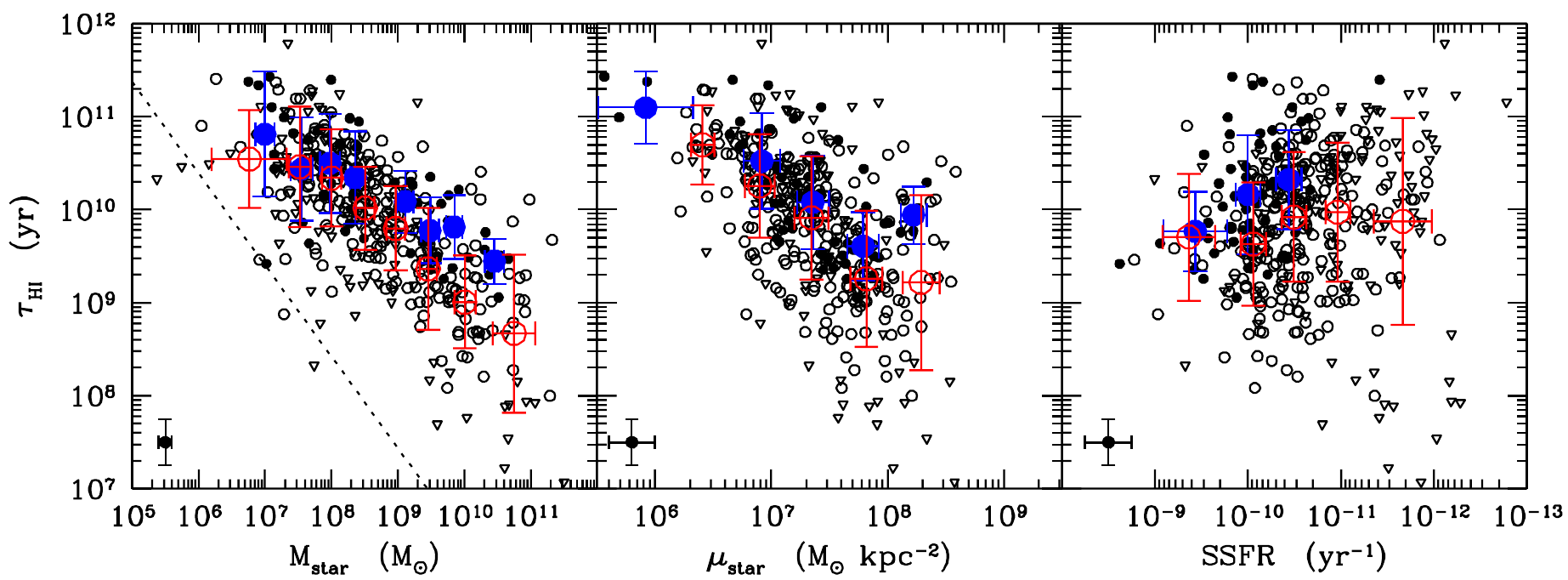}
\caption{Relationship between the HI gas depletion timescale and the stellar mass (left), the stellar mass surface density (centre), 
and the specific star formation rate (right) for HI-normal ($HI-def$ $\leq$0.4; filled dots) and 
HI-deficient ($HI-def$ $>$0.4; empty circles) galaxies. The large filled blue dots indicate the mean values for each morphological class for normal 
gas-rich systems and the empty red ones for cluster HI-deficient galaxies. For the large symbols, the error bar shows the standard deviation of the distribution.
Upper limits are indicated by triangles and are treated as detections in the derivation of the mean values. The dotted line indicates the detection limit of the ALFALFA survey.
The typical uncertainty in the data is shown at the lower left corner of each panel.}
\label{scalinggas}%
\end{figure*}

Figure \ref{scalinggas} shows the variation of the HI gas depletion timescale as a function of  
the stellar mass, the stellar mass surface density, and the specific star formation rate. 
Clearly, Fig. \ref{scalinggas} shows a decrease of the HI gas depletion timescale with increasing stellar mass and stellar mass surface density
while no evident trend is present as a function of the specific star formation rate. Clearly, dwarf galaxies and low surface brightness systems have HI gas reservoirs able to
sustain star formation at the present rate for more than 10 Gyr, while very massive objects ($M_{star}$ $\geq$ 10$^{10}$ M$_{\odot}$) only for a couple of Gyr (see also McGaugh et al. 2017).
Stricking is the similarity in the HI gas depletion timescale measured in unperturbed gas-rich and perturbed gas-poor systems, in particular in the stellar 
mass range 10$^7$ $\lesssim$ $M_{star}$ $\lesssim$ 10$^9$ M$_{\odot}$. We also notice that the scatter in the $\tau_{HI}$ vs. $M_{star}$ relation drawn by 
HI-rich systems ($HI-def$ $\leq$ 0.4) increases with decreasing stellar mass.

\section{Main sequence relation}

\subsection{General properties}

\begin{figure}
\centering
\includegraphics[width=0.5\textwidth]{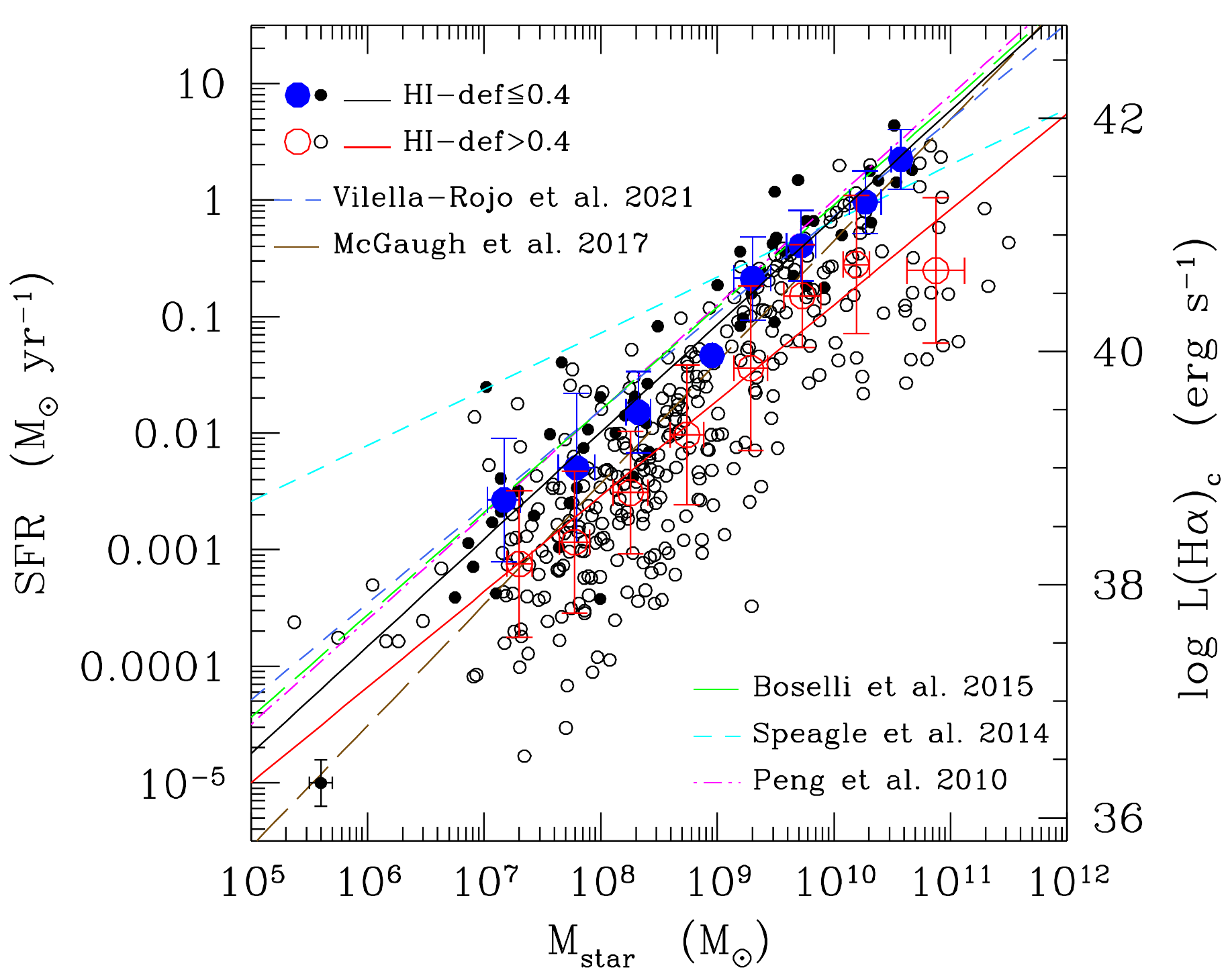}
\caption{Main sequence relation for galaxies in the Virgo cluster coded according to their HI gas content: filled 
dots are for HI gas-rich objects ($HI-def$ $\leq$ 0.4), empty circles for HI-deficient objects ($HI-def$ $>$ 0.4). Big solid blue and
empty red circles are the mean values and standard deviations for HI-normal and HI-deficient objects. Star formation rates 
have been derived assuming stationary conditions. The Y-axis on the right side gives the corresponding H$\alpha$ luminosities corrected for dust attenuation.
The black solid and dotted lines are the bisector and linear fit obtained for gas-rich star forming systems. The best fits are compared 
to those derived for the bright galaxies with $HI-def$ $\leq$ 0.4 included in the HRS (Boselli et al. 2015, bisector fit) and Speagle et al. (2014, 
derived extrapolating the time dependent best fit relation to $z$=0), for the
SDSS sample of Peng et al. (2010) derived for galaxies with $M_{star}$ $\geq$ 10$^{8.5}$ M$_{\odot}$, for the local sample of Vilella-Rojo et a. (2021), 
and for the low surface brightness sample of McGaugh et al. (2017). The typical uncertainty in the data is shown at the lower left corner.
}
\label{mainHIavg}%
\end{figure}

Figure \ref{mainHIavg} shows the main sequence relation derived with the 384 H$\alpha$ detected galaxies analysed in this work. 
The improvement given by the unique set of data gathered thanks to the VESTIGE survey 
is stricking: it allows to extend by $\sim$ 2 orders of magnitude the main sequence relation previously available for the Virgo cluster,
which now ranges bewteen 10$^6$ $\lesssim$ $M_{star}$ $\lesssim$ 3$\times$10$^{11}$ M$_{\odot}$ and 10$^{-5}$ $\lesssim$ $SFR$ $\lesssim$ 10 M$_{\odot}$ yr$^{-1}$, 
a dynamic range never covered so far in any local or high-$z$ sample. 

\begin{figure}
\centering
\includegraphics[width=0.5\textwidth]{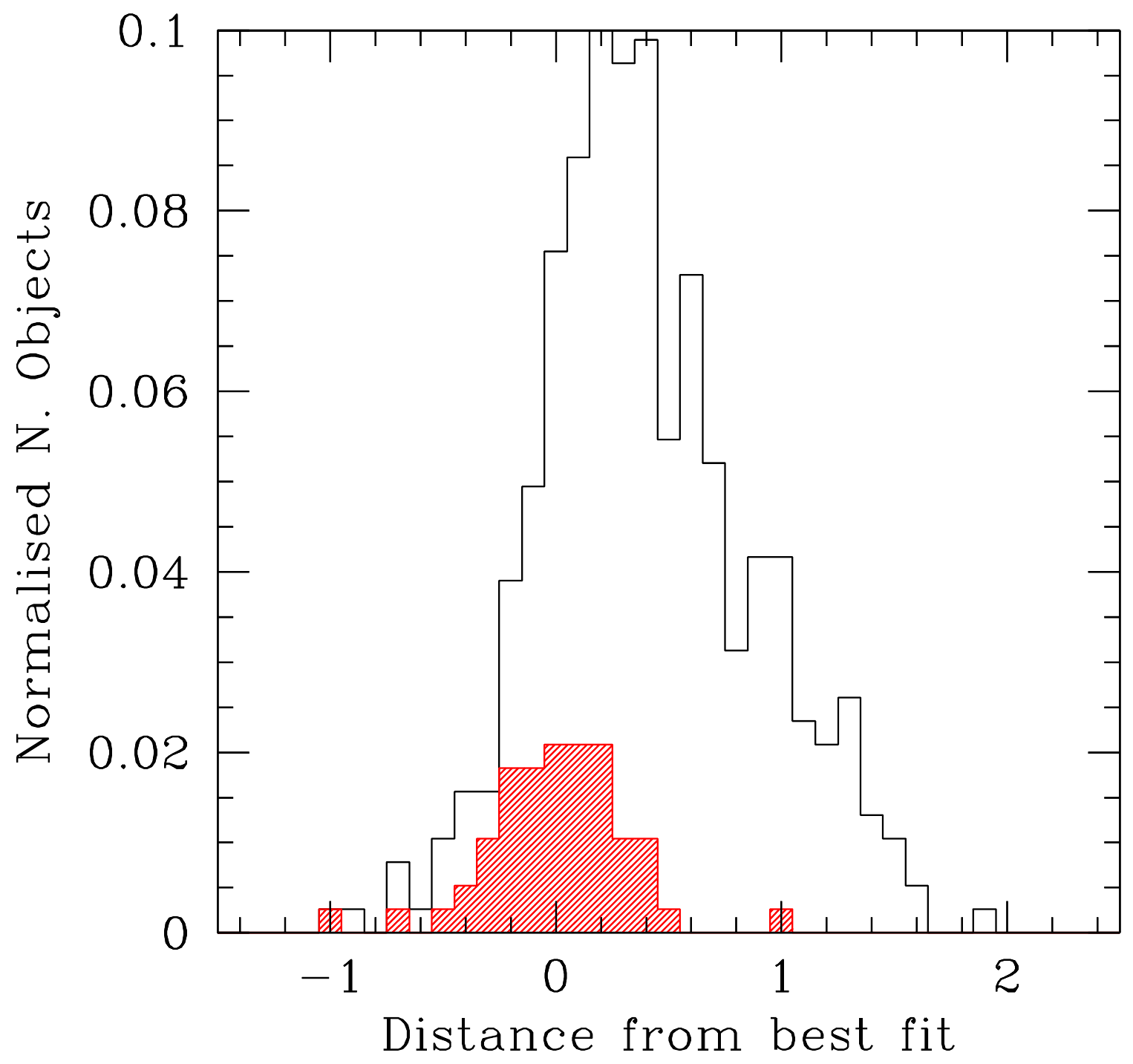}
\caption{Distribution of the orthogonal distance from the best fit of the main sequence relation for the whole sample of galaxies (black solid histogram) and for the 
galaxies with a normal HI gas content ($HI-def$ $\leq$ 0.4, red hatched histogram).
}
\label{scatter}%
\end{figure}

Following Boselli et al. (2015) we can consider as unperturbed systems those galaxies with a normal HI gas content ($HI-def$ $\leq$ 0.4, 56 objects.) We derive the 
main sequence relation by fitting the data using the \textit{linmix} package for python\footnote{https://github.com/jmeyers314/linmix} (Kelly 2007)
which considers in a Bayesian framework the uncertainties on both axis as well as an intrinsic scatter in the relation.
The linear fit is done using a Markov Chain Monte Carlo to quantify the statistical uncertainties on the output parameters (see Fig. \ref{mainHIavg}).
The parameters derived from the fit are given in Table \ref{fit} using two different estimates of the uncertainties on the stellar mass
and on the star formation rate, the first one considering only the uncertainties given by the SED fitting analysis ($M_{star}$) or derived through error propagation ($SFR$), 
the second one adding in quadrature an uncertainty related to the adopted models (see Sec. 3). 
To quantify any possible dependence of the fitted parameters on the adopted sample, we derive also the fitted relation using two different cuts 
in the HI-deficiency parameter ($HI-def$ $\leq$ 0.3 and $HI-def$ $\leq$ 0.5). Figure \ref{scatter} shows the distribution of the orthogonal distance from the derived main sequence relation
for the whole sample and for the HI-normal galaxies ($HI-def$ $\leq$ 0.4). Table \ref{fit} indicates that the slope of the fitted relation is robust vs. the 
adopted cut in the HI-deficiency parameter or the assumed uncertainties on the data. Figure \ref{scatter} shows that  
galaxies with a normal HI gas content ($HI-def$ $\leq$ 0.4) are symmetrically distributed on teh derived best fit relation. 

The slope ($a$=0.92) is
very close to that previously estimated for the HRS galaxies by Boselli et al. (2015) or for a large statistical sample of SDSS galaxies by Peng et al. (2010),
this last sampled only for $M_{star}$ $\geq$ 10$^{8.5}$ M$_{\odot}$. The slope is, however, significantly steeper than the one derived by Speagle et al. (2014)
when extrapolating the time-dependent main sequence relation to $z$=0, suggesting that its determination in the local universe is still highly uncertain.
Its value is in between the one derived from the J-PLUS local sample of Vilella-Rojo et al. (2021) ($a$=0.83) 
and the one of local low surface brightness galaxies of McGaugh et al. (2017) ($a$=1.04) both derived using H$\alpha$ narrow-band imaging data. These two samples span a quite large dynamic
range in stellar mass (10$^7$ $\leq$ $M_{star}$ $\leq$ 10$^{11}$ M$_{\odot}$) but are both dominated by objects with $M_{star}$ $\geq$ 10$^8$ M$_{\odot}$. 

When limited to unperturbed systems ($HI-def$$\leq$0.4), the intrinsic scatter of the relation ($\sigma$ $\simeq$ 0.40) is comparable to that observed in other
samples of local galaxies (e.g. Whitaker et al. 2012; Speagle et al. 2014; Ilbert et al. 2015; Gavazzi et al. 2015; Popesso et al. 2019; Vilella-Rojo et al. 2021).
The relation does not show any evident bending at high masses (e.g. Bauer et al. 2013, Gavazzi et al. 2015, Popesso et al. 2019), although this effect might be 
hampered by the limited number of massive systems present in such a small sampled volume. 
It also does not show any change of slope in the low stellar mass regime ($M_{star}$ $\lesssim$ 10$^8$ M$_{\odot}$) now sampled for the first 
time with an excellent statistics. There is, however, a significant increase of the intrinsic scatter of the relation with decreasing stellar mass: $\sigma$ $\sim$ 0.4 dex 
for the whole sample of HI-normal galaxies, $\sigma$ $\sim$ 0.27 for massive systems ($M_{star}$ $>$ 10$^{9}$ M$_{\odot}$), and $\sigma$ $\sim$ 0.48 for dwarfs
($M_{star}$ $<$ 10$^9$ M$_{\odot}$; see also Table \ref{Tabscaling}).
Interesting is also the fact that the slope of the relation is very close for HI-normal ($a$=0.92) and HI-deficient galaxies ($a$=0.82),
suggesting that the process responsible for the gas depletion and for the subsequent quenching of the star formation process acts indifferently at all stellar masses.
 


\begin{table*}[hbt!]
\caption{Galaxies at more than 1$\sigma$ above the main sequence}
\label{aboveMS}
{
\[
\begin{tabular}{cccccccc}
\hline
Name	     & Type    & log $M_{star}$ & log $SFR$ & log $SSFR$ &  log $\tau_{HI}$ & $HI-def$  & Comments \\
\noalign{\smallskip}
	     &	       & M$_{\odot}$    &  M$_{\odot}$ yr$^{-1}$ & yr$^{-1}$ & yr   &    &     \\
\hline
      VCC207 &       BCD   &   7.661 &    -1.392  &   -9.053 &   9.64	&      0.08&  				\\
      VCC428 &       BCD   &   6.912 &    -1.862  &   -8.773 &   9.46	&      0.74& filaments in a loop\\
      VCC562 &       BCD   &   7.737 &    -1.589  &   -9.326 &   8.33	&   $>$1.54& very asymmetric, tails \\
      VCC683 &        dE   &   7.666 &    -1.613  &   -9.278 &   8.61	&   $>$0.90& merging ring?  \\
      VCC801 &Sa? pec HII  &   9.692 &     0.170  &   -9.522 &   9.26	&      0.06& starburst with clear outflows \\
     VCC1313 &       BCD   &   7.015 &    -1.604  &   -8.618 &   9.42	&      0.26&  \\
     VCC1554 &    IBm HII  &   9.490 &     0.071  &   -9.420 &   9.36	&      0.06& starburst with clear outflows \\
     VCC1744 &       BCD   &   7.286 &    -1.747  &   -9.033 &   8.88	&      1.03&  \\
   AGC224696 &   Im/BCD    &   7.755 &    -1.452  &   -9.208 &   9.58	&      0.48& very asymmetric  \\
   AGC226326 &       Im    &   7.038 &    -2.274  &   -9.312 &  10.19	&      0.47& tail  \\
 \noalign{\smallskip}
\hline
\end{tabular}
\]
Notes: galaxies at more than 1$\sigma$ above the main sequence with stellar mass $M_{star}$ $>$ 8$\times$10$^6$ M$_{\odot}$.
}
\end{table*}

\subsection{Galaxies above the main sequence}

\begin{figure}
\centering
\includegraphics[width=0.5\textwidth]{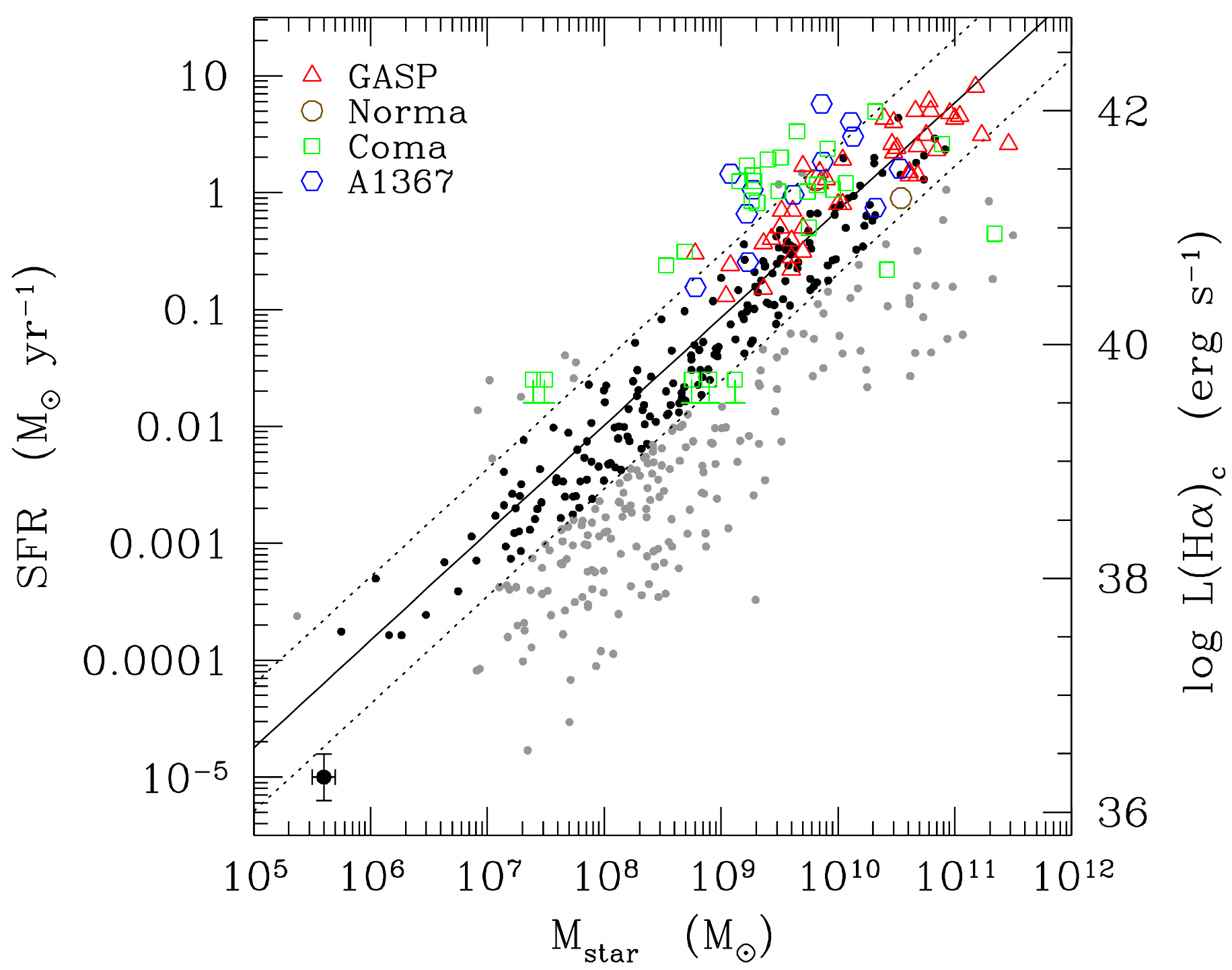}
\caption{Main sequence relation for galaxies coded according to their distance from the bisector fit derived for HI gas-rich late-type systems. Black filled dots
are for galaxies within 1 $\sigma$ from the relation, grey filled dots for galaxies $>$ 1$\sigma$ above or below the relation. The solid line gives the best 
fit obtained for gas-rich star forming systems, the dotted lines the limits 1$\sigma$ above and below the best fit. Galaxies undergoing a ram pressure stripping event
identified in Boselli et al. (2022a) are indicated with large coloured empty symbols: brown circles for those belonging to the Norma cluster, green squares 
for Coma, and blue exagons for A1367. Red triangles indicate the jellyfish sample of Vulcani et al. (2018). Star formation rates and stellar masses are derived 
consistently with those measured within the Virgo cluster. The typical uncertainty in the data is shown at the lower left corner.
}
\label{mainComaA1367GASP}%
\end{figure}

Galaxies located above the main sequence are systems undergoing a starburst event often related to a merging episode (e.g. Rodighiero et al. 2011)
or in clusters to a violent ram pressure stripping event (Vulcani et al. 2018). 
Ten objects (3\% of the whole sample) with stellar mass $M_{star}$ $>$ 8$\times$10$^6$ M$_{\odot}$ are located at more than 1$\sigma$ above the main 
sequence relation (see Table \ref{aboveMS}). This fraction seems very limited when compared to that observed in other samples of local cluster objects 
in Coma and A1367 (Boselli et al. 2022a; Pedrini et al. 2022a) undergoing a ram pressure stripping event, 
or in the GASP sample of jellyfish galaxies (Vulcani et al. 2018; see fig. \ref{mainComaA1367GASP}). 
Quantifying the fraction of galaxies above the main sequence in these other samples, however, is impossible because of the lack of
deep and complete untargetted surveys. VESTIGE is the only one which includes a large number of star forming systems 
of limited activity lacking in the other works, where most (if not all) of the observed galaxies have been selected because of their strong
star formation activity, peculiar morphology etc., and are thus not complete samples in any sense. These other selection critaria obviously favor 
objects with strong star formation activities as those expected for galaxies located above the main sequence relation, making these samples not 
indicated for driving statistical quantities. The low number of galaxies above the main sequence in the VESTIGE data with respect to the other samples, 
however, seems real given the limited number of catalogued late-type galaxies in these clusters, particularly in Coma which is strongly dominated by 
the early-types. This potential difference might have a physical reason.
The Virgo cluster has a dymanical mass $\simeq$ a factor of 10 smaller than other massive 
clusters such as Coma and A1367. It is thus conceivable that the perturbing mechanisms at place in these more massive clusters are more efficient
in perturbing their gas rich members and triggering their star formation activity than in Virgo (e.g. Boselli et al. 2022a). 
There is, indeed, observational evidence that this is the case:
1) the spiral fraction in Virgo is significantly higher than in Coma (Boselli \& Gavazzi 2006); 2) the fraction of HI-deficient galaxies, and the 
mean HI-deficiency of late-type galaxies in Coma and A1367 is higher than in Virgo (Solanes et al. 2001, Boselli \& Gavazzi 2006); 3) the increase of the
radio continuum activity per unit far-infrared emission of galaxies in Coma and A1367 is higher than in Virgo (Gavazzi \& Boselli 1999a, 1999b; 
Boselli \& Gavazzi 2006); 4) the fraction of late-type galaxies with ionised gas tails, witnessing an ongoing ram pressure stripping process,
is also higher in Coma and A1367 ($\sim$ 50\%\, Boselli \& Gavazzi 2014; Yagi et al. 2010, 2017; Gavazzi et al. 2018b) than in Virgo, 
as revealed by the VESTIGE survey (Boselli et al. in prep.). 5) all these selection effects are expected to be even more important in the 
GASP sample of jellyfish galaxies, where the selected objects (generally one or just a few per cluster) are those showing the most extreme 
morphological perturbations.

Observations (e.g. Gavazzi et al. 1995; Vollmer 2003; Vulcani et al. 2018) and simulations (Fujita \& Nagashima 1999; Bekki 2014; Steinhauser et al. 2016
Steyrleithner et al. 2020; Troncoso-Ibarren et al. 2020; Boselli et al. 2021) consistently indicate that under some conditions the star formation activity of cluster
galaxies can be triggered by the compression of the gas during a ram pressure stripping process (Boselli et al. 2022a). The analysis of the ionised gas morphological
properties of the 15 Virgo cluster galaxies more than 1$\sigma$ above the main sequence can help us to quantify how many of these objects are suffering 
this perturbing mechanism. We recall that the intrinsic scatter of the relation is $\sigma$ = 0.40 dex, i.e. these objects have their star formation activity 
increased by more than a factor of 2.5. The continuum subtracted H$\alpha$ images of all these objects are shown in Appendix B, and their 
physical parameters are given in Table \ref{aboveMS}. A large fraction of the sample is gas-rich (4 galaxies with $HI-def$ $\leq$ 0.4 vs. 6 with $HI-def$ $>$ 0.4).
Among them there are two intermediate mass systems, VCC 801 (NGC 4383) and VCC 1554 (NGC 4532).
Both show prominent filaments of ionised gas escaping from the galaxy disc, suggesting that an outflow of gas is triggered by the strong starburst activity.
These morphological peculiarities are not typical of galaxies undergoing a ram pressure stripping event, where the ionised gas is rather compressed along a 
curved region at the leading edge of the interaction with the surrounding ICM and a low surface brightness tail formed at the opposite side
(typical examples are CGCG 97-73 in A1367, Gavazzi et al. 1995, 2001, and IC 3476 in Virgo, Boselli et al. 2021). They are rather typical in merging 
systems such as M82 (Shopbel \& Bland-Hawthorn 1998; Mutchler et al. 2007). The remaining galaxies, 
all dwarfs with 8$\times$10$^6$ $<$ $M_{star}$ $<$ 10$^{9.2}$ M$_{\odot}$, are mainly BCDs (6/10), 
or irregulars with a star formation activity often concentrated in a few bright and giant HII regions. Thier bursty activity is evident but not always
related to an external perturbing mechanism. 
We can thus conclude that in a cluster such as Virgo (a few 10$^{14}$ M$_{\odot}$) ram pressure stripping,
identified as the dominant perturbing mechanism (e.g. Boselli et al. 2014a; 2022a), on average does not significantly increase the activity of star formation
of the perturbed galaxies. If this happens, it occurs on very short timescales (Boselli et al. 2021), so that the probability to see galaxies with an incresed 
activity remains very low.

\subsection{Origin of the scatter}

\begin{figure}
\centering
\includegraphics[width=0.5\textwidth]{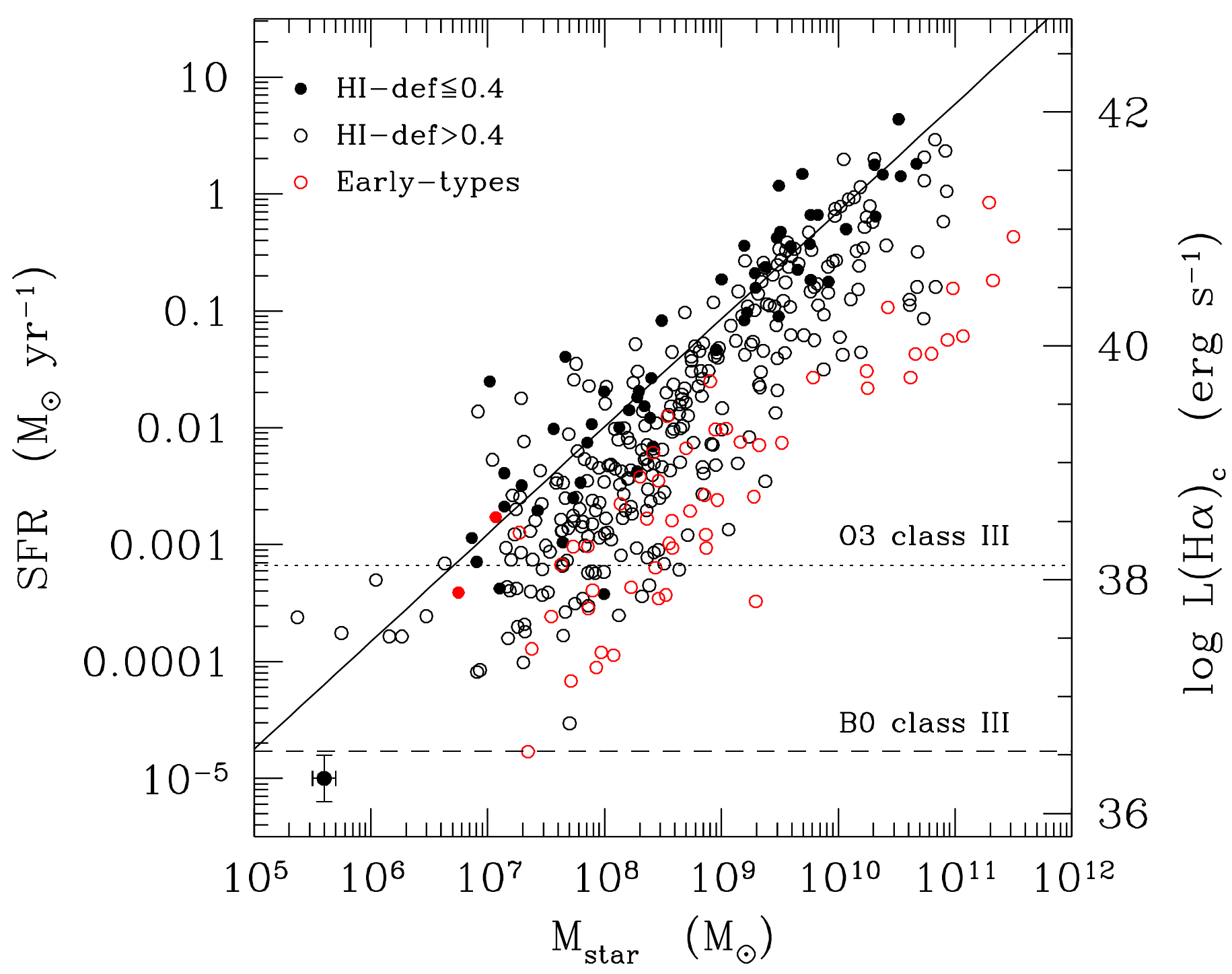}
\caption{Main sequence relation for galaxies coded according to their morphological type and HI gas content: late-type galaxies ($\geq$ Sa) are indicated by black symbols, 
early-type galaxies by red symbols. Filled dots are for HI gas rich objects ($HI-def$ $\leq$ 0.4), while HI-deficient objects ($HI-def$ $>$ 0.4) by empty circles. Star formation rates 
have been derived assuming stationary conditions. The Y-axis on the right side gives the corresponding H$\alpha$ luminosities corrected for dust attenuation.
The black solid and dotted lines are the bisector and linear fit obtained for gas-rich star forming systems. The horizontal dotted and dashed lines indicate the corresponding 
SFR derived using the number of ionising photons produced by an O3 class III and a B0 class III star as derived using the model atmospheres of Sternberg et al. (2003).
The typical uncertainty in the data is shown at the lower left corner.}
\label{mainHI}%
\end{figure}

\begin{figure}
\centering
\includegraphics[width=0.5\textwidth]{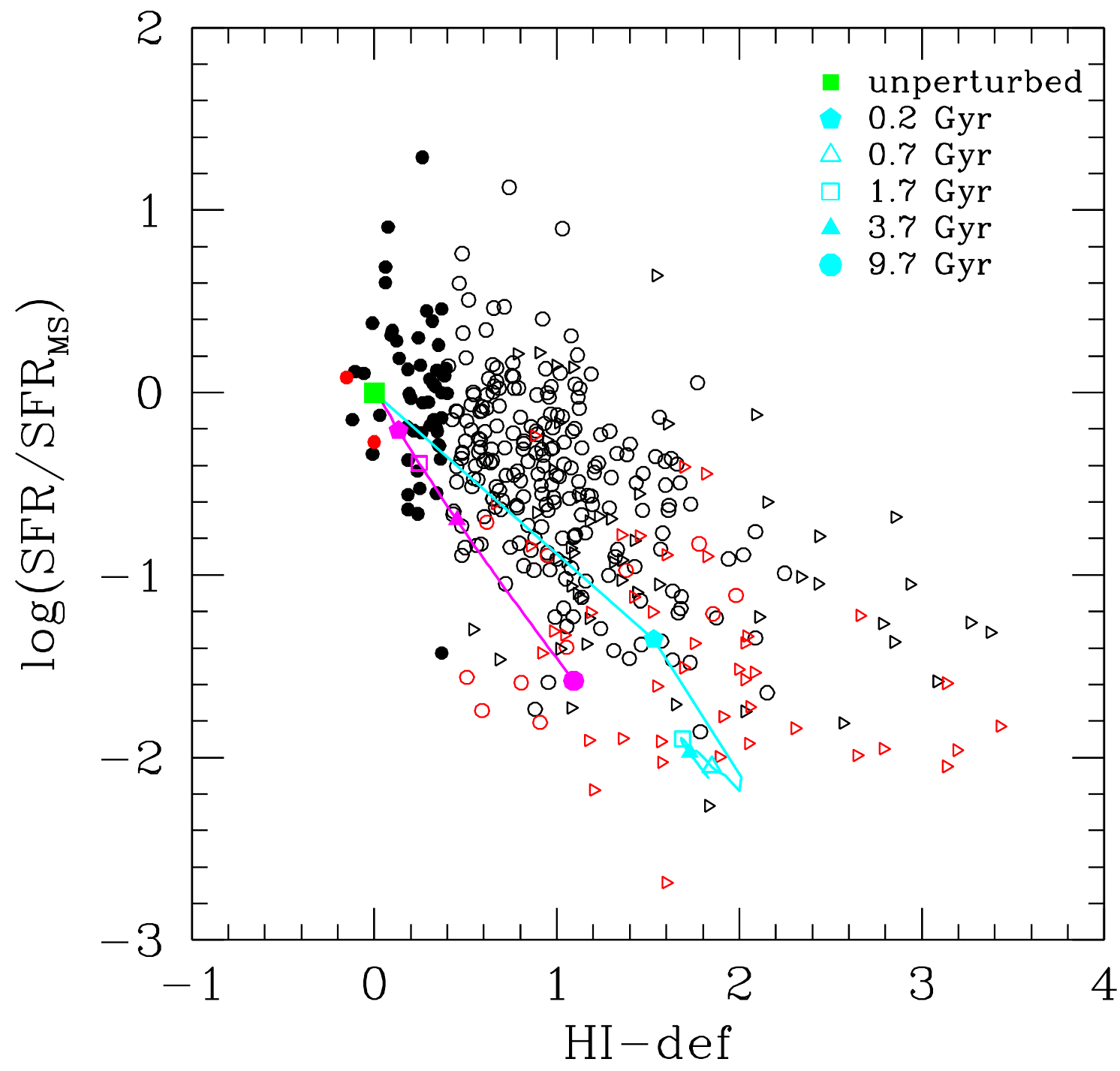}
\caption{Relationship between the offset from the main sequence relation log$SFR/SFR_{MS}$ and the HI-deficiency parameter: late-type galaxies ($\geq$ Sa) are indicated by black symbols, 
early-type galaxies by red symbols, circles for HI detected sources, triangles for HI undetected objects (lower limits to the HI-deficiency parameter). 
Filled symbols are for HI gas rich objects ($HI-def$ $\leq$ 0.4), empty symbols for HI-deficient objects ($HI-def$ $>$ 0.4). The data are compared to the 
predictions of models for starvation (magenta) and ram pressure stripping (cyan) derived for a galaxy of rotational velocity 70 km s$^{-1}$ and spin 
parameter $\lambda$=0.05. The green filled square indicates the prediction for an unperturbed system. Different symbols along the models indicate the position of the model galaxies at a given look-back
time from the beginning of the interaction. 
}
\label{distmainHI}%
\end{figure}

When measured on the whole sample, the observed relation has an important scatter ($\sim$ 1 dex) at all stellar masses, significantly larger than that observed 
in other samples of local galaxies ($\lesssim$ 0.3 dex). This large scatter is mainly due to objects located below the main sequence relation drawn
by representative samples of field galaxies or by unperturbed objects within the Virgo cluster ($HI-def$ $\leq$ 0.4). 
Interesting is the limited number of objects in the starburst sequence formed by merging systems 
observed in other H$\alpha$ selected samples at high redshift (Caputi et al. 2017) and rare in the local universe (Rodighiero et al. 2011; Sargent et al. 2012).
As already noticed by Boselli et al. (2015), the scatter in the relation is strongly related to
the HI-deficiency parameter, suggesting that the reduced activity of star formation typical of these quenched cluster galaxies is tightly
connected to their atomic gas reservoir (see Fig. \ref{mainHI}). This is clear in Fig. \ref{distmainHI}, where the offset from the main sequence
relation drawn by unperturbed HI-rich systems ($HI-def$ $\leq$ 0.4) is plotted against the HI-deficiency parameter. Figure \ref{distmainHI} indeed shows a tight relation
between the offset of galaxies from the main sequence relation and their amount of available HI gas.

The tight connection of the activity of star formation with the total atomic gas content,
both reduced in cluster galaxies, was already noticed in the past (e.g. Gavazzi et al. 2002, 2013; Boselli et al. 2014a). Although the atomic gas
does not directly participate to the star formation process, which rather depends on the molecular gas phase (e.g. Bigiel et al. 2008), 
it is the principal supplier of the giant molecular clouds where star formation takes place. Since a large fraction of the HI gas is located in the outer galaxy regions,
infall of atomic gas into the stellar disc has been invoked to explain the observed strong correlation between the total gas content and star formation in
local galaxies (e.g. Boselli et al. 2001). The tight connection between the star formation activity and the atomic gas content also suggests that the 
conversion rate from atomic to molecular gas is roughly proportional to the atomic gas mass. This is deduced by the fact that a) the star formation rate is tightly related 
at large scales to the total molecular gas content (e.g. Kennicutt 1998b; Boselli et al. 2002) and at smaller scales ($\sim$ 500 pc) to the molecular gas column density 
(e.g. Bigiel et al. 2008), and 
b) the atomic gas consumption time is similar for galaxies irrespective of their HI deficiency (in a limited but broad mass range, see Sec. 4.2).
We recall that an accurate estimate of the gas consumption time scale would require the determination of the molecular gas content, unfortunately 
unknown in this sample dominated by low-mass and low-metallicity systems, where the detection of the CO line emission is challenging. There is some evidence that 
the molecular gas phase is also reduced in HI-deficient objects, although at a lower level than the atomic gas (e.g. Fumagalli et al. 2009; Boselli et al. 2014b; Mok et al. 2017). 

The observed scatter in the relation exceeds that previously observed for Virgo galaxies in the high stellar mass range 
(10$^9$ $\lesssim$ $M_{star}$ $\lesssim$ 10$^{11}$ M$_{\odot}$). The VESTIGE sample includes objects with a star formation activity reduced even more than  
that of the HI-deficient Virgo galaxies belonging to the HRS. Figure \ref{mainHI}
indicates that these most extreme objects are mainly early-type galaxies with a low H$\alpha$ emission. In most of these objects the H$\alpha$ emission
is clearly related to an ongoing star formation activity in the inner regions, mainly located in rotating discs, with only a few cases where the ionised gas emission is associated 
to filaments where the ionising radiation could be also produced by cooling flows or shocks (M87, NGC 4262, NGC 4552; Gavazzi et al. 2000, 2018a; Boselli et al. 2019, 2022b). In the low stellar mass range ($M_{star}$ $\lesssim$ 10$^{9}$ M$_{\odot}$)
these early-type galaxies are dwarf systems with a residual star formation activity in their nucleus, as first noticed by Boselli et al. (2008a).
These early-types with a residual star formation activity were mainly lacking in previous H$\alpha$ surveys such as the follow-up observations of the HRS (Boselli et al. 2015), 
which were targetting only late-type systems, but are present and frequent in blind surveys such as VESTIGE.

\subsection{Comparison with models of ram pressure stripping and starvation}

Several perturbing mechanisms have been proposed in the literature to explain the reduced activity of star formation in cluster galaxies (e.g. Boselli \& Gavazzi 2006, 2014). 
Among these, the one generally invoked to explain the observed properties of nearby cluster galaxies, such as the presence of cometary tails of gas in its different phases without
any associated stellar structure, truncated gaseous discs vs. unperturbed stellar discs etc., is ram pressure stripping (Gunn \& Gott 1972, see the recent review of 
Boselli et al. 2022a for details). This stripping process seems dominant also within the Virgo cluster as indicated by several statistical studies 
(e.g. Vollmer et al. 2001a, Vollmer 2009, Boselli et al. 2008a,b, 2014a, 2014b, Gavazzi et al. 2013) or observations of representative objects 
(Vollmer et al. 1999, 2000, 2001b, 2004a,b, 2006, 2008a,b, 2012, 2018, 2021; Vollmer 2003; Yoshida et al. 2002, 2004; Kenney et al. 2004, 2014; Crowl et al. 2005; 
Boselli et al. 2006, 2016a, 2021; Chung et al. 2007; Pappalardo et al. 2010; Abramson et al. 2011, 2016; Abramson \& Kenney 2014; Sorgho et al. 2017; Fossati et al. 2018;
Cramer et al. 2020). 

\begin{figure}
\centering
\includegraphics[width=0.5\textwidth]{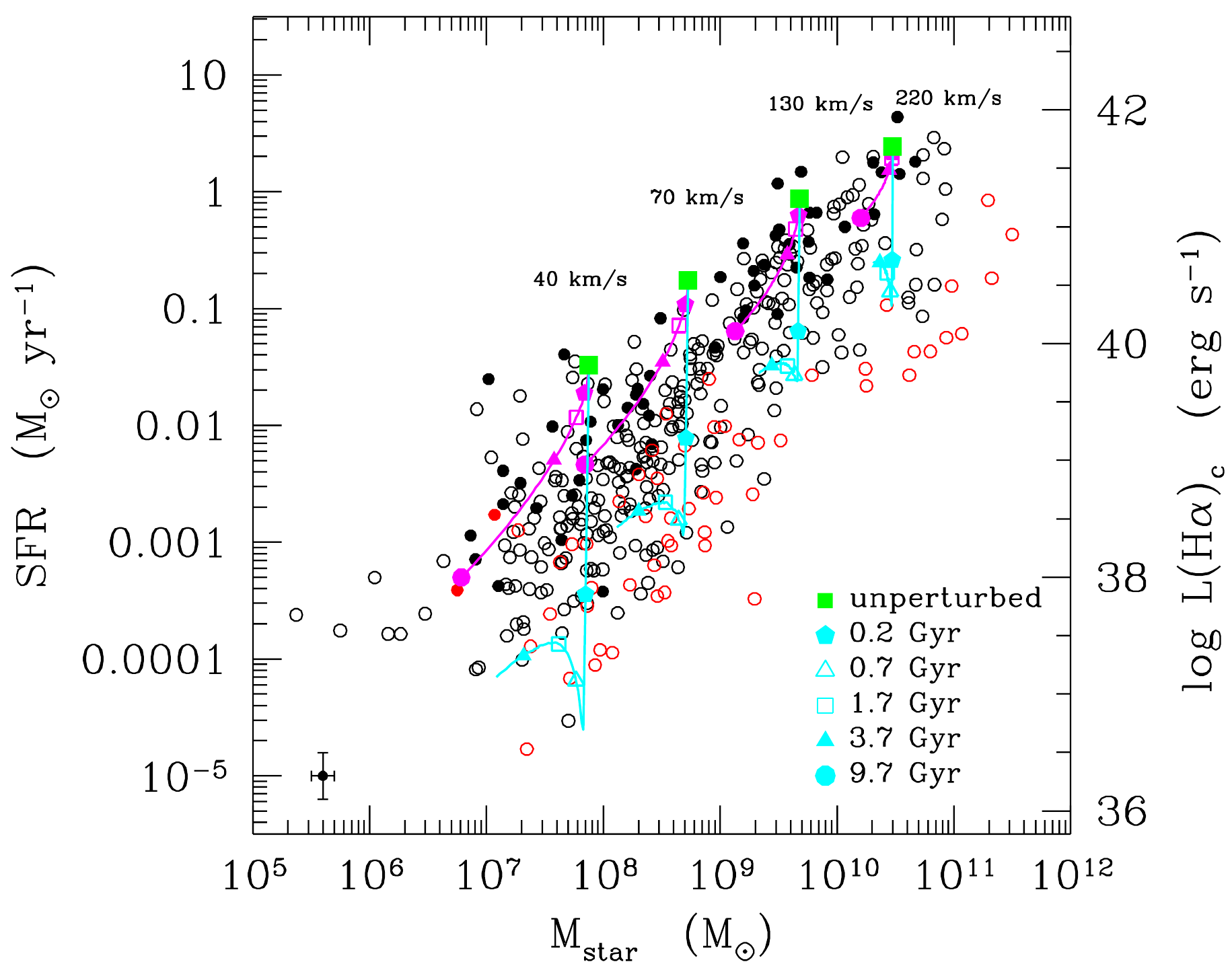}
\caption{Comparison of the observed main sequence relation with the predictions of ram pressure stripping modeles for galaxies
of spin parameter $\lambda$=0.05 and rotational velocities 40, 70, 130, and 220 km s$^{-1}$. Unperturbed models are indicated with filled green squares, 
starvation models by the magenta lines, and ram pressure stripping models by the cyan lines. 
Different symbols along the models indicate the position of the model galaxies at a given look-back
time from the beginning of the interaction. 
Black (red) filled circles are for late-type (early-type) galaxies with a normal HI content ($HI-def$ $\leq$ 0.4), black (red) empty
circles for HI-deficient ($HI-def$ $>$ 0.4) late-type (early-type) systems. The typical uncertainty in the data is shown at the lower left corner.
}
\label{mainmodel}%
\end{figure}

The observed scatter 
in the main sequence relation is also consistent with this picture, as indeed indicated by the prediction of tuned ram pressure stripping models (see Fig. \ref{mainmodel}).
These physically motivated models, which have been first introduced by Boselli et al. (2006) to explain the truncated gas and star forming disc of NGC 4569, have been succesfully used to predict 
the evolution of the chemo-spectrophotometric 2D properties of dwarf systems (Boselli et al. 2008a,b), diffuse galaxies (Junais et al. 2021, 2022), and the normal 
quenched galaxy population of the Virgo cluster (Boselli et al. 2014a, 2016b). These models, which are extensively described in these references, are based on the multizone
chemo-spectrophotometric models of galaxy evolution of Boissier \& Prantzos (2000) and updated with an empirically determined star formation law (Boissier et al. 2003)
relating the star formation rate to the total gas surface density, this last modulated by the galaxy rotational velocity:

\begin{equation}
{\Sigma_{SFR} = \alpha \Sigma_{gas}^{1.48}V(R)/R}
\end{equation}

\noindent
where $V(R)$ is the rotational velocity at the radius $R$. Galaxies are treated as exponential discs
embedded in a spherical dark matter halo, characterised by two free parameters, the rotational velocity of the system, which is tightly connected with its total dynamical mass, 
and the spin parameter $\lambda$, which rather traces the distribution of the angular momentum. The models also include a radially dependent infall rate of pristing gas
which decreases exponentially with time. The parameters regulating the infall rate have been calibrated to reproduce the colour and metallicity gradients of 
isolated galaxies (Prantzos \& Boissier 2000, Munoz-Mateos et al. 2011). 

To reproduce the perturbations induced by the surrounding environment, we simulated a ram pressure stripping process as in Boselli et al. (2006). 
The ram pressure stripping event is simulated by assuming a gas-loss rate inversely proportional to the gravitational potential of the
galaxy $\epsilon\Sigma_{gas}/\Sigma{potential}$, with an efficiency $\epsilon$ depending 
on the IGM gas density radial profile of the Virgo cluster, here taken from Vollmer et al. (2001a).

To limit the number of free parameters, here we consider only models with a fixed spin parameter of $\lambda$=0.05, the typical value of normal late-type galaxies 
predicted by cosmological simulations (Mo et al. 1998) and derived from observations of unperturbed field objects (Munoz-Mateos et al. 2011), while we vary the rotational velocity
of galaxies from 20 to 300 km s$^{-1}$ to consider a wide range in dynamical mass. Adopting a wider range of spin parameters more representative of a complete galaxy distribution would
slightly encrease the scatter of the main sequence relation for unperturbed systems. For the ram pressure stripping model we assume the same stripping efficiency derived for NGC 4569 
($\epsilon_0$ = 1.2 M$_{\odot}$ kpc$^{-2}$ yr$^{-1}$) and proved to well reproduce the 2D properties of large samples of Virgo cluster objects.

Since the gravitational potential well of the galaxies decreases with galactocentric distance, the models predict that the gas is stripped outside-in, producing truncated gas discs
as those typically observed in the HI maps of Virgo galaxies (e.g. Warmels 1986, Cayatte et al. 1990, 1994; Chung et al. 2009), or in the other components of
the ISM (molecular gas, Fumagalli et al. 2009, Boselli et al. 2014b, Mok et al. 2017; dust, Cortese et al. 2010, 2014, Longobardi et al. 2020, 
Boselli et al. 2022a). The lack of gas reduces the 
activity of star formation, producing similar truncated discs in the youngest stellar populations (e.g. Koopmann \& Kenney 2004a,b, Koopmann et al. 2006,  Fossati et al. 2013, 
Boselli et al. 2015).
The perturbed galaxies become redder in colour because of the aging of the stellar populations, not supplied by the formation of new stars on the 
discs (Boselli et al. 2014a). We recall that these models assume a face-on stripping process. This is a very simplified approach since
the inclination of galaxies along their parabolic orbit through the ICM changes. Since the efficiency of stripping is known to vary as a function  
of the inclination (e.g. Roediger \& Br\"uggen 2006), this makes the reconstruction of the star formation history of the perturbed galaxies quite uncertain.  

For comparison we also include a starvation scenario (e.g. Larson et al. 1980, Balogh et al. 2000, Treu et al. 2003), modeled by stopping the infall of pristine gas necessary 
to reproduce the observed colour and metallicity gradients of star forming discs. The starved galaxy then becomes 
anemic simply because it exhausts its gas reservoir through ongoing star formation. As defined, starvation is a passive phenomenon, where 
the gas is exhausted via star formation. For this reason, the simulated starvation process is much less efficient than ram pressure
to quench the activity of star formation of the perturbed galaxies (Boselli et al. 2006, 2014a).

The prediction of the models are compared to the observations in Fig. \ref{mainmodel}. Unperturbed model galaxies follow the main sequence relation, albeit with
a slightly higher star formation activity than that observed in Virgo late-type systems with a normal HI gas content. 
When crossing the cluster, however, ram pressure can remove a sufficient amount of atomic gas to significantly reduce their activity of star formation. 
At 0.7 Gyr after the beginning of the interaction the star formation rate can be reduced by an order of magnitude in massive systems 
($M_{star}$ $\geq$ 10$^{10}$ M$_{\odot}$), while up to more than two
orders of magnitude in the low mass regime ($M_{star}$ $\lesssim$ 10$^{9}$ M$_{\odot}$). The models indicate that this quenching episode occurs on relatively 
short timescales ($\lesssim$ 700 Myr) compared to the typical crossing time of the cluster ($\simeq$ 1.7 Gyr), in line with previous observations of statistical samples
(e.g. Crowl \& Kenney 2008, Boselli et al. 2008a, 2016b, Ciesla et al. 2021) or of representative Virgo cluster objects (Vollmer et al. 1999, 2004, 2006, 2008a,b,  2012, 2018, 2021; 
Boselli et al. 2006; Fossati et al. 2018). On the contrary, starvation, which is a mild process,
is not able to significantly reduce the activity of star formation of the perturbed galaxies as the one observed in the VESTIGE sample, and this in galaxies of all stellar masses.
In this scenario the star formation activity can be reduced by at most a factor of 2-3, but this requires very long timescales, of the order of 
several Gyr (Boselli et al. 2006, 2014a).

\begin{figure}
\centering
\includegraphics[width=0.5\textwidth]{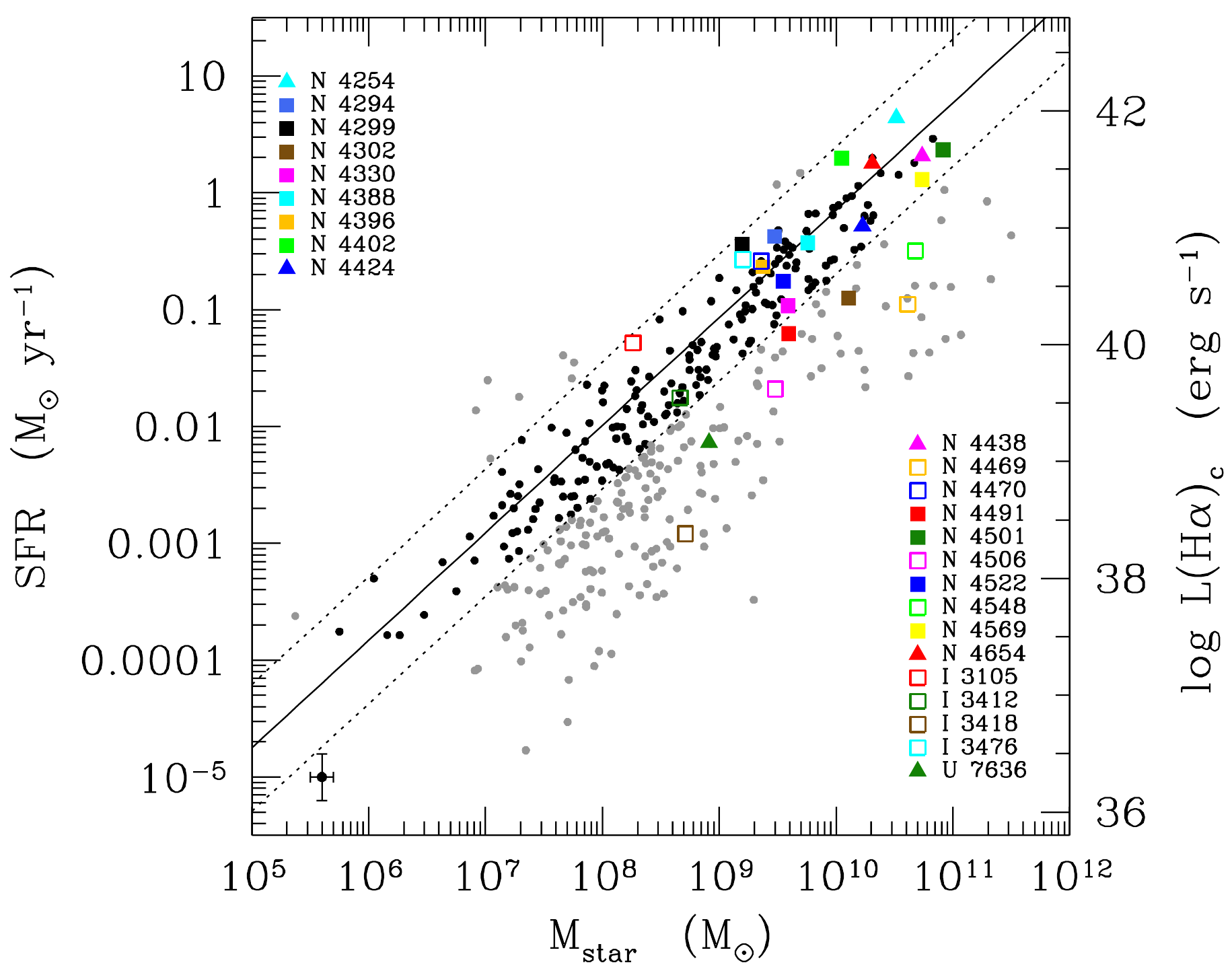}
\caption{Main sequence relation for galaxies coded according to their distance from the bisector fit derived for HI gas-rich late-type systems. Black filled dots
are for galaxies within 1 $\sigma$ from the relation, grey filled dots for galaxies $>$ 1$\sigma$ above or below the relation. The solid line gives the bisector 
fit obtained for gas-rich star forming systems, the dotted lines the limits 1$\sigma$ above and below the best fit. Galaxies undergoing a ram pressure stripping event
identified in Boselli et al. (2022a) are indicated with coloured squares, while those suffering a combined effect of ram pressure and harassment by triangles.
The typical uncertainty in the data is shown at the lower left corner.}
\label{mainname}%
\end{figure}

Since high sensitivity multifrequency observations, models and simulations of selected targets unambigously identified several objects undergoing a ram pressure stripping event 
in the Virgo cluster (see Table 2 in Boselli et al. 2022a for details), we can see how these objects are located within the main sequence relation (see Fig. \ref{mainname}). 
Figure \ref{mainname} confirms that the scatter of the relation is dominated by galaxies with a lower than expected star formation activity. Among those identified as suffering 
a ram pressure stripping event (24 objects), 17 are located within 1$\sigma$ from the main sequence, 0 above, and 7 below. Although this subsample of objects is not 
statistically complete, it confirms that overall the activity of star formation of ram pressure stripped galaxies, if perturbed, is reduced.
Figure \ref{mainname}, however, also shows that in a few objects the star formation activity might be enhanced with respect to 
that of typical main sequence galaxies. This occurs mainly
at low stellar masses, and might be due to i) an increased activity due to the compression of the gas on the disc of the galaxy occuring at a particular and
specific phase of the stripping process (e.g. Boselli et al. 2021, 2022a), or ii) an inaccurate determination of the star formation rate at these
low mass regimes for the several reasons mentioned in Sec. 3.4. The detailed analysis of these galaxies presented in see Sec. 5.2 and in Appendix B suggests that their enhanced activity
is often, but not always due to a ram pressure stripping event.
Finally, we stress that the fact that several star forming galaxies are located below the main sequence relation, or equivalently in the green valley located between 
the red sequence and the blue cloud (e.g. Boselli et al. 2014a), is not a direct indication that the quenching mechanism acts on long timescales (e.g. Paccagnella et al. 2016)
just because it does not take into account the high infall rate of galaxies within the cluster, for Virgo estimated at $\sim$ 200-300 objects Gyr$^{-1}$ 
for galaxies of $M_{star}$ $\geq$ 10 $^9$ M$_{\odot}$ (Boselli et al. 2008a; Gavazzi et al. 2013; Sorce et al. 2021).

\subsection{Evolutionary state of the perturbed galaxies}

The evolutionary path of galaxies within a cluster environment is very complex and depends on several parameters, such as the properties of the high density region (ICM
distribution, density, temperature), of the impact parameters of galaxies along their orbits within the cluster (radial, circular, infalling as isolated objects 
or as members of small groups), on the properties of galaxies (total mass, gas content and distribution), and on the epoch when the infall occurs.
Furthermore, different mechanisms can contribute at different epochs to modify the evolution of galaxies. The reconstruction of the orbit of a single galaxy within a cluster,
the measurement of the main impact parameters, the identification of the dominant stripping process and the reconstruction of the following evolution of the different stellar populations 
require accurate models and simulations expressly tuned to reproduce the observed 2D spectrophotometric and kinematical properties of the perturbed objects.
As mentioned in Sec. 5.3, this has been done only for a few objects in Virgo. A general evolutionary picture, however, can be driven by comparing the statistical
properties of a complete sample of galaxies as the one analysed in this work with the prediction of cosmological simulations. This is generally done by means of the phase-space diagram
showing the position of the selected galaxies defined as the relative velocity vs. the mean velocity of the cluster normalised to the velocity dispersion of the cluster itself 
as a function of the clustercentric distance (e.g. Mahajan et al. 2011, Haines et al. 2015, 
Jaff\'e et al. 2015, 2018, Rhee et al. 2020). 
This exercise has been already done for galaxies within the Virgo cluster (e.g. Vollmer et al. 2001, Boselli et al. 2014a, Yoon et al. 2017, 
Morokuma-Matsui et al. 2021).

\begin{figure}
\centering
\includegraphics[width=0.5\textwidth]{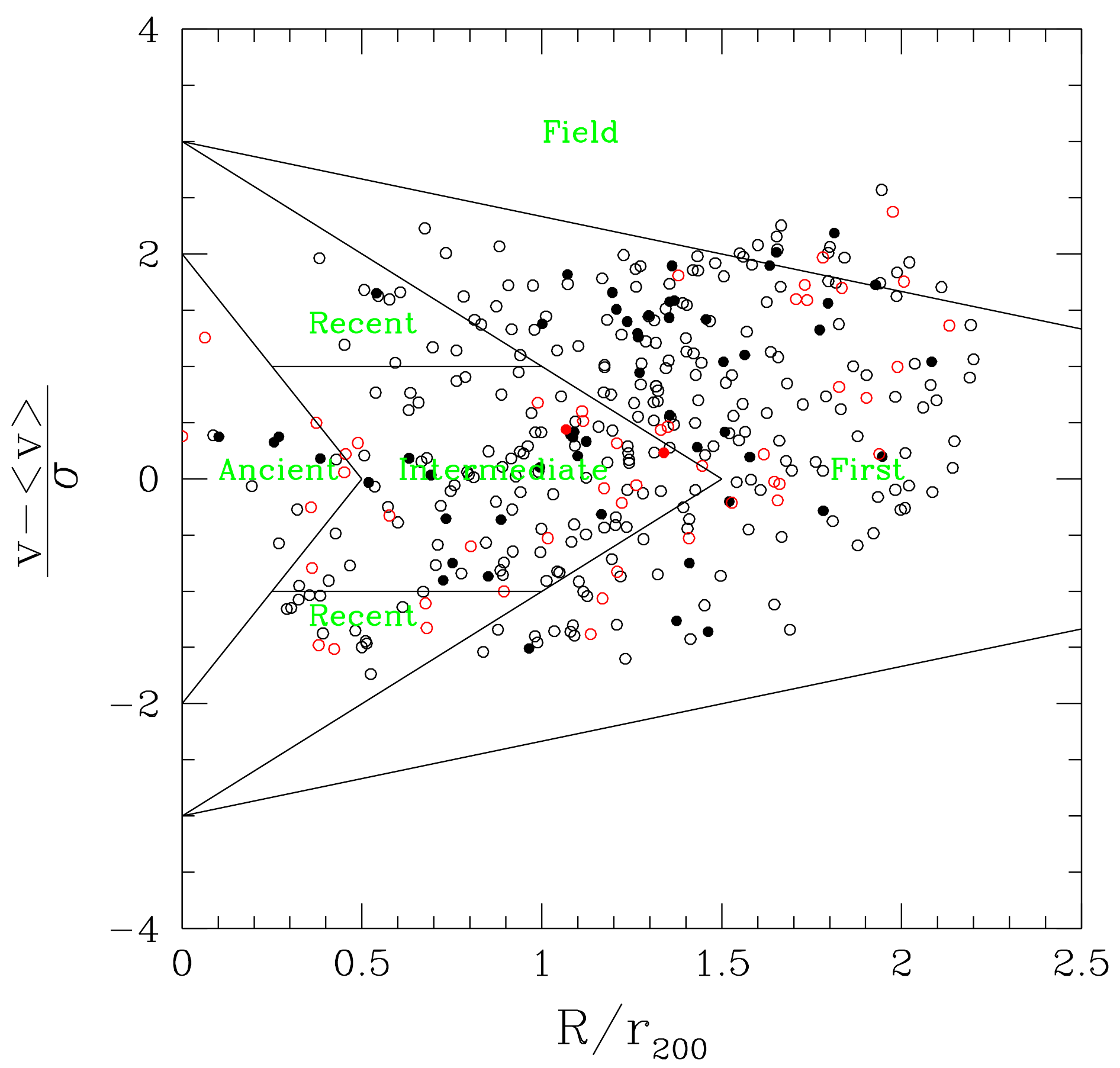}
\caption{Phase space diagram for the H$\alpha$ detected galaxies within the Virgo cluster. For all galaxies, the X-axis indicates the projected distance 
from the central galaxy M87 in units of $r_{200}$, the Y-axis the excess in velocity with respect to the mean velocity of cluster A ($<v>$ = 955 km s$^{-1}$) normalised
to its mean velocity dispersion ($\sigma$ = 799 km s$^{-1}$; Boselli et al. 2014a). Late-type galaxies ($>$ Sa) are indicated by black symbols, 
early-type galaxies by red symbols. Filled dots are for HI gas-rich objects ($HI-def$ $\leq$ 0.4), while HI-deficient objects ($HI-def$ $>$ 0.4) by empty circles.
The solid lines delimit the different regions extracted from the simulations of Rhee et al. (2017) to identify galaxies in different phases of their infall into the cluster: 
first (not fallen yet), recent (0$<$ $\tau_{inf}$ $<$3.6 Gyr), intermediate (3.6$<$ $\tau_{inf}$ $<$6.5 Gyr), and ancient (6.5$<$ $\tau_{inf}$ $<$13.7 Gyr) infallers. 
}
\label{phasespacebox}%
\end{figure}

Figure \ref{phasespacebox} shows the distribution of the H$\alpha$ detected galaxies within the phase-space diagram of the Virgo cluster, with galaxies coded 
according to their atomic gas content and morphological type as in Fig. \ref{mainHI}. The distribution of galaxies within this diagram is compared to that predicted by
the cosmological simulations of Rhee et al. (2017), which give a general trend as a function of the epoch of first infall. We recall that 
these simulations give just a statistically weighted trend, and can thus be considered in the comparison only in these terms. Figure \ref{phasespacebox}
does not show any statistically significant trend in the distribution of the HI-rich and HI-deficient galaxies within the phase-space diagram.

\begin{figure}
\centering
\includegraphics[width=0.5\textwidth]{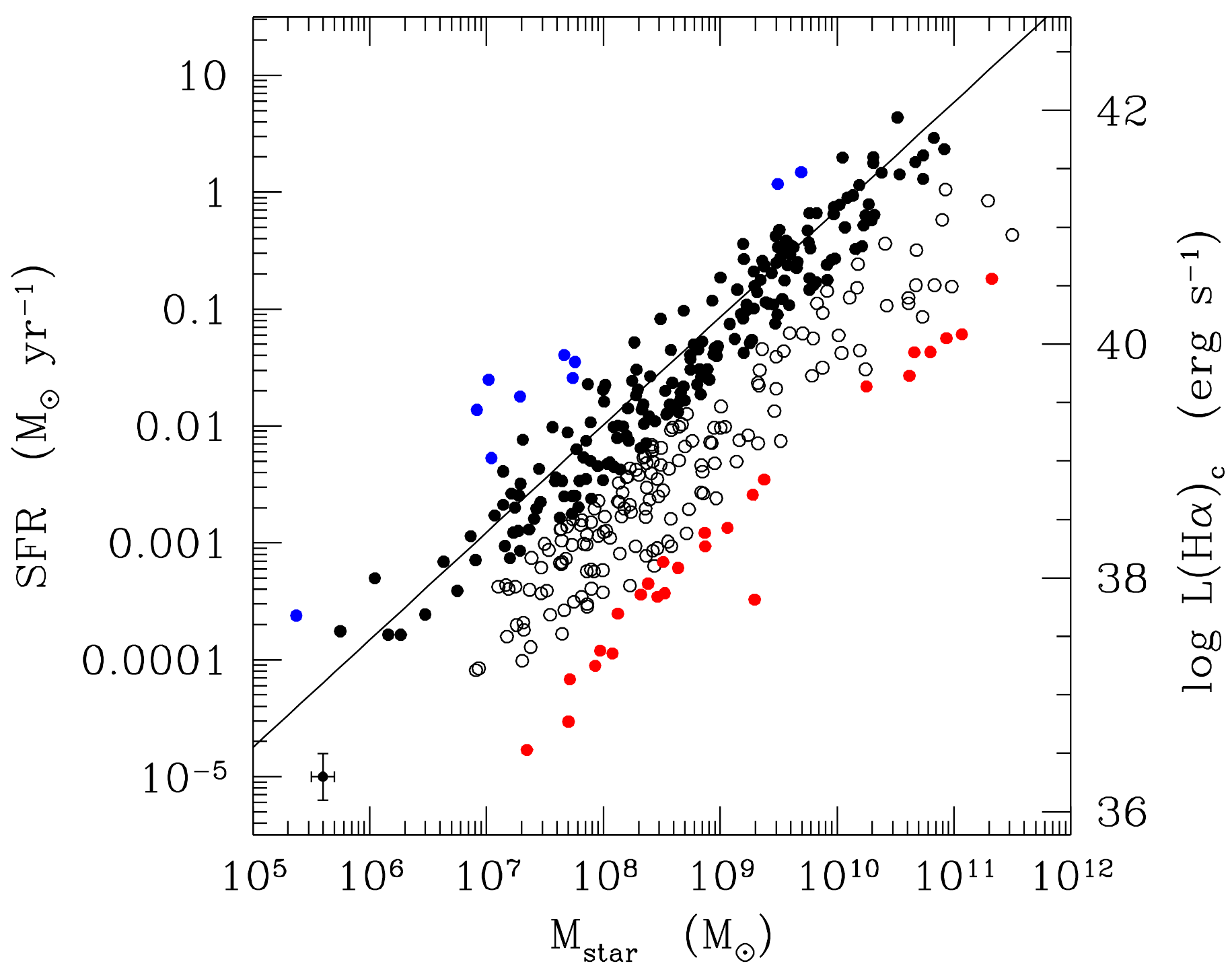}
\caption{Main sequence relation for galaxies coded according to their distance from the bisector fit derived for HI gas-rich late-type systems. Black filled dots
are for galaxies within 1 $\sigma$ from the relation, blue filled dots for galaxies $>$ 1$\sigma$ above the relation, black empty circles for galaxies 1$\sigma$ to 3$\sigma$
below the relation, and red filled dots for galaxies $>$ 5$\sigma$ below the relation. The solid line gives the bisector fit obtained for gas-rich star forming systems.
The typical uncertainty in the data is shown at the lower left corner.}
\label{maintest}%
\end{figure}

We made the same exercise by dividing galaxies according to their position along the main sequence. For this purpose, we divided the main sequence relation 
in four main subsamples, as indicated in Fig. \ref{maintest}: normal galaxies, located within 1$\sigma$ from the best fit derived for the subsample of HI-normal galaxies 
($HI-def$ $\leq$ 0.4), where $\sigma$ is the dispersion of the relation (0.40 dex); starburst galaxies, located at more than 1$\sigma$ above the relation, partly quenched galaxies
(1$\sigma$ $<$ $s$ $\leq$ 3$\sigma$), and quenched systems ($s$ $>$ 3$\sigma$), where $s$ is the distance perpendicular to the main sequence relation.

\begin{figure}
\centering
\includegraphics[width=0.5\textwidth]{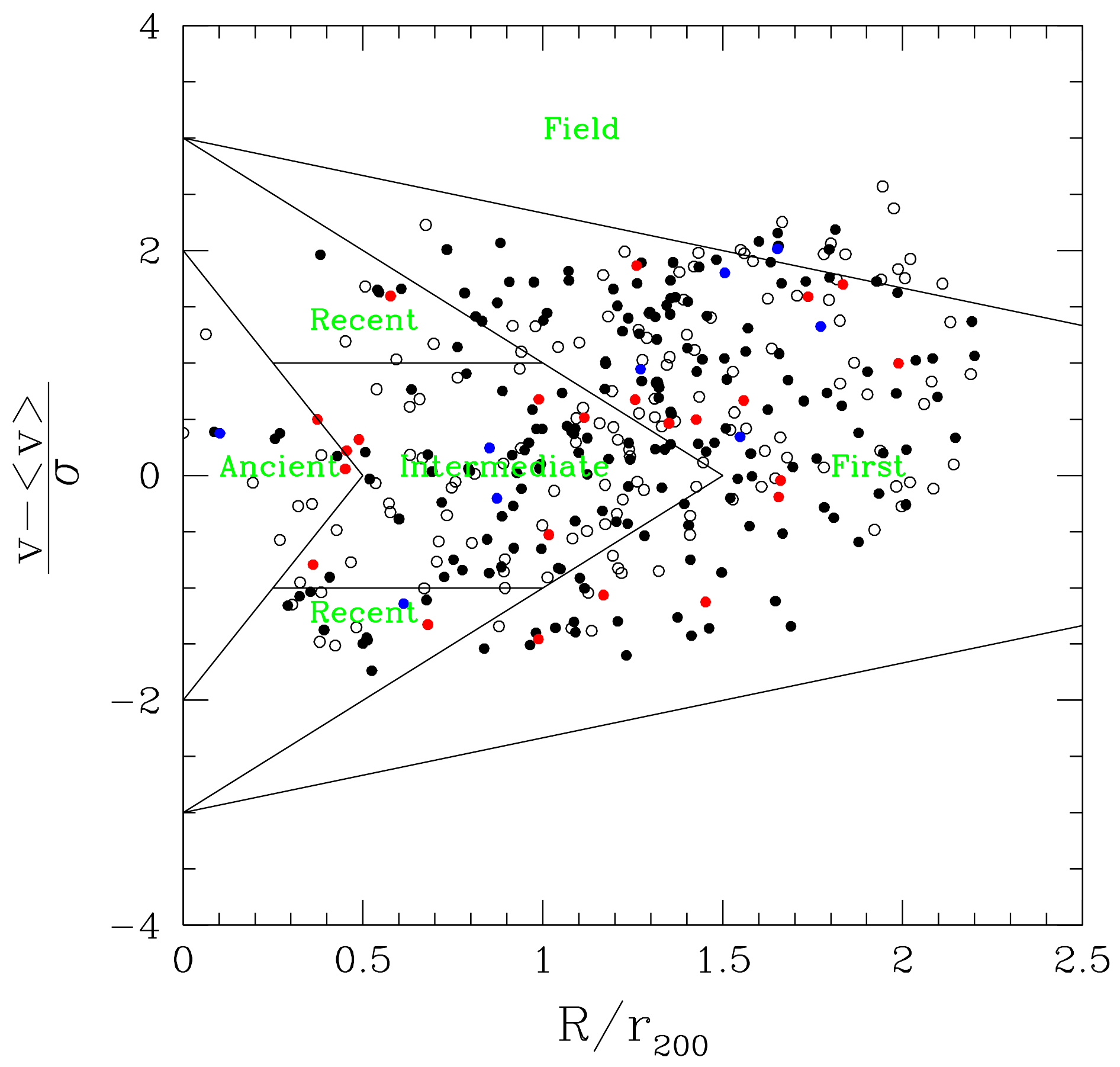}
\caption{Phase space diagram for galaxies coded according to their distance from the bisector fit derived for HI gas-rich late-type systems. Black filled dots
are for galaxies within 1 $\sigma$ from the relation, blue filled dots for galaxies $>$ 1$\sigma$ above the relation, black empty circles for galaxies 1$\sigma$ to 3$\sigma$
below the relation, and red filled dots for galaxies $>$ 5$\sigma$ below the relation. 
The solid lines delimit the different regions extracted from the simulations of Rhee et al. (2017) to identify galaxies in different phases of their infall into the cluster: 
first (not fallen yet), recent (0$<$ $\tau_{inf}$ $<$3.6 Gyr), intermediate (3.6$<$ $\tau_{inf}$ $<$6.5 Gyr), and ancient (6.5$<$ $\tau_{inf}$ $<$13.7 Gyr) infallers. 
}
\label{phasespaceboxsfr}%
\end{figure}

\begin{figure*}
\centering
\includegraphics[width=1.0\textwidth]{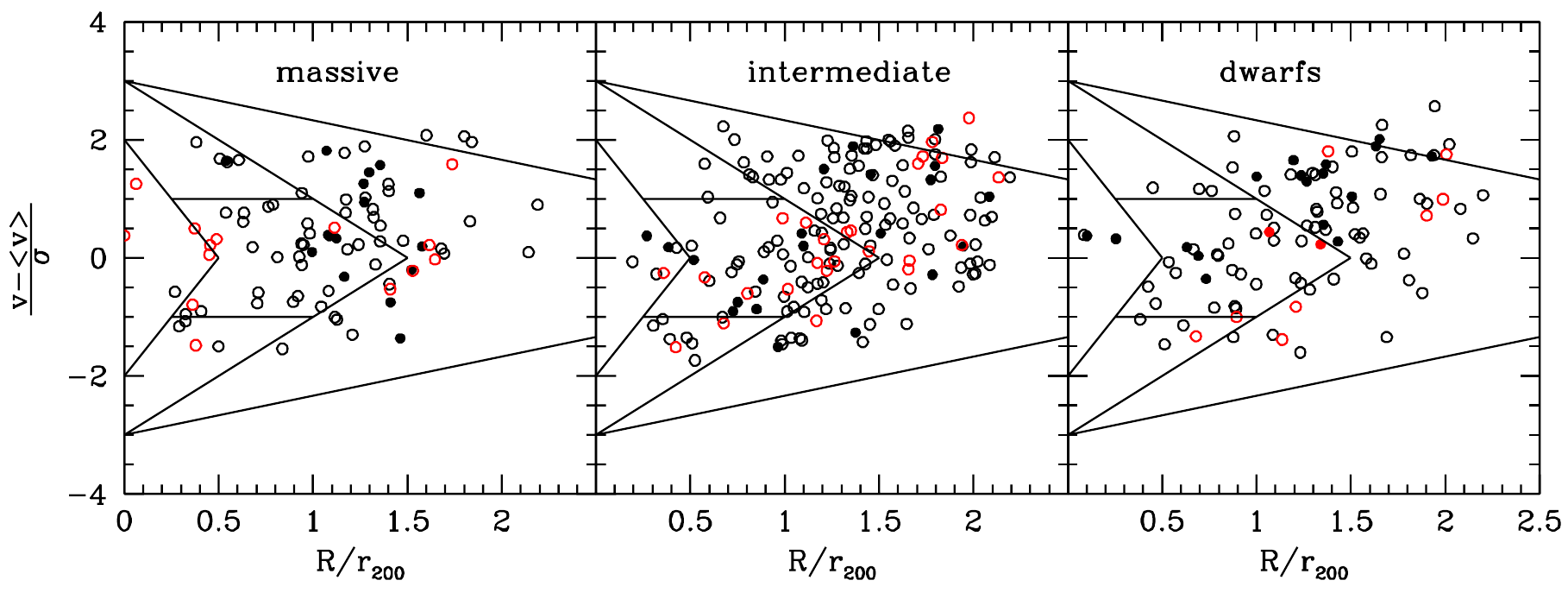}
\caption{Phase space diagram in three different bins of stellar mass ($M_{star}$ $>$ 10$^{9.5}$ M$_{\odot}$, massive; 10$^8$ $<$ $M_{star}$ $\leq$ 10$^{9.5}$ M$_{\odot}$, intermediate;
$M_{star}$ $\leq$ 10$^{8}$ M$_{\odot}$, dwarfs) for galaxies coded according to their morphological type (black symbols for late-types, red symbols for early-types) and atomic gas content 
(filled dots for HI gas-rich objects ($HI-def$ $\leq$ 0.4), empty circles for HI-deficient objects ($HI-def$ $>$ 0.4)). The solid lines delimit the different regions extracted from the simulations of Rhee et al. (2017) to identify galaxies in different phases of their infall into the cluster: 
first (not fallen yet), recent (0$<$ $\tau_{inf}$ $<$3.6 Gyr), intermediate (3.6$<$ $\tau_{inf}$ $<$6.5 Gyr), and ancient (6.5$<$ $\tau_{inf}$ $<$13.7 Gyr) infallers. 
}
\label{phasespaceboxmass}%
\end{figure*}

\begin{figure*}
\centering
\includegraphics[width=1.0\textwidth]{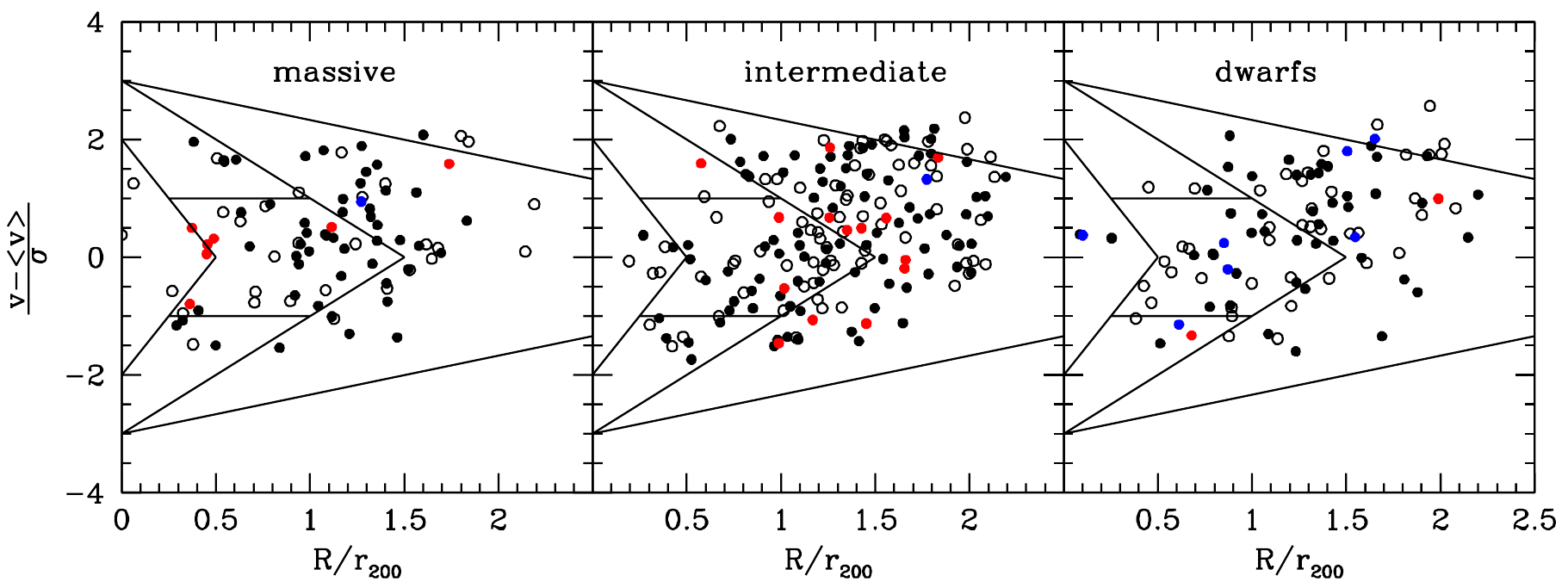}
\caption{Phase space diagram in three different bins of stellar mass ($M_{star}$ $>$ 10$^{9.5}$ M$_{\odot}$, massive; 10$^8$ $<$ $M_{star}$ $\leq$ 10$^{9.5}$ M$_{\odot}$,  intermediate;
$M_{star}$ $\leq$ 10$^{8}$ M$_{\odot}$, dwarfs) for galaxies coded according to their distance from the main sequence relation derived for gas-rich late-type systems: black filled dots
are for galaxies within 1 $\sigma$ from the relation, blue filled dots for galaxies $>$ 1$\sigma$ above the relation, black empty circles for galaxies 1$\sigma$ to 3$\sigma$
below the relation, and red filled dots for galaxies $>$ 3$\sigma$ below the relation. The solid lines delimit the different regions extracted from the simulations of Rhee et al. (2017) to identify galaxies in different phases of their infall into the cluster: 
first (not fallen yet), recent (0$<$ $\tau_{inf}$ $<$3.6 Gyr), intermediate (3.6$<$ $\tau_{inf}$ $<$6.5 Gyr), and ancient (6.5$<$ $\tau_{inf}$ $<$13.7 Gyr) infallers. }
\label{phasespaceboxmasssfr}%
\end{figure*}

Figure \ref{phasespaceboxsfr} shows the distribution of galaxies within the phase-space diagram with galaxies coded according their distance from the main sequence
as indicated in Fig. \ref{maintest}. Figure \ref{phasespaceboxsfr} does not show any evident trend in the distribution of the different class of star forming objects. 
Since ram pressure stripping, identified in Virgo as the dominant
gas stripping mechanism, strongly depends on the total mass of the perturbed galaxies (in massive systems the deep gravitational potential well can more
efficiently keep the cold gas anchored to the stellar disc than in dwarf systems, where the gas is easily and completely stripped on short timescales), we plotted in 
Fig. \ref{phasespaceboxmass} and \ref{phasespaceboxmasssfr} the distribution of the selected galaxies in the phase-space diagram coded according to their atomic 
gas content and star formation activity in three bins of stellar mass: $M_{star}$ $>$ 10$^{9.5}$ M$_{\odot}$, massive systems; 10$^8$ $<$ $M_{star}$ $\leq$ 10$^{9.5}$ M$_{\odot}$, 
intermediate mass systems; and $M_{star}$ $\leq$ 10$^{8}$ M$_{\odot}$, dwarfs.

As for the previous plots, Fig.  \ref{phasespaceboxmass} and \ref{phasespaceboxmasssfr} do not show any strong systematic segregation in the distribution of the HI-deficient 
and HI-rich galaxies or star forming and quenched objects within the phase-space diagram. Worth noticing, however, is the fact that Virgo is a cluster still 
in formation as suggested by the presence of several subgroups. Its 3D structure is thus very complex and hardly schematised in a spherical geometry. 
Under this configuration any systematic effect in the galaxy distribution within this diagram can be heavily smeared out by projection effects. Interesting, however, is the fact that 
HI-rich galaxies are already rare in the infalling regions (15\%\ of the star forming systems, with respect to $\sim$ 70\%\ in the field, Cattorini et al. 2022).  
This means that the dominant perturbing mechanism is 1) already active at the periphery of the cluster ($R/r_{200}$ $\geq$ 1), and 2) efficiently removes the atomic gas 
content on short timescales ($\lesssim$ 1 Gyr), otherwise galaxies would have retained their gas reservoir along their orbits up to the inner regions. The similar distribution 
of star forming galaxies within the phase-space diagarm consistently indicate that, as for the gas removal, the star formation process is also quenched on short timescales,
as expected in a ram pressure stripping scenario (e.g. Boselli et al. 2008a, 2014a, 2016b).

\subsection{Extension to lower stellar mass}

\begin{figure*}
\centering
\includegraphics[width=0.48\textwidth]{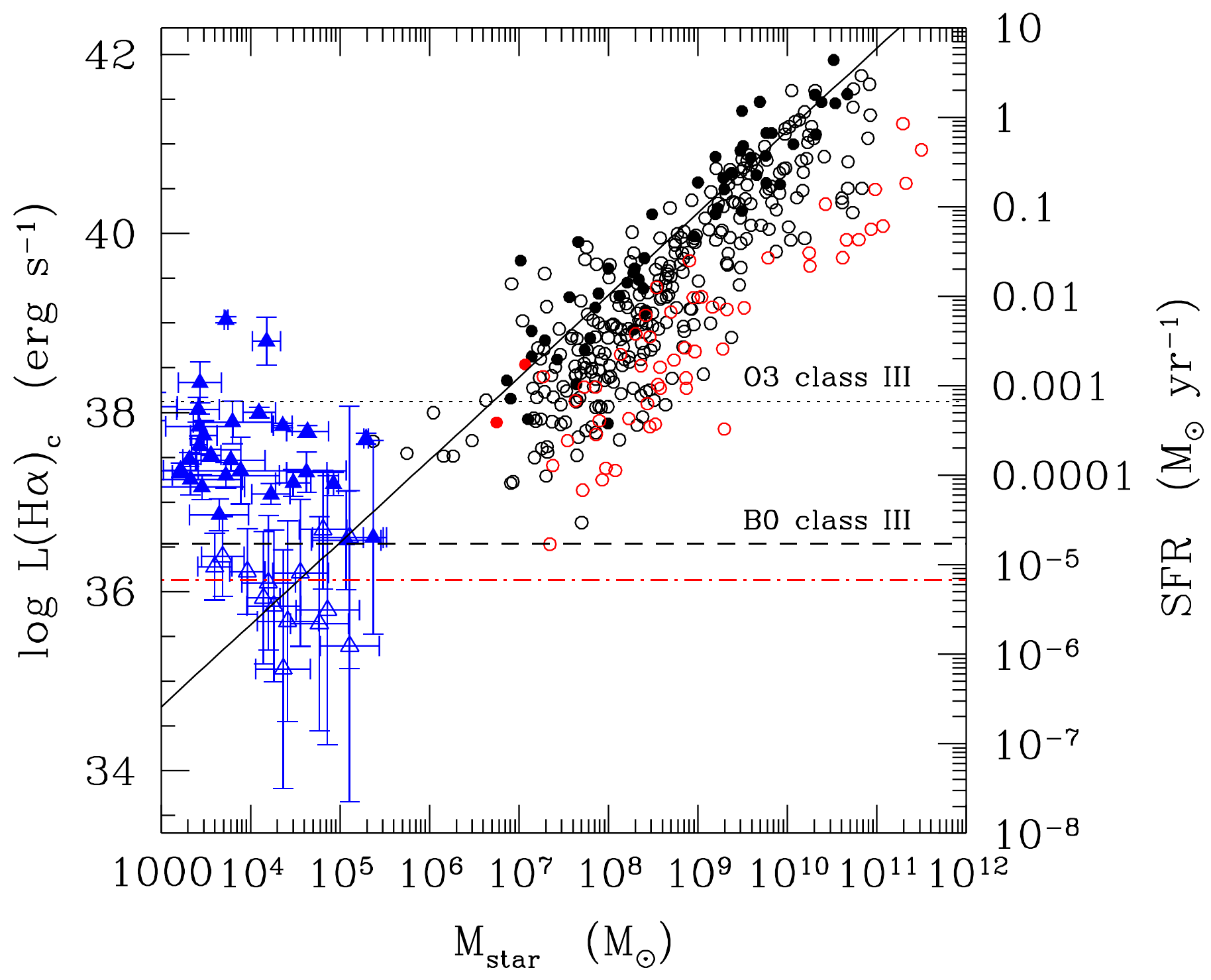}
\includegraphics[width=0.48\textwidth]{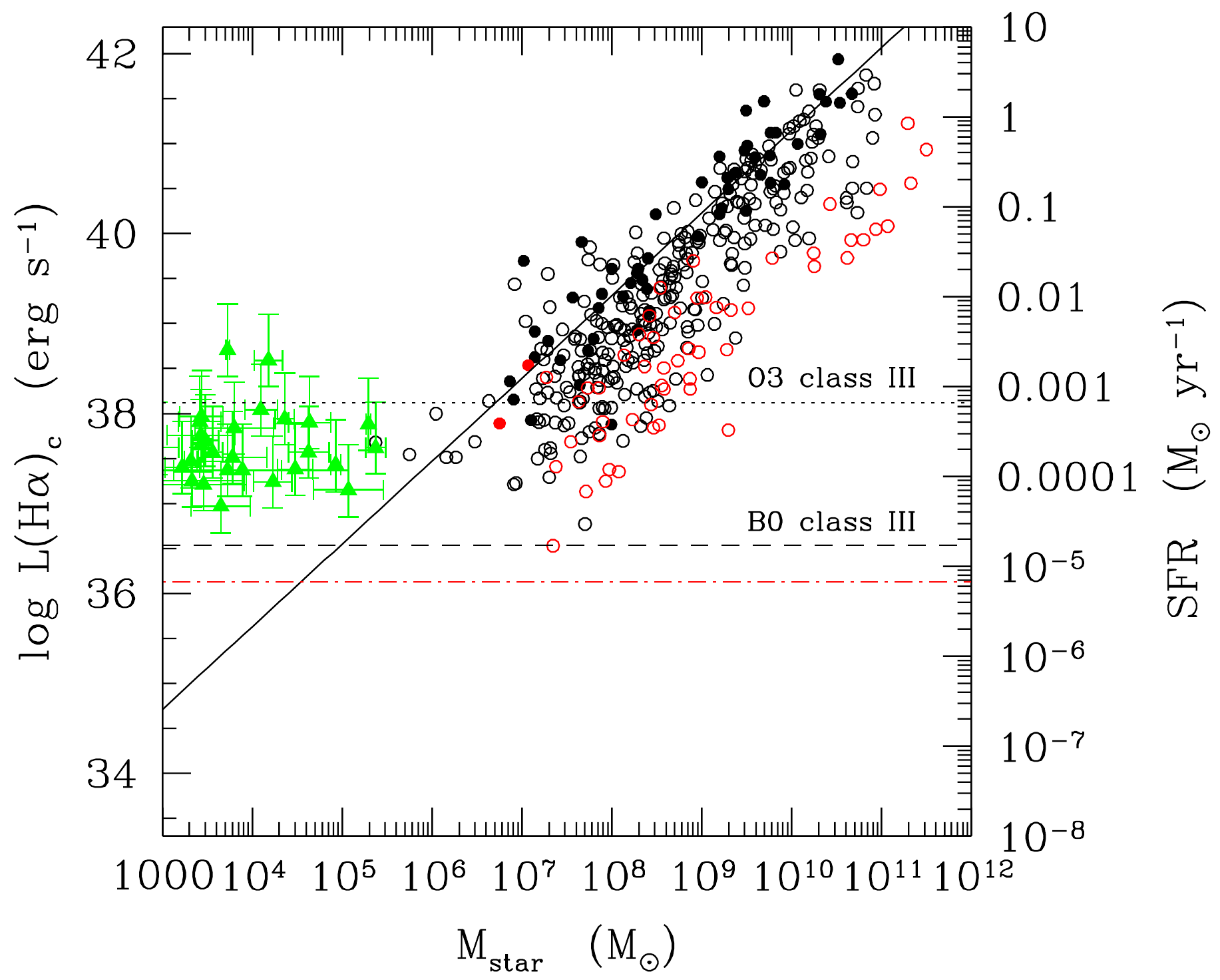}
\caption{Left panel: relation between the H$\alpha$ luminosity (corrected for dust attenuation) and the stellar mass 
for galaxies coded according to their morphological type and HI gas content: late-type galaxies ($>$ Sa) are indicated by black symbols, 
early-type galaxies by red symbols. Filled dots are for HI gas rich objects ($HI-def$ $\leq$ 0.4), while HI-deficient objects ($HI-def$ $>$ 0.4) by empty circles.
The black solid and dotted lines are the bisector and linear fit obtained for gas-rich star forming systems. Blue triangles are the extraplanar HII 
regions detected in NGC 4254 (Boselli et al. 2018b), where stellar masses and star formation rates have been determined using the SED fitting code CIGALE. Filled
triangles are for H$\alpha$ detected regions, empty triangles for the undetected ones. The errorbars of the HII regions of NGC 4254 are those given by the SED fitting analysis.
The horizontal dotted and dashed lines indicate the corresponding SFR derived using the number of
ionising photons produced by an O3 class III and a B0 class III star as derived using the model atmospheres of Sternberg et al. (2003). 
The red dotted-dashed line indicates the corresponding completness limit of VESTIGE at the mean distance of Virgo (16.5 Mpc). 
The right Y-axis gives the corresponding SFR (main sequence relation) derived assuming stationary conditions.
Right panel: similar plot where the green filled triangles are the extraplanar HII regions of NGC 4254 detected in H$\alpha$, where the stellar 
masses are derived using the SED fitting analysis with CIGALE, while the star formation rates are measured using the same calibration based on the H$\alpha$ 
luminosity as for the galaxy sample, assuming $A(H\alpha)$ = 0.7 mag and [NII]$\lambda$6584\AA/H$\alpha$ = 0.2, as in Boselli et al. (2018b).
The uncertainties on the stellar mass of the HII regions of NGC 4254 are given by the fitting code CIGALE, while those on their star formation rate indicate the dynamic range 
in this parameter assuming either $A(H\alpha)$ = 0 mag or $A(H\alpha)$ = 1.5 mag.}
\label{mainHIext}%
\end{figure*}

The VESTIGE survey has also detected several HII regions located well outside the star forming discs of Virgo cluster galaxies formed within the gas stripped during the 
interaction with the surrounding environment. Typical examples of these extraplanar star forming regions are those detected in the tail of IC3418 probably
formed after a ram pressure stripping event (VCC1217; Hester et al. 2010, Fumagalli et al. 2011b, Hota et al. 2021), those located in between UGC 7636 and M49 formed by a 
combined effect of ram pressure stripping withn the halo of M49 and gravitational interactions between the two galaxies (VCC 1249; Arrigoni-Battaia et al. 2012), 
or those located along the long HI tail of NGC 4254 formed after a gravitational perturbation in a high speed fly-by encounter with another cluster member (Vollmer et al. 2005, Duc \& Bournaud 2008, 
Boselli et al. 2018b). It has been claimed that these regions, characterised by stellar masses of 10$^3$ $\lesssim$ $M_{star}$ $\lesssim$ 10$^5$ M$_{\odot}$ and star
formation rates of 10$^{-5}$ $\lesssim$ $SFR$ $\lesssim$ 10$^{-3}$ M$_{\odot}$ yr$^{-1}$, might become gravitationally unbound systems and thus be at the origin of
compact objects such as globular clusters or ultra compact dwarf galaxies (UCDs) typical of rich environments\footnote{As stressed in Boselli et al. (2018b), UCDs have stellar masses
higher than those observed in these star forming regions (Liu et al. 2020). They might have been formed in the past, when galaxies were much more gas-rich than in the local universe, by a similar 
perturbing mechanism able to remove larger quantities of gas.}. If we consider these objects as unbound systems representative of the compact galaxy 
population, we can see how they populate the main sequence relation in comparison to more extended systems. This exercise can be done for the extraplanar regions of NGC 4254, 
where star formation rates and stellar masses are available for $\simeq$ 60 objects (Boselli et al. 2018b) (see Fig. \ref{mainHIext}). These extraplanar HII regions
have been identified in the FUV GALEX band. Their stellar masses and star formation rates have been derived using CIGALE on their observed FUV-to-$z$ SEDs as done for the VESTIGE 
galaxies, but adopting a realistic delayed exponentially declining star formation history with parameters representative of individual HII regions 
(see Boselli et al. 2018b for details)\footnote{Stellar masses and star formation rates given in Boselli et al. (2018b) have been divided by 1.64 and 1.58, respectively, 
to convert the original estimates based on a Salpeter IMF to a Chabrier IMF.}.  To avoid any possible systematic bias
in the comparison, this plot is done using either the star formation rates derived from the SED fitting modeling or those directly measured from the VESTIGE H$\alpha$ fluxes 
corrected for [NII] contamination and dust attenuation as for the rest of the sample galaxies.

Figure \ref{mainHIext} clearly shows that these extraplanar star forming regions do not follow the main sequence relation drawn by the rest of the Virgo cluster galaxies.
They are characterised by much higher star formation rates than those expected for their low stellar mass, which should be of the order of 
10$^{-8}$ $\lesssim$ $SFR$ $\lesssim$ 10$^{-5}$ M$_{\odot}$ yr$^{-1}$. We recall, however, that these rates of star formation are too low to produce a statistically 
significant number of ionising stars because of a poor sampling of the IMF in the high mass stellar regime. As a consequence, they cannot be measured using an H$\alpha$ survey.
This is indeed confirmed by the lack of any H$\alpha$ emitting source with a luminosity $L(H\alpha)$ $\lesssim$ 10$^{36.5}$ erg s$^{-1}$, the typical H$\alpha$ luminosity of 
an early B star\footnote{The regions with star formation rates derived with the SED fitting analysis of $SFR$ $\lesssim$ 2 $\times$ 10$^{-5}$ M$_{\odot}$ yr$^{-1}$ shown in the
left panel of Fig. \ref{mainHIext}, indeed, are not detected in H$\alpha$.}. 
Despite this strong selection effect which can justify the lack of H$\alpha$ emitting sources at low star formation rates ($SFR$ $\lesssim$ 2 $\times$ 10$^{-5}$ M$_{\odot}$ yr$^{-1}$), 
these extraplanar HII regions have star formation activities 
two-to-three orders of magnitudes higher per unit stellar mass than those of normal star forming galaxies populating the main sequence. This spectacular difference can be explained 
if we consider these HII regions as short-living starbursts, as indeed suggested by their star formation history derived in the SED fitting analysis (Boselli et al. 2018b).
Their activity lasts only for a few million years ($\lesssim$ 50 Myr), gradually stopping to form quenched systems. They are thus unstable objects, and their presence 
above the main sequence is a temporary effect lasting only for short timescales. As extensively discussed in Boselli et al. (2018b), many of these regions have already 
completed their starburst phase and are becoming redder and redder, witnessing an aging stellar population.

\section{Discussion}

\subsection{Completeness}

The exceptional depth of the VESTIGE survey allows us to trace the main sequence relation and other important scaling relations 
for a nearby rich cluster of galaxies down to stellar masses of a few 10$^6$ M$_{\odot}$
and star formation rates of a few 10$^{-5}$ M$_{\odot}$ yr$^{-1}$, sampling thus a dynamic range never reached so far in any other sample of isolated or cluster galaxies.
In particular, the sensitivity of the narrow-band imaging data ($f(H\alpha)$ $\simeq$ 4 $\times$ 10$^{-17}$ erg s$^{-1}$ cm$^{-2}$ (5$\sigma$) for point sources)
corresponds to a limit in the H$\alpha$ luminosity of $L(H\alpha)$ $\simeq$ 10$^{36}$ erg s$^{-1}$ at the distance of Virgo, a value slightly lower than the
expected luminosity due to the ionising photons produced by a single early B star. This means that the VESTIGE survey samples the whole range of H$\alpha$ luminosity
of galaxies down to the lowest possible star formation rates physically measurables with H$\alpha$ data. Since at this low luminosities the emission associated to dwarf galaxies
is generally do to a very few ionising stars ($\lesssim$ 3), it is possible that the measured fluxes are contaminated by background line emitters. Following Sun et al. (2007) we can estimate 
that the number of Ly$\alpha$ line emitters at $z$$\sim$4.1 detectable within one square degree is of the order of 200 at the sensitivity of VESTIGE.
If we consider that at the distance of the cluster the typical size of dwarf galaxies of M$_{star}$ $\simeq$10$^6$-10$^7$ M$_{\odot}$ is of the order of $\sim$ 1\arcmin\ in diameter, 
the probability that a background line emitter falls within the projected image of a dwarf galaxy is of $\simeq$ 4\%. The contamination of background galaxies is thus
certainly negligible. On the contrary we might lose some Virgo cluster objects if their H$\alpha$ emission is limited to a point source without any extended emission and 
the galaxy has such a peculiar morphology not to be identified as a cluster member using the typical scaling relations derived in the NGVS (Lim et al. 2020). Since we detected only
five emitting sources over the 384 identified by VESTIGE with an extended H$\alpha$ morphology suggesting their cluster membership but not included in the NGVS 
catalogue of 3689 Virgo members, we expect that the number of H$\alpha$ emitting sources associated to the cluster and missed by VESTIGE is very limited.

\subsection{Accuracy in the SFR determination}

The narrow-band imaging data used in this work have been corrected for [NII] contamination and dust attenuation before being transformed into star formation rates.
These corrections are important sources of uncertainties in the determination of an accurate star formation rate of massive galaxies, where dust attenuation and 
[NII] contamination are relevant. For all the massive galaxies we estimated the [NII] contamination using long slit integrated spectroscopy and we applied two different 
dust attenuation corrections based on integrated spectroscopy and far-IR emission. As extensively discussed in Boselli et al. (2015), the combination of these
independent techniques are optimal for minimising the uncertainties and provide the most accurate results (see also Calzetti et al. 2007, 2010, Kennicutt et al. 2009).
The same set of data, however, was only sporadically available for dwarf systems, were we applied (if any) corrections based on scaling relations.
In this metal poor galaxy population, however, both [NII] contamination and dust attenuation, if present, are very low (e.g. Boselli et al. 2009),
thus their possible impact on the determination of the star formation rate of galaxies is negligible.

A more important source of uncertainty in this
low star formation rate regime might be the non stationary conditions and the stochastic sampling of the IMF. The stationary conditions are reached whenever 
the number of massive stars responsible for the H$\alpha$ emission of a galaxy formed per unit time equals that of the same stars leaving the main sequence, 
i.e. that the star formation rate of the galaxy is constant on timescales longer that the typical age of the ionising stellar population ($\lesssim$ 10 Myr).

This stationary condition is certainly not satisfied in dwarf systems, where the total H$\alpha$ emission is due to a very limited number of ionising stars and/or 
star clusters which might turn on or off in a stochastic way during the star formation history of the galaxy (e.g. Boselli et al. 2009). Indeed, at these very 
low star formation rates, the gaps occurring in between episodes of formation of individual clusters can lead to irregular and highly bursty star formation 
histories, as suggested by models that rely on sampling of the cluster mass function (Fumagalli et al. 2011a; da Silva et al. 2012), numerical simulations 
(Flores Vel{\'a}zquez et al. 2021), and observations (Emami et al. 2019). Although somewhat sub-dominant compared to the burstiness, the stochastic sampling 
of the IMF could lead to further scatter in the luminosity of low mass galaxies. This effect is instead largely relevant for individual HII regions 
(Elmegreen 2000; Cervino et al. 2003, 2013; Cervino \& Luridiana 2004; Koda et al. 2012). The intrinsic burstiness and the sampling of the IMF are 
expected not only to introduce scatter in the correlations presented in this work, but also to induce a bias in the estimates of the SFR for most of the 
range under exam in this work (see e.g. da Silva et al. 2014). We also note that, due to the presence of HI deficient galaxies that down-scatter 
from the main sequence, the effects of burstiness and stochasticity are not readily noticeable as an increased scatter in the main sequence 
measured on the whole sample. They might, however, be at the origin of the observed increase in the scatter at low stellar masses 
in the HI gas-rich population (see Table \ref{Tabscaling}). Indeed, we recall that the intrinsic scatter of the relation is $\sim$ 0.40 dex
for the full sample of galaxies with $HI-def$ $\leq$ 0.4, $\sim$ 0.27 dex when limited to massive objects ($M_{star}$ $>$ 10$^9$ M$_{\odot}$) and $\sim$ 0.48 in
dwarf systems ($M_{star}$ $\leq$ 10$^9$ M$_{\odot}$).
We can also add that, whenever the few emitting sources are located far away each other on an extended stellar disc, the adopted flux extraction procedure,
which uses a large aperture, gives large uncertainties because of a possible variation of the background within the aperture.
Overall these effects might increase the scatter of the main sequence relation at low luminosities.

\subsection{An observational constrain for galaxy evolution at low stellar mass}

The main sequence relation derived in this work is not optimal to study the mass quenching phenomenon (e.g. Peng et al. 2010) nor the downsizing effect observed in
nearby late-type galaxies (e.g. Cowie et al 1996; Gavazzi et al. 1996; Boselli et al. 2001) just because a) the volume of the local universe sampled by VESTIGE 
is too limited to include a sufficient number of massive ($M_{star}$ $\geq$ 10$^{11}$) objects and b) it includes a massive cluster, where environmental quenching is at place.
The perturbations induced by the cluster environment can be identified and removed by limiting the analysis to the HI gas-rich systems 
($HI-def$ $\leq$ 0.4, see Boselli et al. (2022a) for details). Indeed, when limited to this subsample of objects, the main sequence relation has properties (slope, zero point,
and dispersion, see Table \ref{fit}) comparable to those observed in other local samples. Despite its very limited statistics (56 objects), however, the VESTIGE sample of unperturbed galaxies 
is the only one which covers with sufficient statistics the very low stellar mass range of dwarfs ($M_{star}$ $\leq$ 10$^8$ M$_{\odot}$), posing thus important constraints 
to models of galaxy formation and evolution. It confirms that the efficiency in transforming the atomic gas reservoir into stars is reduced in these objects as already 
inferred from the analysis of limited samples of local systems (e.g. Roychowdhury et al. 2009, 2014; Filho et al. 2016). It also shows that the main sequence relation has a 
constant slope down to $M_{star}$ $\simeq$ 5 $\times$ 10$^6$ M$_{\odot}$, consistent with the predictions of the most recent Illustris TNG50 simulations, now able 
to trace this relation down to $M_{star}$ $\simeq$ 10$^7$ M$_{\odot}$ (Pillepich et al. 2019). Finally, it also suggests that the feedback of supernovae often invoked
to reduce gas infall and inhibit the formation of new stars in dwarf systems (e.g. Chevalier \& Clegg 1985; Kauffmann et al. 1993; Cole et al. 1994), if present, does not have a spectacular effect 
as the one due to AGN feedback proposed to explain the bending of the main sequence at high stellar masses (Bouche et al. 2013; Lilly et al. 2013; 
see however Gavazzi et al. 2015; Erfanianfar et al. 2016)\footnote{However, a possible bias might be present: if the feedback of supernovae is very efficient and
able to remove all the gas, the H$\alpha$ emission of these dwarf galaxies would become null in $\simeq$ 10 Myr, the typical age of ionising stars.
These objects would not be detected by VESTIGE.}.

\subsection{An evolutionary picture in a rich environment}

The VESTIGE sample is perfectly suited to study the quenching process in a high density environment.
The analysis presented in Sec. 5.3 clearly indicates that the scatter of the main sequence relation is tightly related to the total amount of atomic hydrogen, 
confirming previous results in more limited samples (e.g. Gavazzi et al. 2013; Boselli et al. 2015). The atomic gas, being located on a disc much more extended than the 
stellar disc (e.g. Cayatte et al. 1994) and thus weakly bound to the gravitational potential well of the galaxy is a baryonic component easily removed during the dynamical 
interaction of galaxies with the surrounding intracluster medium (ram pressure). Indeed, analytical considerations (Domainko et al. 2006; Hester 2006) 
and hydrodynamic simulations (Abadi et al. 1999; Quilis et al. 2000; Roediger \& Hensler 2005; Roediger \& Br{\"u}ggen 2007) consistently indicate 
that the stripping process occurs outside-in, producing truncated gaseous discs. The lack of gas quenches the activity of star formation in the outer regions, 
producing truncated stellar discs in the youngest stellar populations, including those able to ionise the surrounding medium (Boselli et al. 2006; Crowl \& Kenney 2008; Fossati et al. 2018).
The pertubed galaxies thus quench their activity of star formation, moving below the main sequence relation. The increasing difference in the gas consumption timescale
with stellar mass observed between HI-normal and HI-deficient galaxies might suggest that, thanks to their deep gravitational potential well,
massive systems are stripped of their losely bound diffuse HI gas while still retain their molecular gas component while in dwarf systems both gas phases are efficiently removed.
The position of galaxies within the phase space diagram
suggests that the gas stripping process and the following quenching of the star formation activity are rapid and able to perturb the evolution of galaxies at
the periphery of the cluster. A rapid quenching of the star formation activity ($\tau$ $\lesssim$ 1 Gyr),
suggested by our ram pressure stripping models (see Fig. \ref{mainmodel}), is consistent with all the timescales estimates derived from the analysis of stellar populations
of statistical samples of Virgo galaxies (Crowl \& Kenney 2008; Boselli et al. 2008a, 2014a, 2016b) or with the detailed study of selected perturbed objects 
(e.g. Vollmer et al. 2004, 2008a,b, 2012, 2018; Boselli et al. 2006, 2021; Fossati et al. 2018). The presence of gas-deficient objects in the infalling regions,
however, might also indicate that part of the gas has been removed while the galaxies were members of infalling groups or within the filaments (pre-processing;
Dressler 2004; Fujita 2004; Cybulski et al. 2014). Indeed, HI-poor galaxies have been detected in several filaments around the Virgo cluster (e.g. Castignani et al. 2022;
Cattorini et al. 2022).

The dominant effect of ram pressure stripping is thus that of reducing the activity of star formation in perturbed galaxies. As extensively discussed in
Boselli et al. (2022a), however, under some particular conditions, the star formation activity can be enhanced during the interaction
as indeed suggested by hydrodynamic simulations (Fujita \& Nagashima 1999; Bekki \& Couch 2003; Bekki 2014; Steinhauserer et al. 2012, 2016; 
Henderson \& Bekki 2016; Steyrleithner et al. 2020; Lee et al. 2020; Troncoso-Iribarren et al. 2020).
This occurs when the impact parameters of the galaxy during its orbit within the cluster are such to keep the compressed gas on the disc of the galaxy, favoring star formation.
The very limited number of galaxies above the main sequence suggests that a) these conditions are very rare in a cluster such as Virgo ($\simeq$ 10$^{14}$ M$_{\odot}$), and b) if this occurs, 
the starburst phase is a short living phenomenon. The detailed analysis of some representative objects such as IC 3476, which is an active star forming galaxy located above the main
sequence (see  Fig. \ref{mainname}), also suggests that the compression of gas favoring star formation is a local phenomenon which thus can hardly increase
the overall activity of the perturbed object (Boselli et al. 2021). Since the efficiency of ram pressure stripping increases with the mass of the cluster, 
we expect both a more rapid and drastic quenching phenomenon and a higher fraction of objects above the main sequence in clusters as massive as A1367 and 
Coma ($\simeq$ 10$^{15}$ M$_{\odot}$), as indeed observed (e.g. Boselli et al. 2022a; Pedrini et al. 2022). The fact that most of the jellyfish galaxies of the GASP sample (Poggianti et al. 2017)
are located above the main sequence (Vulcani et al. 2018) rather suggests that they form a particular population of objects where the interaction with the surrounding
environment trigger rather than decrease the star formation activity. The star forming regions formed in the tails of perturbed galaxies such as those observed in NGC 4254 
(Boselli et al. 2018b) might be a representative case of these unstable systems. Created within the turbulent regions of the stripped gas, they form stars during a single, 
short-living ($\lesssim$ 10 Myr) burst episode. They rapidly become quiescent systems once the molecular gas is transformed into stars, decoupled from
the newly formed stars since still suffering ram pressure, or changing of phase because of heat conduction and mixing with the surrounding IGM 
(Fumagalli et al. 2011b; Arrigoni-Battaia et al. 2012; Jachym et al. 2014; Boselli et al. 2018b; Cramer et al. 2019). Indeed, most of these regions are not detected in H$\alpha$, 
but are bright in the FUV and NUV bands.

A final consideration from the present analysis is that the observed scatter in the main sequence relation cannot be done by a slow quenching of the star formation activity
do to the gas exhaustion via star formation (starvation). Indeed, the stop of gas infall necessary to sustain star formation in isolated galaxies and invoked in 
the cluster environment when a galaxy becomes satellite of a larger halo requires extremely long timescales (several Gyr) to have a measurable effect. At the same time it
cannot produce the most quenched objects observed within the Virgo cluster. This is probably due to the fact that a) the total gas available on the disc, including the one locked 
up in the molecular phase, can sustain star formation rate for several Gyr as shown in Sec. 4 (see also Kennicutt 1989, 1998; Wong \& Blitz 2002; Leroy et al. 2008, 2013; 
Bigiel et al. 2008; Boselli et al. 2014c), b) an important fraction of recycled gas available to sustauin star formation is produced by
the evolved stellar populations inhabiting the disc (e.g. Kennicutt 1994), and c) the gas is not in equilibrium within the disc, 
and is thus not always available to form new stars (e.g. Semenov et al. 2017). This gas can contribute to feed the star formation process during the slow evolution of starved galaxies.
We recall, however, that the consumption of the available gas can be speed up in the presence of outflows. The efficiency of outflows, indeed, could be increased once galaxies become
satellites of a larger halo if some of the infalling gas is removed during the interaction (McGee et al. 2014; Balogh et al. 2016; Trussler et al. 2020). 
The external pressure exerted by the surrounding ICM, however, could limit this effect by preventing outflows to leave the disc.

\section{Conclusion}

The VESTIGE survey, a H$\alpha$ NB imaging survey of the Virgo cluster up to its virial radius ($\sim$ 104${\degr}$$^2$), detected the ionised hydrogen emission
in 384 Virgo cluster members.
We used this unique set of data to derive star formation rates, specific star formation rates, and HI gas consumption timescales for all these objects
and reconstruct their typical scaling relations for galaxies with 10$^6$ $\lesssim$ $M_{star}$ $\lesssim$ 3 $\times$ 10$^{11}$ M$_{\odot}$ and 
10$^{-5}$ $\lesssim$ $SFR$ $\lesssim$ 10 M$_{\odot}$ yr$^{-1}$, a dynamic range never explored so far. After identifying unperturbed galaxies 
according to their atomic gas content, we show that the main sequence relation has a slope $a$ = 0.92$\pm$0.06 constant over the whole dynamic range of 
stellar mass, and has an intrinsic dispersion $\sigma$ = 0.40.

The data clearly show a segregation of galaxies
within the main scaling relation as a function of their atomic gas content, with HI gas-poor systems having systematically lower star formation rates and 
specific star formation rates than objects of similar stellar mass or stellar mass surface density. On the contrary, HI gas-poor and gas-rich systems have,
on average, comparable consumption timescales. HI gas-poor systems lie below the main sequence relation drawn by unperturbed, gas rich systems, with a distance
increasing with the HI-deficiency parameter. All these results suggest that the lack of HI gas, removed during the interaction of galaxies with their
hostile surrounding environment, induces a decrease in the activity of star formation. Although not directly connected to the star formation process
within HII regions, the HI gas is the principal gas reservoir feeding giant molecular clouds where star formation takes place. Its removal from the outer disc
thus indirectly affects the overall star formation activity of the perturbed systems.

By comparing the main sequence relation with the prediction of 2D spectro-photometric physical models of galaxy evolution expressly tuned to take into account 
the effects induced by the interaction of galaxies with their surrounding environment we showed that the observed scatter in the relation cannot be reproduced
by a mild decrease of the star formation activity due to the passive consumtion via star formation of the total gas reservoir not replenished by external gas infall 
(starvation). It rather requires a more active process able to actively remove the gas reservoir from the galaxy disc, quenching the activity of star 
formation outside-in on relatively short ($\lesssim$ 1 Gyr) timescales (ram pressure). These short timescales for gas removal and star formation quenching are 
also consistent with the position of galaxies within the phase-space diagarams, where HI-deficient objects are already dominant in the first infall region.

We have also shown that the compact HII regions formed outside the galaxy discs within the tail of stripped material do not follow the main sequence relation since 
have star formation rates 2-to-3 orders of magnitude above those expected for their stellar mass. This enhanced star formation activity is due to a 
short living ($\lesssim$ 50 Myr) starburst episode intended to fade producing passive compact objects such as globular clusters and UCD galaxies typical in rich environments.

\begin{acknowledgements}

We thank the anonymous referee for the accurate reading of the manuscript and for constructive comments and suggestions.
We are grateful to the whole CFHT team who assisted us in the preparation and in the execution of the observations and in the calibration and data reduction: 
Todd Burdullis, Daniel Devost, Bill Mahoney, Nadine Manset, Andreea Petric, Simon Prunet, Kanoa Withington.
We acknowledge financial support from ``Programme National de Cosmologie and Galaxies" (PNCG) funded by CNRS/INSU-IN2P3-INP, CEA and CNES, France,
and from ``Projet International de Coop\'eration Scientifique" (PICS) with Canada funded by the CNRS, France.
This research has made use of the NASA/IPAC Extragalactic Database (NED) 
which is operated by the Jet Propulsion Laboratory, California Institute of 
Technology, under contract with the National Aeronautics and Space Administration
and of the GOLDMine database (http://goldmine.mib.infn.it/) (Gavazzi et al. 2003).
This work was partially funded by the ANID BASAL project FB210003. MB acknowledges support from FONDECYT regular grant 1211000.
AL is supported by Fondazione Cariplo, grant No 2018-2329.

\end{acknowledgements}

\begin{appendix}

\section{Variations with morphological type}

\begin{figure}
\centering
\includegraphics[width=0.5\textwidth]{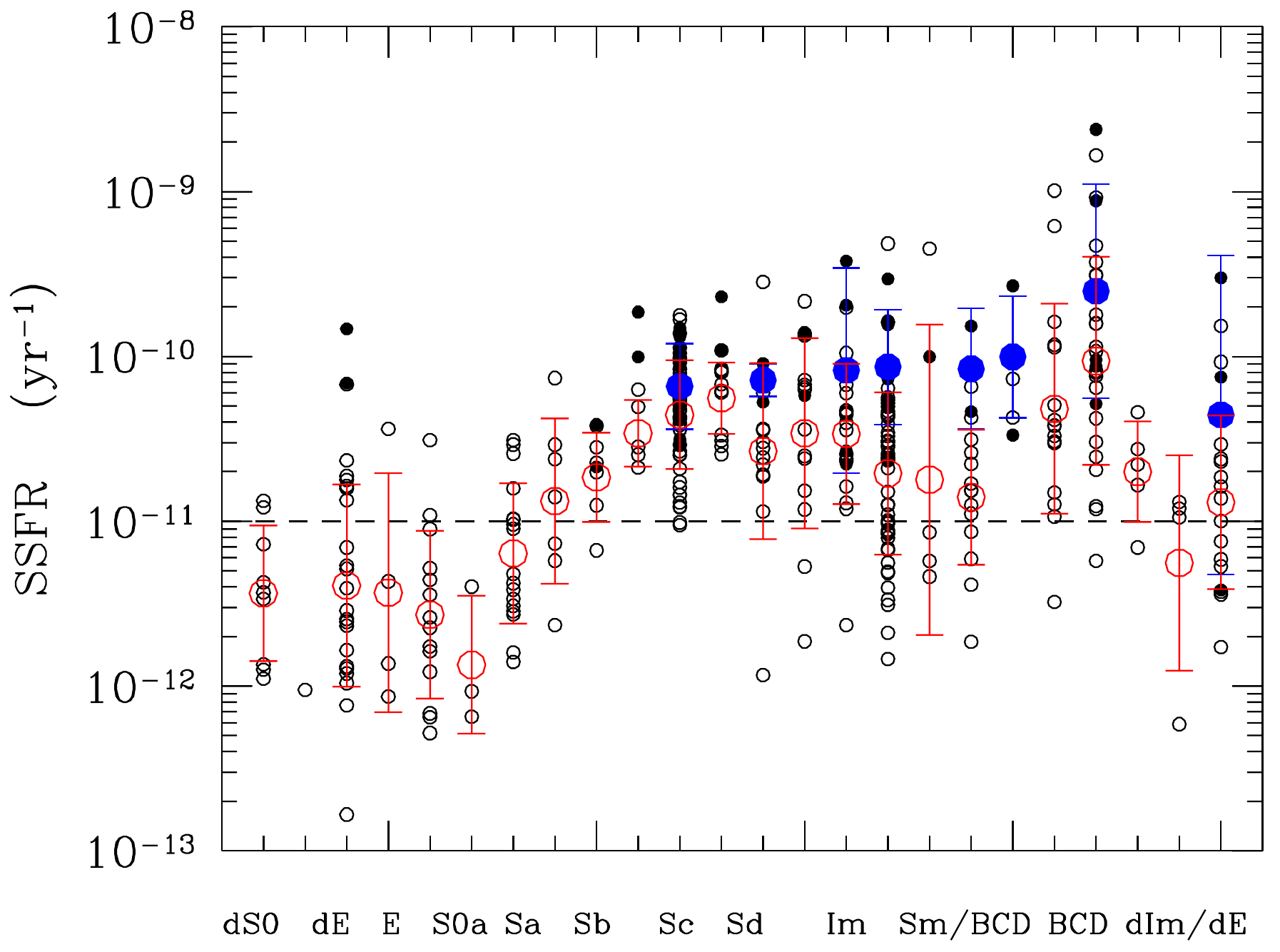}
\caption{Relationship between the specific star formation rate and the morphological type for HI-normal ($HI-def$ $\leq$0.4; filled dots) and 
HI-deficient ($HI-def$ $>$0.4; empty circles) galaxies. The large filled blue dots indicate the mean values for each morphological class for normal 
gas-rich systems and the empty red ones for cluster HI-deficient galaxies. For the large symbols, the error bar shows the standard deviation of the distribution.
The dashed line shows the limit between star forming and quiescent galaxies.
}
\label{sfrtype}%
\end{figure}

\begin{figure}
\centering
\includegraphics[width=0.5\textwidth]{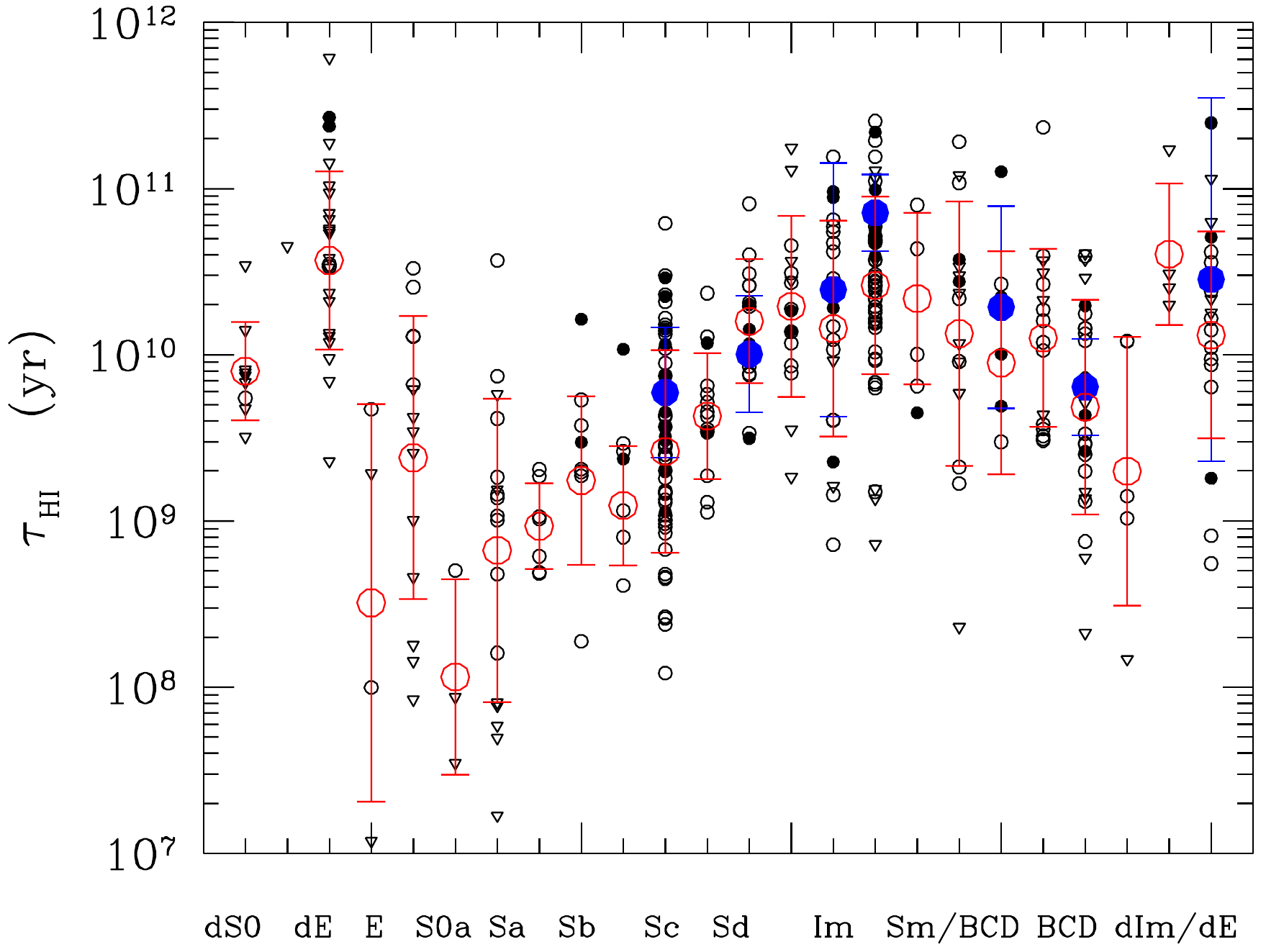}
\caption{Relationship between the HI gas depletion timescale and the morphological type for HI-normal ($HI-def$ $\leq$0.4; filled dots) and 
HI-deficient ($HI-def$ $>$0.4; empty circles) galaxies. The large filled blue dots indicate the mean values for each morphological class for normal 
gas-rich systems and the empty red ones for cluster HI-deficient galaxies. For the large symbols, the error bar shows the standard deviation of the distribution.
Upper limits are indicated by triangles and are treated as detections in the derivation of the mean values.}
\label{tauHItype}%
\end{figure}

The extraordinary quality of the images available for Virgo galaxies allows an accurate determination of 
the morphological type (e.g. Binggeli et al. 1985). Since in the local universe the star formation properties of galaxies
have been historically studied in different morphological classes (e.g. Kennicutt et al. 1994; Roberts \& Haynes 1994), 
we plot here the variation of the specific star formation rate
(Fig. \ref{sfrtype}) and of the gas depletion timescale (Fig. \ref{tauHItype}) as a function of the morphological type.
They can be compared to those derived for the \textit{Herschel} Reference Survey presented in Boselli et al. (2014b, 2015).

As expected, Fig. \ref{sfrtype} shows that the activity of star formation per unit stellar mass is significantly higher in spiral galaxies than in H$\alpha$ detected 
early-type systems. These early-types are just a small fraction of the quiescent galaxy population dominant in the Virgo cluster (Sandage et al. 1988). They are characterised by
a residual star formation activity and are probably representative of the population of galaxies at an evolved stage of their transformation from the blue cloud
to the red sequence (e.g. Boselli et al. 2014a). The same figure also shows that, on average, the specific star formation rate of gas-rich systems of similar morphological type
is higher than that of gas-poor objects. Finally, despite the much improved quality of the data, of the adopted dust attenuation corrections, and of the morphological classification, 
it is clear that also among the unperturbed sample the intrinsic scatter within each morphological type is huge, and confirms the fact that galaxies of similar morphology undergo very
different histories of star formation (e.g. Kennicutt et al. 1994). 
The lack of gas-rich systems with morphological type earlier than Sc and the large fraction of HI undetected systems among the early-types hampers
the identification of any clear trend between the gas depletion timescale and the morphological type (Fig. \ref{tauHItype}). For galaxies of type $>$ Sc 
the difference between $\tau_{HI}$ in gas-rich and gas-poor systems is marginal. 

\section{Galaxies above the main sequence relation}

\begin{figure*}
\centering
\includegraphics[width=0.85\textwidth]{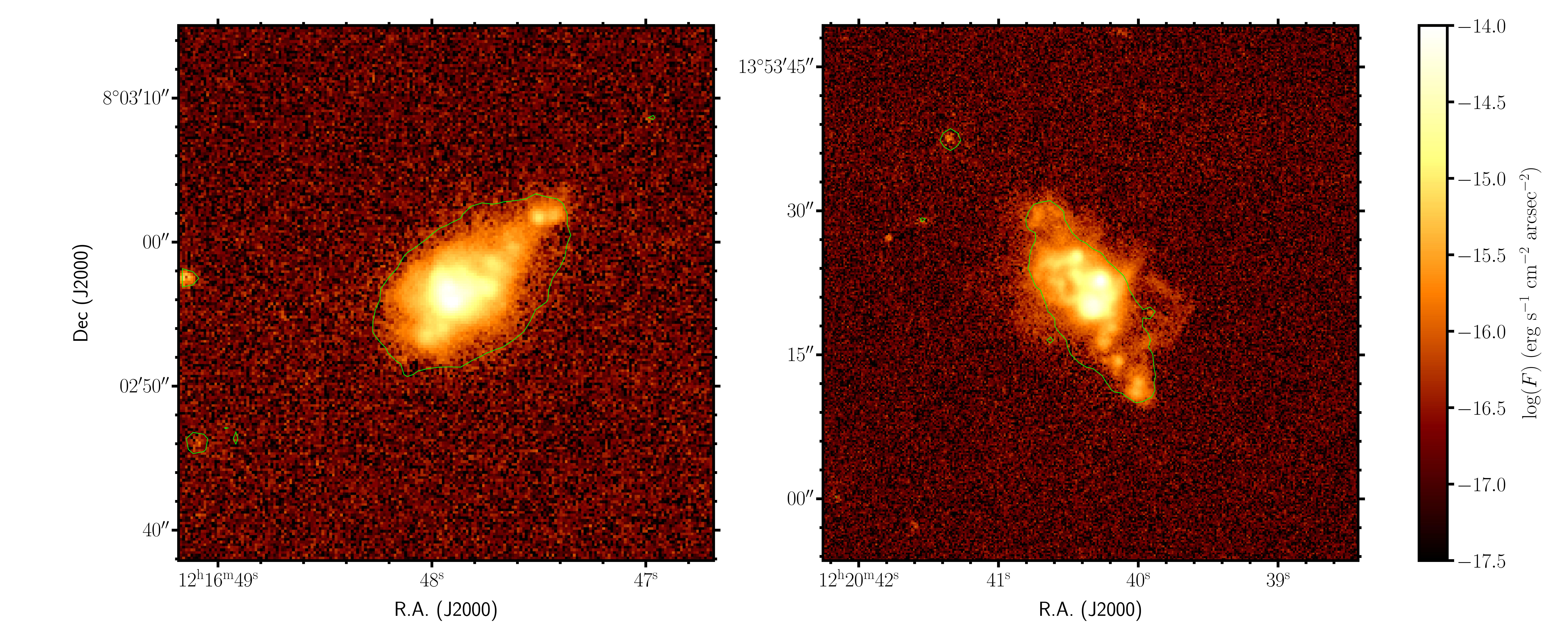}\\
\includegraphics[width=0.85\textwidth]{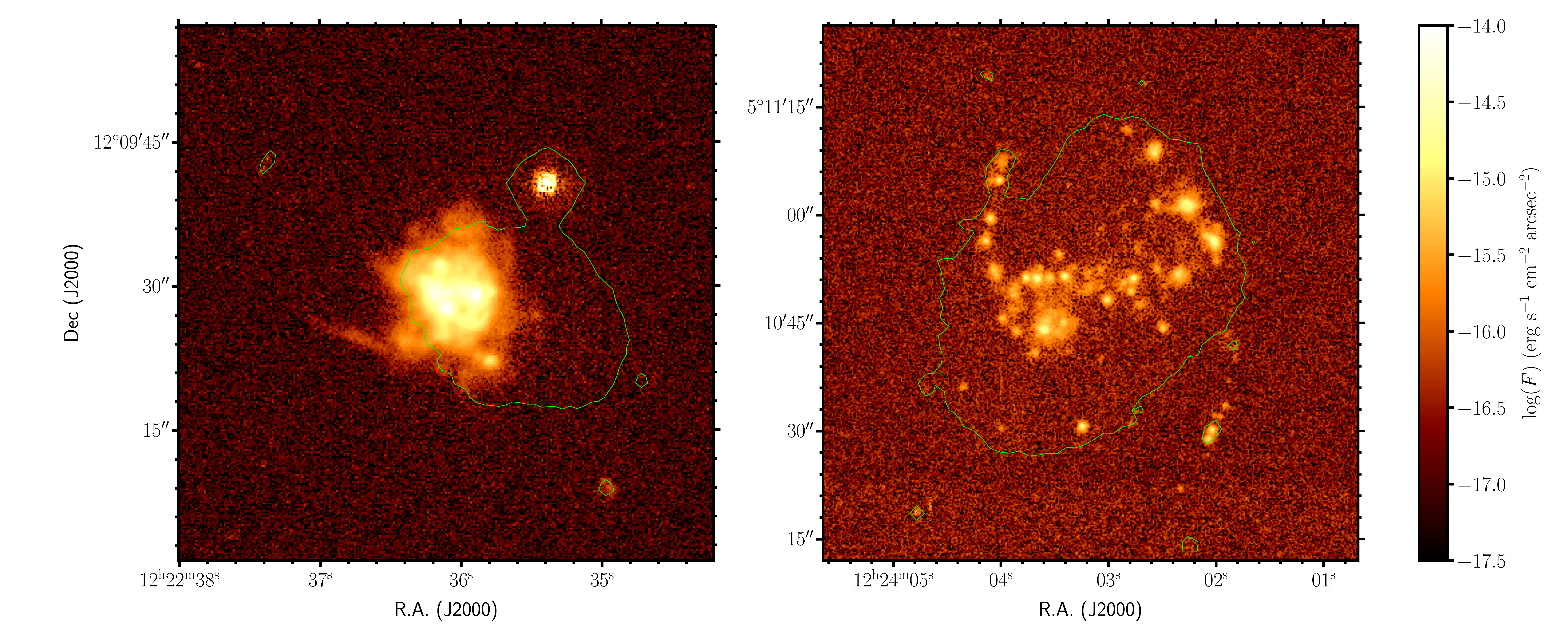}\\
\includegraphics[width=0.85\textwidth]{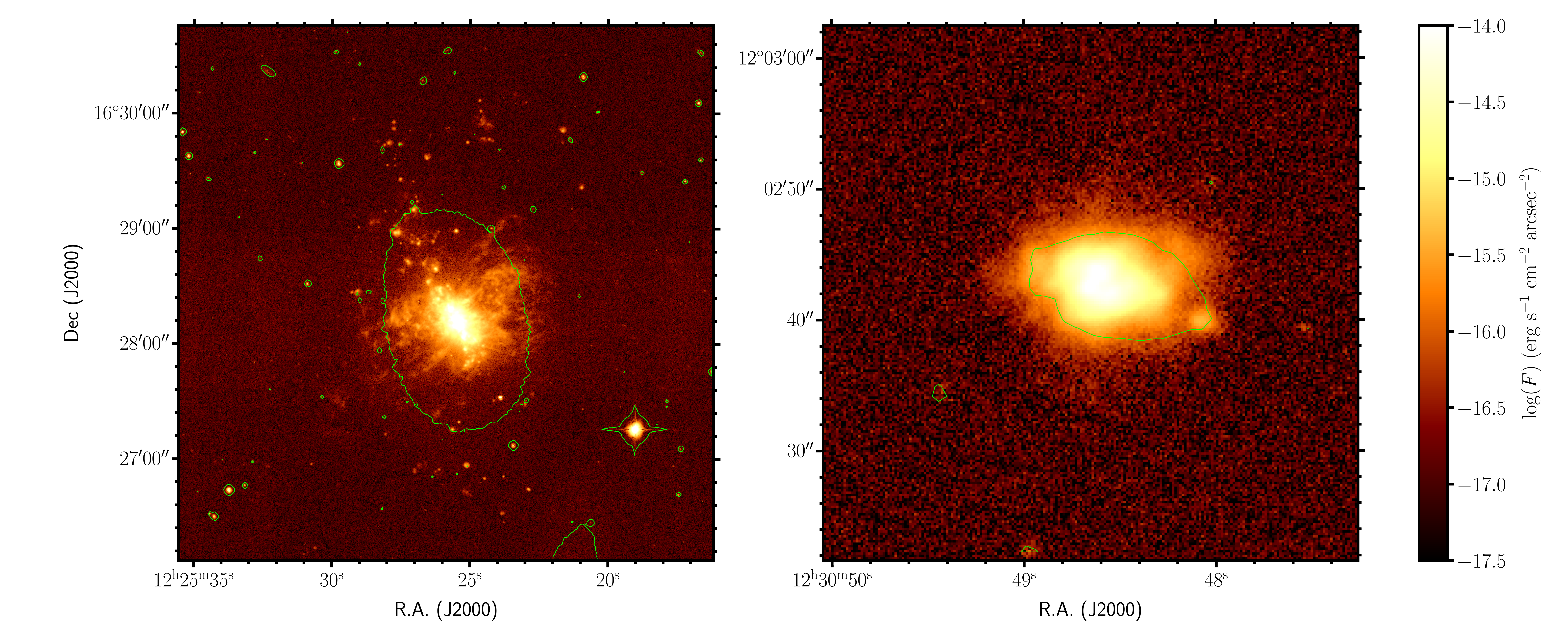}\\
\includegraphics[width=0.85\textwidth]{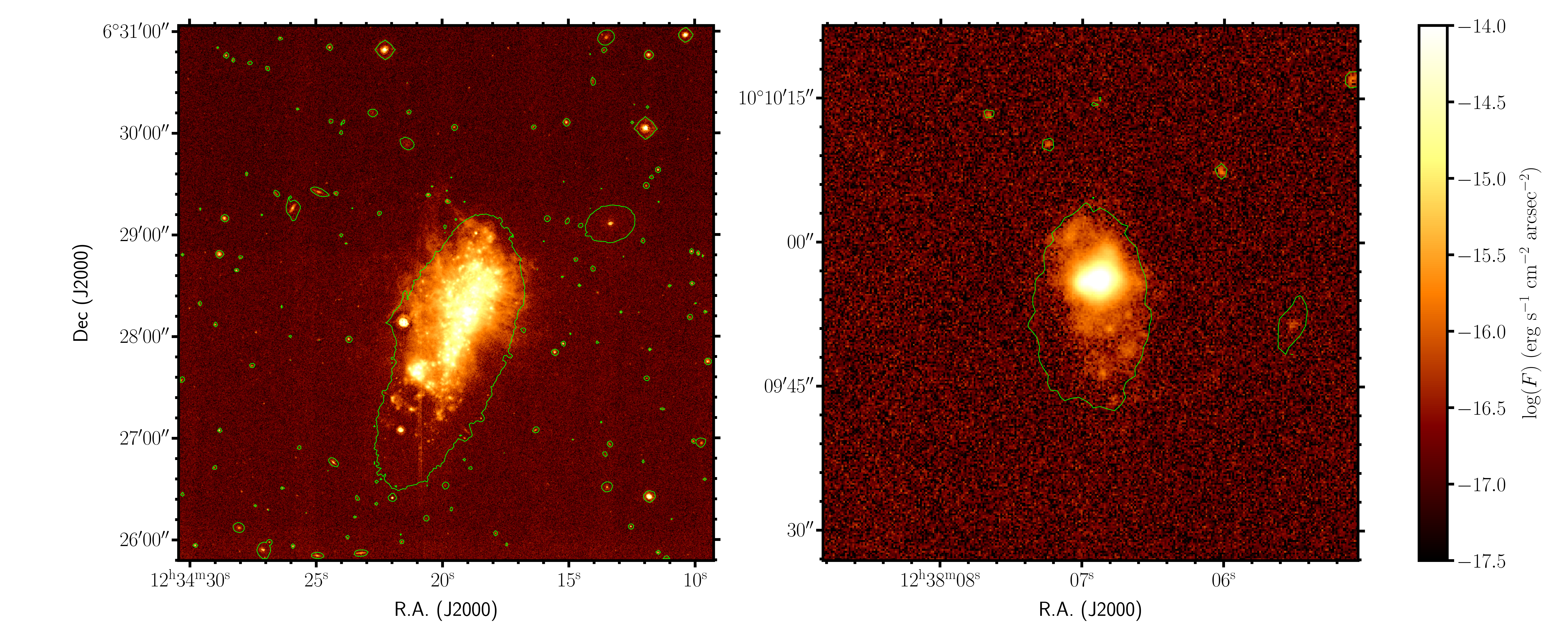}\\
\caption{Continuum-subtracted H$\alpha$ images of the galaxies VCC 207, VCC 428, VCC 562, VCC 683, VCC 801, VCC 1313, VCC 1554, VCC 1744 (from top left to bottom right) 
with overplotted in green contours at the $r$-band 24 mag arcsec$^{-2}$ isophote tracing the distribution of the stellar continuum.}
\label{figaboveMS}%
\end{figure*}

Figure \ref{figaboveMS} shows the continuum-subtracted H$\alpha$ images of the 10 galaxies located more than 1$\sigma$ above the main sequence relation
and listed in Table \ref{aboveMS}. In most of them the star formation activity is dominated by one or a few bright HII regions typical of starburst. 
Some of them (VCC 428, VCC 801, VCC 1554) show filaments or loops of ionised gas escaping from these bright HII regions and extending out of the stellar disc, 
suggesting the presence of outflows. In others the asymmetric distribution of the star forming regions and the presence of tails of ionised gas rather suggest
an ongoing ram pressure stripping event (VCC 562, AGC224696, AGC 226326).

\begin{figure*}
\centering
\includegraphics[width=0.85\textwidth]{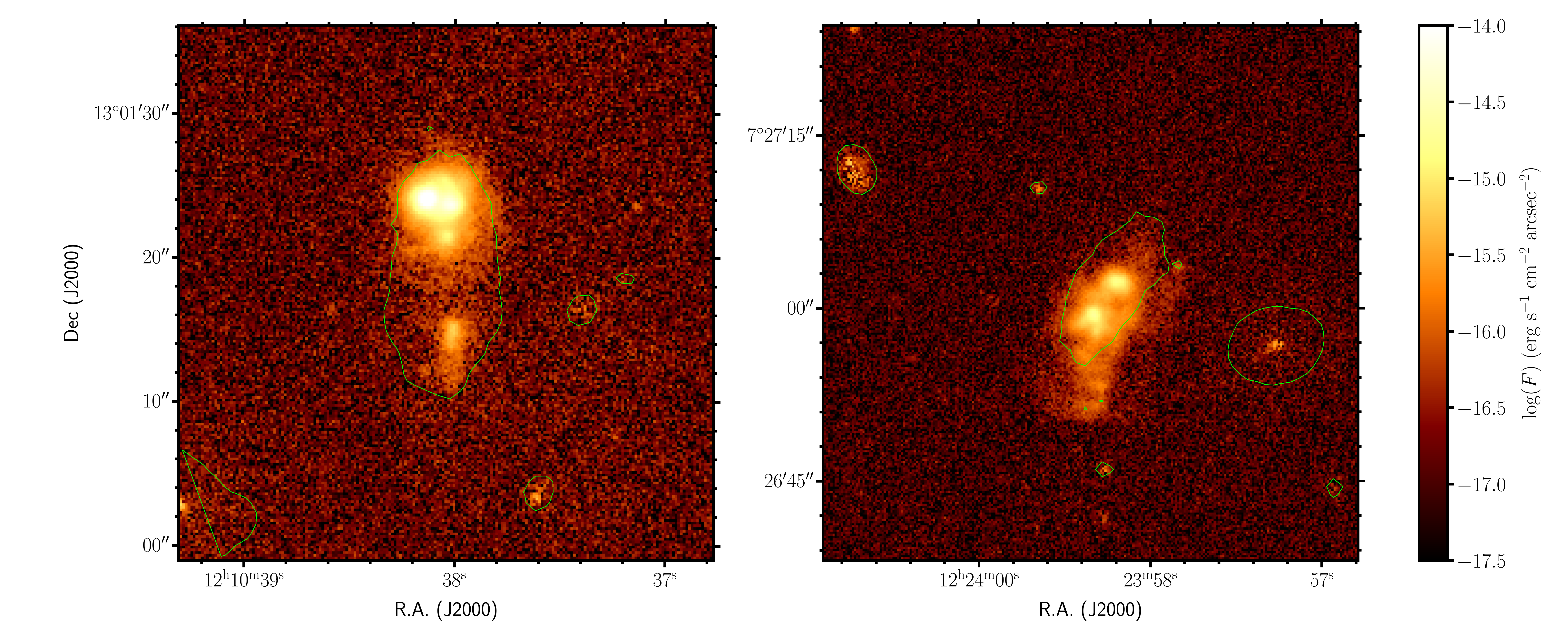}\\
\caption{Continuum-subtracted H$\alpha$ images of the galaxies AGC 224696, AGC 226326 (from left to right)
with overplotted in green contours at the 
$r$-band 24 mag arcsec$^{-2}$ isophote tracing the distribution of the stellar continuum.}
\end{figure*}

\end{appendix}

\end{document}